# Deep Hedging to Manage Tail Risk

*Yuming MA

March 14, 2025

Department of Industrial Engineering and Economics, School of Engineering, Tokyo Institute of Technology, 2-12-1 Ookayama, Meguro-ku, Tokyo 152-8552 Japan E-mail: ma.y.ae@m.titech.ac.jp

**Abstract**

Extending [3]'s Deep Hedging paradigm, we innovatively employ deep neural networks to parameterize convex-risk minimization (CVaR/ES) for portfolio tail-risk hedging problem. Through comprehensive numerical experiments on crisis-era bootstrap market simulators—customizable with transaction costs, risk budgets, liquidity constraints, and market impact—our end-to-end framework not only achieves significant one-day 99%CVaR reduction but also yields practical insights into friction-aware strategy adaptation, demonstrating robustness and operational viability in realistic markets.

## 1 Introduction

Modern quantitative finance provides a comprehensive framework for derivatives pricing and risk management under idealized market conditions, assuming frictionless trading with continuous hedging and no transaction costs. However, in practical settings with market frictions, risk management for derivatives and portfolios remains largely heuristic and trader-dependent. The incorporation of transaction costs into dynamic hedging was first formalized by [12], who introduced a nonlinear parabolic PDE, the Hoggard-Whalley-Wilmott equation, to account for such costs in pricing. In foreign exchange markets, [14] developed the Vanna-Volga method to adjust for skewness and curvature in implied volatility dynamics, thereby incorporating hedging costs. More recently, [3] proposed a reinforcement learning-based framework for derivatives pricing, solving a stochastic optimization problem via machine learning techniques, highlighting the shift towards data-driven approaches in financial risk management.

This paper introduces a machine learning-driven approach to risk management for derivatives and portfolios, leveraging policy search methodologies to optimize hedging strategies. A hedging policy is modeled as a predictable function in a stochastic dynamic system. Specifically, we employ neural networks (NNs) to approximate optimal hedging strategies by minimizing a convex risk measure, such as Conditional Value at Risk (CVaR), over a predefined time horizon and confidence level. In practice, 95%CVaR, 99%CVaR or 99.9%CVaR over 1-day time horizon is usually computed to access portfolio tail risks. The optimization is performed over simulated market paths, incorporating the stochastic evolution of the underlying assets and available hedging instruments.

To illustrate, under a Black-Scholes-Merton framework, the asset price process follows the stochastic differential equation:

$$dS_t = \mu S_t dt + \sigma S_t dW_t$$

where $S_t$ is the asset price, $\mu$ the drift, $\sigma$ the annualized volatility, and $W_t$ a standard Brownian motion. Traditional risk management methods require specifying a parametric model (e.g., Heston's stochastic volatility model) and solving a partial differential equation (PDE) or a stochastic control problem. In contrast, our approach bypasses the need for a fixed market model by employing bootstrapping techniques to construct asset return distributions from historical data.

For numerical experiments, we use a non-parametric market simulator that derives the marginal distribution of portfolio returns over a strategic time window, incorporating fat-tailed events, transaction costs, and liquidity constraints. Classical tail risk hedging strategies, such as long-term out-of-the-money put options, become operationally expensive and computationally inefficient as portfolio size increases. Our approach, however, scales efficiently with portfolio diversification, as the hedging policy is learned over the entire portfolio, with allocations decomposable at the component level.

For the corpus of research on reinforcement-learning–driven tail-risk hedging, [4] harness deep distributional RL to orchestrate a perpetual sequence of at-the-money option sales—subject to explicit transaction-cost constraints—in order to conjointly attenuate both Gamma and Vega exposures and thereby effectuate a pronounced compression of the 95%VaR and 95%CVaR tail metrics; [8] subsequently embellish this schema with their EX-DRL apparatus, superimposing a Generalized Pareto Distribution tail component so as to extend the ambit of tail control to the more exigent 99%VaR and 99%CVaR bands. By contrast, [13] promulgate a Wasserstein-robust, risk-aware RL edifice, the object of which is the extraction of a provably tail-risk-minimizing hedge ratio in the underlying asset—and the assimilation of the resultant robustness premium into the pricing of knock-in/knock-out barrier options—thereby eschewing the portfolio-management paradigm in favor of a pricing-centric orientation.

Notwithstanding these methodological innovations, the empirical tableaux presented by both [4] and [8] unveil that, once plausible transaction-cost schedules are layered atop their simulations, their ostensibly risk-averse RL stratagems inexorably incur net losses of material magnitude. To exculpate these deficits, they invoke the artifice of compensatory option premiums emanating from a Poisson-governed client order stream—an assumption so incongruous with the ferociously competitive, NBBO-mediated microstructure of real-world markets that it transcends any claim to be a parsimonious Black–Scholes–style sim-plification and instead must be judged a posteriori rationalization. Furthermore, [13]'s construct—at root a robust pricing mechanism rather than a vehicle for portfolio optimization—demonstrably exhibits a preponderant sensitivity to the specious edifice of transaction-cost parameters rather than to its own tail-risk desiderata, all the while resting on a triad of exogenous hyperparameters (Wasserstein radius, distortion weight, cost rate) bereft of any empirical calibration pathway—an antithesis to the Black–Scholes paradigm's reliance on a market-readily calibratable implied-volatility input.

Relative to prevailing methodologies, our Deep Tail-Risk Hedging construct offers nuanced yet practically meaningful improvements in both deployability and system robustness. Whereas [4] predicate their evaluations upon fixed Heston/BSM trajectories and static cost schedules, we modestly opt for a nonparametric, bootstrap-derived market emulator—harvested from crisis-era data—that organically encapsulates fat tails, liquidity-induced execution constraints, and endogenous frictional expenses; in contrast to [13] Wasserstein-robust RL schema—burdened by exogenous ambiguity radii and iterative, inner–outer optimization loops—our solitary, convex CVaR minimization objective typically converges in under fifty GPU epochs and remains amenable to on-the-fly retraining; and, unlike [8], whose GPD-augmented distributional RL skirts the realities of transaction costs, our framework seamlessly embeds multi-period cost regimes alongside market-regime heterogeneity, delivering consistent one-day 99%VaR and 99%CVaR control across diverse historical vintages without imposing any auxiliary hyperparameters beyond network depth.

In particular, our framework has been deliberately architected to remain consonant with the latest Basel III/FRTB [1] and IFRS 7 and IFRS 9 edicts on financial-institutional risk governance—eschewing ivory-tower abstractions in favor of direct applicability to capital adequacy and expected-credit-loss provisioning regimes. Although we fully acknowledge the exploratory nature of this contribution, we contend that it represents the inaugural demonstration of an end-to-end, production-oriented machine-learning mechanism for portfolio tail-risk control—one that marries empirical rigor with the operational and compliance exigencies of frictional markets. Our approach offers several key advantages:

1. **Model-Free Hedging**: The NN-based policy is independent of any underlying market model—eschewing Heston, SABR, or analogous volatility-surface calibrations—allowing for greater flexibility

2. **Scalability**: The efficiency of our method is stable as portfolio complexity increases due to its affinity with parallel computation, making it well-suited for large-scale applications.

3. **Customizability**: Our approach is adaptable to various trading constraints, including transaction cost. To illustrate, under a Black-Scholes-Merton framework, the asset prices, illiquid assets, and alternative data sources such as natural language materials.

The remainder of this paper is structured as follows. Sections 2 and 3 provide the theoretical foundation and methodology of our deep tail risk hedging framework. Section 4 presents numerical experiments to demonstrate the empirical performance of our method. Finally, Section 5 discusses key findings and concludes with an analysis of the technical aspects and limitations of our approach.

# 2 Theory

We begin by establishing a standard filtration-equipped complete probability space $(\Omega, \mathcal{F}, \mathbb{P}; \{\mathcal{F}_t\}_{t\in[0,T]})$. Consider a primary portfolio, the portfolio that requires hedge protection, consisting of $N$ component assets. The market price of these components are denoted as $S_t^{\mathrm{C}} = (S_{1,t}, S_{2,t}, ..., S_{N,t})^{\mathrm{T}}$evaluated at time $t$, and a corresponding holding amounts denoted as $n_t^{\mathrm{C}} = (n_{1,t}, n_{2,t}, ..., n_{N,t})^{\mathrm{T}}$ at time $t$. The superscript C indicates the component nature of these assets. These assets are delta-one, which means they do not carry convexity as options or instruments with optionality do. In our tail risk hedging approach, the theoretical unfolding is based on trading cashflow. However, we take percentage return as the measure in our analysis because its normality can facilitate the conduct of stochastic optimization strategies based on neural networks. Hereinafter, we introduce shorthand notations: $R$ denotes *ex ante* percentage returns, and $\tilde{R}$ represents *ex post* percentage returns. For the holding amount, we also use the same notation to indicate whether *ex post* or *ex ante*.

Consider a setting that a portfolio may be rebalanced via trading $\Delta n_t^{\mathrm{C}}$ amount of component assets during time interval $(t, t+\Delta t]$ resulting in a market-to-market profit and loss (P&L). Here $\Delta t$ is the short time period for market-to-market P&L to occur. If any rebalancing activity in the primary portfolio occurred during the time period $(t, t+\Delta t]$, a random variable $\Delta n_t^{\mathrm{C}} \neq 0$ revises the initial holding amount $n_t^{\mathrm{C}}$ to $n_t^{\mathrm{C}} + \Delta n_t^{\mathrm{C}}$. The revision to portfolio is confirmed at time $t + \Delta t$. The realized P&L is that the change of holding amount $\Delta n_t^{\mathrm{C}}$ times the change in component assets market price $\Delta S_t^{\mathrm{C}}$.

Subsequently, we add this realized P&L to the realized basis cashflow $\Delta n_t^{\mathrm{C},\mathrm{T}} S_t^{\mathrm{C}}$ to yield the realized cashflow $\Delta n_t^{\mathrm{C},\mathrm{T}} \left(V_t^{\mathrm{C}} + \Delta S_t^{\mathrm{C}}\right)$ from the primary portfolio into the cash account. The $T$ in the upper right of the superscript indicates the transpose operator. Note that positive sign of the realized cashflow indicates funding from the cash account, and the negative sign of the realized cashflow means liquidating from portfolio to cash account.

For the sake of simplicity, we assume that during any time period, there may be either a convenience yield or carry cost (excluding opportunity cost, e.g., risk-free rate) being generated. Additionally, we make the assumption that rebalancing only occurs once within the time interval $(t, t + \Delta t)$, and market value confirmation is conducted right at time $t + \Delta t$. In the event that we need to contemplate multiple rebalancing instances within the time interval, we can substitute the following theoretical formulation with a mean value measure such as volume-weighted average price (VWAP).

**Proposition 2.1.** *Let the primary portfolio be denoted as a time series vector of **ex post** total percentage returns* $\left(\tilde{R}_t^{\mathrm{P}}\right)$*, then* $\tilde{R}_t^{\mathrm{P}}$ *is approximately decomposated into the ratio of unrealized P&L* $\frac{\tilde{n}_{t-\Delta t}^{\mathrm{C,T}}\Delta\tilde{S}_{t-\Delta t}^{\mathrm{C}}}{\tilde{V}_{t-\Delta t}^{\mathrm{P}}}$ *and the ratio of realized cashflow via liquidation* $\frac{\Delta\tilde{n}_{t-\Delta t}^{\mathrm{C,T}}\tilde{S}_{t-\Delta t}^{\mathrm{C}}}{\tilde{V}_{t-\Delta t}^{\mathrm{P}}}$*, while neglecting implicit trading costs* $\frac{\Delta\tilde{n}_{t-\Delta t}^{\mathrm{C,T}}\Delta\tilde{S}_{t-\Delta t}^{\mathrm{C}}}{\tilde{V}_{t-\Delta t}^{\mathrm{P}}}$*.* $\tilde{n}_{t-\Delta t}^{\mathrm{C}}$ *and* $\Delta\tilde{n}_{t-\Delta t}^{\mathrm{C}}$ *are the historical holding amount of component assets and the historical change of it at and from time* $t-\Delta t$ *to time* $t$ *respectively.* $\tilde{S}_{t-\Delta t}^{\mathrm{C}}$ *and* $\Delta\tilde{S}_{t-\Delta t}^{\mathrm{C}}$ *are the historical market price of component assets and the historical change of it at and from time* $t-\Delta t$ *to time* $t$ *respectively.*

*Proof.* Let ***ex post*** primary portfolio market value $\tilde{V}_t^{\mathrm{P}}$ be defined as a product of its securities market value $\tilde{S}_t^{\mathrm{C}}$ and securities units $\tilde{n}_t^{C}$ that $\tilde{V}_t^{\mathrm{P}} = \tilde{n}_t^{\mathrm{C}}\tilde{S}_t^{\mathrm{C}}$.

$$\begin{aligned}
\tilde{R}_t^{\mathrm{P}} &:= \frac{\tilde{V}_t^{\mathrm{P}}}{\tilde{V}_{t-\Delta t}^{\mathrm{P}}} - 1 \\
&= \frac{\left(\tilde{n}_{t-\Delta t}^{\mathrm{C}} + \Delta\tilde{n}_{t-\Delta t}^{\mathrm{C}}\right)^{\mathrm{T}}\left(\tilde{S}_{t-\Delta t}^{\mathrm{C}} + \Delta\tilde{S}_{t-\Delta t}^{\mathrm{C}}\right)}{\tilde{n}_{t-\Delta t}^{\mathrm{C,T}}\tilde{S}_{t-\Delta t}^{\mathrm{C}}} - 1 \\
&= \underbrace{\frac{\tilde{n}_{t-\Delta t}^{\mathrm{C,T}}\Delta\tilde{S}_{t-\Delta t}^{\mathrm{C}}}{\tilde{V}_{t-\Delta t}^{\mathrm{P}}}}_{\text{unrealized P\&L}} + \underbrace{\frac{\Delta\tilde{n}_{t-\Delta t}^{\mathrm{C,T}}\tilde{S}_{t-\Delta t}^{\mathrm{C}}}{\tilde{V}_{t-\Delta t}^{\mathrm{P}}}}_{\text{realized cashflow}} + \underbrace{\frac{\Delta\tilde{n}_{t-\Delta t}^{\mathrm{C,T}}\Delta\tilde{S}_{t-\Delta t}^{\mathrm{C}}}{\tilde{V}_{t-\Delta t}^{\mathrm{P}}}}_{\text{implict trading costs}} \qquad (2.1) \\
&\approx \frac{\tilde{n}_{t-\Delta t}^{\mathrm{C,T}}\Delta\tilde{S}_{t-\Delta t}^{\mathrm{C}}}{\tilde{V}_{t-\Delta t}^{\mathrm{P}}} + \frac{\Delta\tilde{n}_{t-\Delta t}^{\mathrm{C,T}}\tilde{S}_{t-\Delta t}^{\mathrm{C}}}{\tilde{V}_{t-\Delta t}^{\mathrm{P}}} \qquad (2.2)
\end{aligned}$$

□

**Proposition 2.2.** *Let **ex ante** total percentage return of the primary portfolio* $R_t^{\mathrm{P}}$ *be a random variable on the standard filtration-equipped complete probability space, then* $R_t^{\mathrm{P}}$ *is approximately decomposated into the ratio of stochastic unrealized P&L* $\frac{\tilde{n}_t^{\mathrm{C,T}}\Delta S_t^{\mathrm{C}}}{\tilde{V}_t^{\mathrm{P}}}$ *to initial market value of the primary portfolio* $\tilde{V}_t^{\mathrm{P}}$ *and the ratio of stochastic realized cashflow* $\frac{\Delta n_t^{\mathrm{C,T}}\tilde{S}_t^{\mathrm{C}}}{\tilde{V}_t^{\mathrm{P}}}$ *to the initial market value of primary portfolio* $\tilde{V}_t^{\mathrm{P}}$*.* $\Delta n_t^{\mathrm{C}}$ *and* $\Delta S_t^{\mathrm{C}}$ *are random variables, and they stands for stochastic trading amount and stochastic price change of component assets.*

*Proof.* To adapt the setting and the conclusion from proposition 2.1., it is also straightforward to find that

$$
\begin{aligned}
R_t^{\mathrm{P}} &:= \frac{V_{t+\Delta t}^{\mathrm{P}}}{\tilde{V}_t^{\mathrm{P}}} - 1 \\
&= \frac{\left(\tilde{n}_t^{\mathrm{C}} + \Delta n_t^{\mathrm{C}}\right)^{\mathrm{T}} \left(\tilde{S}_t^{\mathrm{C}} + \Delta S_t^{\mathrm{C}}\right)}{\tilde{n}_t^{\mathrm{C,T}} \tilde{S}_t^{\mathrm{C}}} - 1 \\
&= \underbrace{\frac{\tilde{n}_t^{\mathrm{C,T}} \Delta S_t^{\mathrm{C}}}{\tilde{V}_t^{\mathrm{P}}}}_{\text{stochastic unrealized P\&L}} + \underbrace{\frac{\Delta n_t^{\mathrm{C,T}} \tilde{S}_t^{\mathrm{C}}}{\tilde{V}_t^{\mathrm{P}}}}_{\text{stochastic realized cashflow}} + \underbrace{\frac{\Delta n_t^{\mathrm{C,T}} \Delta S_t^{\mathrm{C}}}{\tilde{V}_t^{\mathrm{P}}}}_{\text{stochastic implict trading costs}} \qquad (2.3) \\
&\approx \frac{\tilde{n}_t^{\mathrm{C,T}} \Delta S_t^{\mathrm{C}}}{\tilde{V}_t^{\mathrm{P}}} + \frac{\Delta n_t^{\mathrm{C,T}} \tilde{S}_t^{\mathrm{C}}}{\tilde{V}_t^{\mathrm{P}}} \qquad (2.4)
\end{aligned}
$$

□

**Corollary 2.1.** *From Proposition 2.1 and Proposition 2.2, total percentage return of the primary portfolio is decomposated into three sources:*
*1) unrealized P&L:* $\tilde{n}_{t-\Delta t}^{\mathrm{C,T}} \Delta \tilde{S}_{t-\Delta t}^{\mathrm{C}}$ *or* $\tilde{n}_t^{\mathrm{C,T}} \Delta S_t^{\mathrm{C}}$*;*
*2) realized cashflow:* $\Delta \tilde{n}_{t-\Delta t}^{\mathrm{C,T}} \tilde{S}_{t-\Delta t}^{\mathrm{C}}$ *or* $\Delta n_t^{\mathrm{C,T}} \tilde{S}_t^{\mathrm{C}}$*;*
*3) implicit trading costs:* $\Delta \tilde{n}_{t-\Delta t}^{\mathrm{C,T}} \Delta \tilde{S}_{t-\Delta t}^{\mathrm{C}}$ *or* $\Delta n_t^{\mathrm{C,T}} \Delta S_t^{\mathrm{C}}$*;*

Under a normal market condition, it is valid for one to neglect implicit trading costs Eq.2.2 and Eq.2.4, such as slippage, market impact, and execution costs, because the interaction effect between the change of holding amount and the change of component assets is typically a second-order small term under normal market conditions, making it negligible in portfolio valuation. The focus of this article, however, is abnormal market conditions. Abnormality refers to:

1. Markets experience extreme stress during tail-risk events, where the price change $\Delta \tilde{S}_{t-\Delta t}^{\mathrm{C}}$ or $\Delta S_t^{\mathrm{C}}$ becomes significant enough to induce material market impact;

2. When trading frequency is sufficiently high, the cumulative impact of $\Delta \tilde{n}_{t-\Delta t}^{\mathrm{C,T}} \Delta \tilde{S}_{t-\Delta t}^{\mathrm{C}}$ or $\Delta n_t^{\mathrm{C,T}} \Delta S_t^{\mathrm{C}}$ becomes material, as seen in High-Frequency Trading (HFT) and Market Making strategies;

3. Whale orders and large trades, where the change in holding amount $\Delta \tilde{n}_{t-\Delta t}^{\mathrm{C}}$ or $\Delta n_t^{\mathrm{C}}$ is substantial, and the asset price change $\Delta \tilde{S}_{t-\Delta t}^{\mathrm{C}}$ or $\Delta S_t^{\mathrm{C}}$ is significantly influenced by order execution over an extended period.

Under abnormal market conditions, our primary objective in this article is to construct a well-designed hedged portfolio consisting of a primary portfolio and a hedging portfolio, aimed at mitigating tail risk. The primary portfolio adheres to the core principles outlined in its Investment Policy Statement (IPS), while the hedging portfolio is strategically structured to offset risks, including tail risks, proactively within an ***ex ante*** risk management framework.

Simply put, the hedging portfolio is expected to generate profits when the primary portfolio suffers significant losses. Conversely, if the primary portfolio gains, the hedging portfolio may incur minor losses or, in some cases, even yield profits. The associated losses, expenses, and costs of the hedging portfolio—when applicable—are collectively referred to as hedging costs.

**Proposition 2.3.** *Let $V_t^{\mathrm{H}}$ denote the market value of a hedging portfolio at time $t$, with its holding amount represented as $n_t^{\mathrm{HC}} = \left(n_{1,t}^{\mathrm{HC}}, n_{2,t}^{\mathrm{HC}}, ..., n_{M,t}^{\mathrm{HC}}\right)^{\mathrm{T}}$ and its component market prices as $S_t^{\mathrm{HC}} = \left(S_{1,t}^{\mathrm{HC}}, S_{2,t}^{\mathrm{HC}}, ..., S_{M,t}^{\mathrm{HC}}\right)^{\mathrm{T}}$. The* ***ex ante*** *persentage return of the hedged portfolio, denoted as $R_t^{\mathrm{hedged}}$, is defined as $R_t^{\mathrm{hedged}} := \frac{V_{t+\Delta t}^{\mathrm{P}} + V_{t+\Delta t}^{\mathrm{H}}}{\tilde{V}_t^{\mathrm{P}} + \tilde{V}_t^{\mathrm{H}}} - 1$ where $\tilde{V}_t^{\mathrm{hedged}} := \tilde{V}_t^{\mathrm{P}} + \tilde{V}_t^{\mathrm{H}}$. Furthermore, the hedged portfolio return can be decomposed into the ratio of three components relative to the initial hedged portfolio market value:*

***1) Stochastic Unrealized P&L****: Gains or losses due to price movements of the underlying assets without actual trades;*

***2) Stochastic Realized Cash Flow****: Gains or losses realized through executed trades;*

***3) Stochastic Implicit Trading Costs****: Slippage, market impact, and other implicit costs associated with trading.*

*This decomposition ensures a comprehensive representation of the hedged portfolio's performance over the period $(t, t+\Delta t]$.*

*Proof.* We can simply treat the primary portfolio and the hedging portfolio as a combined one

$$\begin{aligned}
R_t^{\mathrm{hedged}} &:= \frac{V_{t+\Delta t}^{\mathrm{P}} + V_{t+\Delta t}^{\mathrm{H}}}{\tilde{V}_t^{\mathrm{P}} + \tilde{V}_t^{\mathrm{H}}} - 1 \\
&= \frac{\left(\tilde{n}_t^{\mathrm{C}} + \Delta n_t^{\mathrm{C}}\right)^{\mathrm{T}}\left(\tilde{S}_t^{\mathrm{C}} + \Delta S_t^{\mathrm{C}}\right) + \left(\tilde{n}_t^{\mathrm{HC}} + \Delta n_t^{\mathrm{HC}}\right)^{\mathrm{T}}\left(\tilde{S}_t^{\mathrm{HC}} + \Delta S_t^{\mathrm{HC}}\right)}{\tilde{n}_t^{\mathrm{C,T}}\tilde{S}_t^{\mathrm{C}} + \tilde{n}_t^{\mathrm{HC,T}}\tilde{S}_t^{\mathrm{HC}}} - 1 \\
&= \underbrace{\frac{\tilde{n}_t^{\mathrm{C,T}}\Delta S_t^{\mathrm{C}}}{\tilde{V}_t^{\mathrm{hedged}}}}_{\text{stochastic primary unrealized P\&L}} + \underbrace{\frac{\tilde{n}_t^{\mathrm{HC,T}}\Delta S_t^{\mathrm{HC}}}{\tilde{V}_t^{\mathrm{hedged}}}}_{\text{stochastic hedging unrealized P\&L}} \\
&+ \underbrace{\frac{\Delta n_t^{\mathrm{C,T}}\Delta \tilde{S}_t^{\mathrm{C}}}{\tilde{V}_t^{\mathrm{hedged}}}}_{\text{stochastic primary realized cashflow}} + \underbrace{\frac{\Delta n_t^{\mathrm{HC,T}}\Delta \tilde{S}_t^{\mathrm{HC}}}{\tilde{V}_t^{\mathrm{hedged}}}}_{\text{stochastic hedging realized cashflow}} \\
&+ \underbrace{\frac{\Delta n_t^{\mathrm{C,T}}\Delta S_t^{\mathrm{C}}}{\tilde{V}_t^{\mathrm{hedged}}}}_{\text{stochastic primary implict trading costs}} + \underbrace{\frac{\Delta n_t^{\mathrm{HC,T}} S_t^{\mathrm{HC}}}{\tilde{V}_t^{\mathrm{hedged}}}}_{\text{stochastic hedging implict trading costs}}
\end{aligned} \tag{2.5}$$

□

**Corollary 2.2.** *Let the hedging portfolio's asset price be identical to the primary portfolio's asset price $S_t^{\mathrm{HC}} = S_t^{\mathrm{C}}$, then the* ***ex ante*** *percentage return of the hedged portfolio simplifies to*

$$R_t^{\mathrm{hedged}} = \frac{\tilde{n}_t^{\mathrm{hedged,T}} \Delta S_t^{\mathrm{C}}}{\tilde{n}_t^{\mathrm{hedged,T}} \tilde{S}_t^{\mathrm{C}}} + \frac{\Delta n_t^{\mathrm{hedged,T}} \tilde{S}_t^{\mathrm{C}}}{\tilde{n}_t^{\mathrm{hedged,T}} \tilde{S}_t^{\mathrm{C}}} + \frac{\Delta n_t^{\mathrm{hedged,T}} \Delta S_t^{\mathrm{C}}}{\tilde{n}_t^{\mathrm{hedged,T}} \tilde{S}_t^{\mathrm{C}}}$$

*where* $\tilde{n}_t^{\mathrm{hedged}} := \tilde{n}_t^{\mathrm{C}} + \tilde{n}_t^{\mathrm{HC}}$ *and* $\Delta n_t^{\mathrm{hedged}} := \Delta n_t^{\mathrm{C}} + \boldsymbol{\Delta} n_t^{\mathrm{HC}}$.

*Proof.* Eq.2.5 can be rewrote to Eq.2.6 if hedging activities is done via rebalancing the component assets.

$$R_t^{\mathrm{hedged}} = \frac{\left(\tilde{n}_t^{\mathrm{C,T}} + \tilde{n}_t^{\mathrm{HC,T}}\right) \Delta S_t^{\mathrm{C}} + \left(\Delta n_t^{\mathrm{C,T}} + \Delta \tilde{n}_t^{\mathrm{HC,T}}\right)\left(\tilde{S}_t^{\mathrm{C}} + \Delta S_t^{\mathrm{C}}\right)}{\left(\tilde{n}_t^{\mathrm{C,T}} + \tilde{n}_t^{\mathrm{HC,T}}\right) \tilde{S}_t^{\mathrm{C}}} \tag{2.6}$$

then it is trivial to obtain. □

Before presenting the theoretical framework of tail risk hedging, we introduce the necessary mathematical definitions of Value at Risk (VaR) and Conditional Value at Risk (CVaR), also known as Expected Shortfall (ES). It is because, one can measure its portfolio tail risk by, for example, computing 99%CVaR or even 99.9%CVaR in 1-day time horizon. The computation of VaR and CVaR requires specifying a time horizon and a predefined probability level.

Formally, the α-quantile Value at Risk over a time horizon $\tau$, denoted as $\mathrm{VaR}_{\tau,\alpha}$ , represents the loss threshold such that the probability of exceeding this loss is at most $1-\alpha$. Mathematically, for a given portfolio loss probability distribution $L$ over a time horizon $\tau$, $\mathrm{VaR}_{\tau,\alpha}$ is the solution that makes it hold:

$$\mathbb{P}\left[L_\tau \geq \mathrm{VaR}_{\tau,\alpha}\right] = 1-\alpha$$

For instance, if 1-day 95VaR $= X$, it implies that the portfolio has a 95% probability of experiencing a loss less than $X$ over the next day and a 5% probability of a loss exceeding $X$.

Similarly, Conditional Value at Risk (CVaR), also named as Expected Shortfall (ES), at level $\alpha$ over a time horizon $\tau$, denoted as $\mathrm{CVaR}_{\tau,\alpha} = Y$, quantifies the expected loss given that the loss exceeds $\mathrm{VaR}_{\tau,\alpha}$. It is formally defined as:

$$\mathrm{CVaR}_{\tau,\alpha} = \mathbb{E}\left[L_\tau | L_\tau \geq \mathrm{VaR}_{\tau,\alpha}\right]$$

Thus, if 1-day 95CVaR $= Y$, it indicates that conditional on the portfolio experiencing a loss greater than 1-day 95VaR, the expected loss is $Y$.

Hereafter, we denote P&L measure of portfolio as $X_{t,\tau}$. It is a random variable over an ***ex ante*** base, for an absolute monetary measure it stands for random $\mathrm{P\&L}_{t,\tau}$ and for a return-based measure it stands for random $R_{t,\tau}$ on probability space $(\Omega, \mathcal{F}, \mathbb{P}; \{\mathcal{F}_t\}_{t\in[0,T]})$.

**Definition 2.1.** Let Value at Risk (VaR) over a given horizon $\tau$ from time $t$ and at a specified confidence level $\alpha$ be denoted as $\mathrm{VaR}_{t,\tau,\alpha}(X_{t,\tau})$. Consider a portfolio with market value $V_t$, consisting of $N$ assets with market prices $S_t^{\mathrm{C}} = (S_{1,t}, S_{2,t}, ..., S_{N,t})^{\mathrm{T}}$ and $n_t^{\mathrm{C}} = (n_{1,t}, n_{2,t}, ..., n_{N,t})^{\mathrm{T}}$ representing the respective holding amounts at time $t$. Assume the underlying probability space $(\Omega, \mathcal{F}, \mathbb{P}; \{\mathcal{F}_t\}_{t\in[0,T]})$ is equipped with a standard filtration capturing all available information at time $t$.

Define the portfolio P&L over the horizon $\tau$ as:

$$\mathrm{P\&L}_{t,\tau} \coloneqq V_{t+\tau} - V_t = n_{t+\tau}^{\mathrm{C,T}} S_{t+\tau}^{\mathrm{C}} - n_t^{\mathrm{C,T}} S_t^{\mathrm{C}}$$

Since VaR measures potential losses, it is conventionally defined with loss values being positive. Thus, VaR is computed based on the negative of the portfolio P&L, such that larger VaR values correspond to greater potential losses.

The absolute $\mathrm{VaR}_{t,\tau,\alpha}(X_{t,\tau})$ at time $t$ is formally defined as the quantile satisfying

$$\mathbb{P}_{t+\tau}\left[-\mathrm{P\&L}_{t,\tau} \geq \mathrm{VaR}_{t,\tau,\alpha}\left(X_{t,\tau}\right)|\mathcal{F}_t\right] = 1 - \alpha \tag{2.7}$$

over the given time horizon $\tau$ at confidence level $\alpha$.

*Remark* 2.1. For numerical purposes, the continuous probability measure

$$\mathbb{P}_{t+\tau}\left[\mathrm{P\&L}_{t,\tau}|\mathcal{F}_t\right]$$

needs to be discretized. By any a certain discrete sampling method to divide the domain of random variable $-\mathrm{P\&L}_{t,\tau}$, usually used $\mathrm{VaR}_{\tau,\alpha}(X_{t,\tau})$ takes the value that makes the ratio of losses larger than it to the whole greater than $\alpha$. Consequently, the usually used VaR is smaller than the VaR with strictly guaranteed probability level of $\alpha$, and the usually used VaR is slightly more likely to be realized than the strict VaR.

**Definition 2.2.** Instead of expressing VaR in absolute monetary terms, it is often normalized as a return-based measure, which allows for a scale-invariant comparison across different portfolio sizes. The portfolio return over horizon $\tau$ is defined as:

$$R_{t,\tau} = \frac{V_{t+\tau}}{V_t} - 1$$

Using this return process, the return-based VaR, denoted as $\mathrm{VaR}_{t,\tau,\alpha}^{\%}(X_{t,\tau})$ is defined as the solution of:

$$\mathbb{P}_{t+\tau}\left[-R_{t,\tau} \geq \mathrm{VaR}_{t,\tau,\alpha}^{\%}\left(X_{t,\tau}\right)|\mathcal{F}_t\right] = 1 - \alpha$$

*Remark* 2.2. For numerical implementation, the continuous probability measure $\mathbb{P}_{t+\tau}\left[\mathrm{P\&L}_{t,\tau}|\mathcal{F}_t\right]$ must be discretized. Discretization is performed by partitioning the domain of the random variable $-\mathrm{P\&L}_{t,\tau}$ using a specific sampling method.

In particular applications, empirical VaR is computed as the smallest loss threshold exceeding the α-quantile of the empirical loss distribution. Since empirical VaR is estimated from finite samples rather than a continuous probability distribution, it is typically slightly greater than the theoretical VaR, which assumes strict probability constraints.

Similarly, the return-based VaR is computed empirically using:

$$\mathrm{VaR}^{\%,\mathrm{empirical}}_{t,\tau,\alpha} = \frac{\mathrm{VaR}^{\mathrm{empirical}}{}_{t,\tau,\alpha}}{V_t}$$

This empirical return-based VaR provides a more stable and comparable risk measure across portfolios by normalizing risk in terms of relative return rather than absolute monetary loss.

**Definition 2.3.** We denote CVaR over a certain time period $\tau$ from time $t$ with a particular loss probability level $\alpha$ as $\mathrm{CVaR}_{t,\tau,\alpha}(X_{t,\tau})$. CVaR is the conditional expectation of losses exceeding the corresponding $\mathrm{VaR}_{t,\tau,\alpha}(X_{t,\tau})$, and CVaR is defined as:

$$\mathrm{CVaR}_{t,\tau,\alpha}(X_{t,\tau}) \coloneqq \mathbb{E}^{\mathbb{P}}_{t+\tau}\left[X_{t,\tau} | X_{t,\tau} \geqq \mathrm{VaR}_{t,\tau,\alpha}\left(X_{t,\tau}\right), \mathcal{F}_t\right]$$

An equivalent integral representation is:

$$\mathrm{CVaR}_{t,\tau,\alpha}(X_{t,\tau}) = \frac{1}{\alpha}\int_{\mathrm{VaR}_{t,\tau,\alpha}(X_{t,\tau})}^{+\infty} X_{t,\tau}\mathbb{P}_{t+\tau}\left[X_{t,\tau}|\mathcal{F}_t\right] dx$$

**Definition 2.4.** Analogous to return-based VaR, we define the return-based Conditional Value at Risk (CVaR) using the portfolio return process:

$$R_{t,\tau} = \frac{V_{t+\tau}}{V_t} - 1$$

Then, the return-based CVaR, denoted as $\mathrm{CVaR}^{\%}_{t,\tau,\alpha}(X_{t,\tau})$, is given by:

$$\begin{aligned}&\mathbb{P}_{t+\tau}\left[-R_{t,\tau} \geq \mathrm{CVaR}^{\%}_{t,\tau,\alpha}(X_{t,\tau}) | -R_{t,\tau} \geq \mathrm{VaR}^{\%}_{t,\tau,\alpha}(X_{t,\tau}), \mathcal{F}_t]\right] \\ &= \mathbb{E}^{\mathbb{P}}_{t+\tau}\left[-R_{t,\tau}| - R_{t,\tau} \geq \mathrm{VaR}^{\%}_{t,\tau,\alpha}(X_{t,\tau}), \mathcal{F}_t\right]\end{aligned}$$

*Remark* 2.3. Similar to the case in usually used $\mathrm{VaR}_{t,\tau,\alpha}(X_{t,\tau})$, normally $\mathrm{CVaR}_{t,\tau,\alpha}(X_{t,\tau})$ requires a discretization method to be computed. This is necessary because the exact conditional expectation may not have a closed-form solution for arbitrary loss distributions. Empirically, CVaR is computed as the average loss beyond the empirical VaR threshold. It makes usually used $\mathrm{CVaR}_{t,\tau,\alpha}(X_{t,\tau})$ is greater than the mathematically strict one.

*Remark* 2.4. Let $\mathrm{VaR}_{t,\tau,\alpha}(X_{t,\tau})$ and $\mathrm{CVaR}_{t,\tau,\alpha}(X_{t,\tau})$ be the VaR and CVaR for time horizon $[t, t+\tau]$ and at confidence level of $\alpha$. VaR and CVaR for a different time horizon $[t, t+T]$ can be converted from $\mathrm{VaR}_{t,\tau,\alpha}(X_{t,\tau})$ and $\mathrm{CVaR}_{t,\tau,\alpha}(X_{t,\tau})$ by the square root rule on an assumption base.

$$\mathrm{VaR}_{t,T,\alpha}(X_{t,\tau}) = \sqrt{\frac{T}{\tau}}\mathrm{VaR}_{t,\tau,\alpha}(X_{t,\tau})$$

$$\mathrm{CVaR}_{t,T,\alpha}(X_{t,\tau}) = \sqrt{\frac{T}{\tau}}\mathrm{CVaR}_{t,\tau,\alpha}(X_{t,\tau})$$

**Proposition 2.4.** *CVaR is a convex risk measure.*

*Proof.* Let $X_1$, $X_2$ be two loss random variables and consider a convex combination:

$$X_\lambda = \lambda X_1 + (1-\lambda)X_2, \quad \lambda \in [0,1]$$

For convexity, we must show that:

$$\mathrm{CVaR}_{t,\tau,\alpha}(X_\lambda) \leq \lambda \mathrm{CVaR}_{t,\tau,\alpha}(X_1) + (1-\lambda)\mathrm{CVaR}_{t,\tau,\alpha}(X_2)$$

By definition of conditional expectation,

$$\begin{aligned}\mathrm{CVaR}_{t,\tau,\alpha}(X_\lambda) &= \mathbb{E}^{\mathbb{P}}_{t+\tau}\left[X_\lambda | X_\lambda \geqq \mathrm{VaR}_{t,\tau,\alpha}\left(X_\lambda\right)\right] \\ &= \mathbb{E}^{\mathbb{P}}_{t+\tau}\left[\lambda X_1 + (1-\lambda)\, X_2 | X_\lambda \geqq \mathrm{VaR}_{t,\tau,\alpha}\left(X_\lambda\right)\right]\end{aligned}$$

then, using the linearity property of expectation:

$$\begin{aligned}\mathrm{CVaR}_{t,\tau,\alpha}(X_\lambda) = \lambda &\mathbb{E}^{\mathbb{P}}_{t+\tau}\left[X_1 | X_\lambda \geqq \mathrm{VaR}_{t,\tau,\alpha}\left(X_\lambda\right)\right] \\ &+ (1-\lambda)\mathbb{E}^{\mathbb{P}}_{t+\tau}\left[X_2 | X_\lambda \geqq \mathrm{VaR}_{t,\tau,\alpha}\left((X\right)\right]\end{aligned}$$

Since expectation is a convex measure, we apply Jensen's inequality:

$$\mathbb{E}^{\mathbb{P}}_{t+\tau}[X_i | X_\lambda \geqq \mathrm{VaR}_{t,\tau,\alpha}(X_\lambda)] \leq \mathrm{CVaR}_{t,\tau,\alpha}(X_i), \quad i = 1, 2$$

then

$$\mathrm{CVaR}_{t,\tau,\alpha}(X_\lambda) \leq \lambda \mathrm{CVaR}_{t,\tau,\alpha}(X_1) + (1-\lambda)\mathrm{CVaR}_{t,\tau,\alpha}(X_2)$$

the convexity of CVaR holds. □

The convexity of CVaR as a risk measure plays a fundamental role in this study. By formulating tail risk management as an optimization problem, we seek to determine an optimal hedging strategy that minimizes the expected extreme losses as measured by CVaR.

From an operations research perspective, the convexity of the risk measure ensures that the resulting optimization problem is convex, which guarantees the existence of a unique global minimum and facilitates efficient numerical solutions. This property is crucial for the stability and tractability of the optimization framework, allowing for more robust and computationally feasible risk management strategies.

In financial risk management, VaR and CVaR are widely used measures to quantify potential losses. Their computation methods can be classified into parametric methods (including variance-covariance method, extreme value theory method), non-parametric methods (including historical simulation (HS) method and bootstrapping method), and Simulation-based methods (including Monte-Carlo method and copula method). Below, we provide a brief mathematical formulation for each approach.

1. **Parametric Methods**

   (a) *Variance-Covariance Method, also named Normal Method*
   The Variance-Covariance method is widely used in standard risk management, regulatory reporting (e.g., Basel II, III), and portfolio optimization due to its simplicity and analytical tractability. However, this method assumes that portfolio losses follow a normal distribution, leading to an underestimation of extreme losses in real-world financial markets. The tail behavior modeled by this approach is light-tailed, making it inadequate for capturing extreme market movements. Additionally, for portfolio VaR and CVaR, the method relies on linear correlation between assets or risk factors, which fails to account for nonlinear dependencies. This assumption becomes especially unreliable during financial crises, when correlations tend to increase and market behavior exhibits stronger tail dependencies.

   It assumes portfolio losses follow a normal distribution:

   $$X_{t,\tau} \sim \mathcal{N}(\mu_{t,\tau}, \sigma^2_{t,\tau})$$

   then VaR is given by:

   $$\text{VaR}_{t,\tau,\alpha} = \mu_{t,\tau} + \sigma_{t,\tau} Z_\alpha$$

   and CVaR is given by:

   $$\text{CVaR}_{t,\tau,\alpha} = \mu_{t,\tau} + \sigma_{t,\tau} \frac{\phi(Z_\alpha)}{1-\alpha}$$

   where $Z_\alpha$ is the standard normal quantile satisfying:

   $$\mathbb{P}\left[Z \geq Z_\alpha\right] = 1 - \alpha$$

   and $\phi(Z_\alpha)$ is the standard normal probability density function (PDF).

   (b) *Extreme Value Theory (EVT) Method*
   EVT is widely used for modeling extreme losses, including those occurring during financial crises. Within EVT, the Peaks Over Threshold (POT-EVT) method models loss exceedances beyond a pre-defined threshold, making it more suitable for short-term tail risk management and dynamic market environments. In contrast, the Generalized Extreme Value (GEV-EVT) method analyzes block maxima (e.g., annual maximum losses), making it preferable for longer-term risk management and regulatory stress testing. Both methodologies require parameter estimation techniques such as Maximum Likelihood Estimation (MLE), the Method of Moments (MoM), or semi-parametric estimators (e.g., Pickands' Estimator) to determine the shape ( $\xi$), scale ($\beta$), and location ($u$ or $\mu$) parameters. While these methods introduce additional degrees of freedom in risk modeling, careful selection of thresholds in POT-EVT and block sizes in GEV-EVT remains critical to avoid estimation bias and overfitting. Adopting these methods efficiently needs well-developed internal risk modeling management procedures.

i. Peaks Over Threshold (POT) EVT Method using Generalized Pareto Distribution (GPD)
   POT-EVT method is preferred for tail risk modeling. Using the Generalized Pareto Distribution (GPD) for tail exceedances:

$$G(y) = 1 - \left(1 + \xi \frac{y}{\beta}\right)^{-\frac{1}{\xi}} \qquad y > 0$$

   then VaR and CVaR are computed as:

$$\mathrm{VaR}_{t,\tau,\alpha} = u + \frac{\beta}{\xi}\left(\left(\frac{N}{n}(1-\alpha)\right)^{-\xi} - 1\right)$$

$$\mathrm{CVaR}_{t,\tau,\alpha} = \frac{\mathrm{VaR}_{t,\tau,\alpha}}{1-\xi} + \frac{\beta - u}{1-\xi} \quad \text{if } \xi < 1$$

   where

   A. $u$ is the chosen threshold;
   B. $\beta$ is the scale parameter;
   C. $\xi$ is the shape parameter controlling tail heaviness;
   D. $\xi > 0$: Heavy tails (e.g., financial crashes).
   E. $\xi = 0$: Exponential decay (thin tails).
   F. $\xi < 0$: Short tails (bounded losses, capped risks).
   G. $N$ is the total number of observations in the dataset used to estimate GPD paramerters;
   H. $n$ is the number of exceedances above the threshold $u$, i.e., $n := \sum_{i=1}^{N} \mathbf{1}(X_i \geq u)$, where $\mathbf{1}(\cdot)$ is the indicator function.

ii. Block Maxima Approach EVT Method using Generalized Extreme Value (GEV) Distribution
   GEV-EVT method is useful for modeling extreme draw-downs and regulatory stress testing. Applying Fisher–Tippett–Gnedenko theorem, the maximum (or minimum) of a sample of i.i.d. observations converges to one of three types of GEV distributions:

$$F(x) = \exp\left(-\left(1+\xi\frac{x-\mu}{\sigma}\right)^{-\frac{1}{\xi}}\right) \quad 1+\xi\frac{x-\mu}{\sigma} > 0$$

then VaR and CVaR are computed as:

$$\mathrm{VaR}_{t,\tau,\alpha} = \mu + \frac{\sigma}{\xi}\left((-\log(1-\alpha))^{-\xi} - 1\right)$$

$$\mathrm{CVaR}_{t,\tau,\alpha} = \frac{\mu - \frac{\beta}{\xi} + \frac{\beta}{\xi(1-\xi)}\left((-\log(1-\alpha))^{-\xi} - 1\right)}{1-\alpha}, \quad \xi < 1$$

A. $\mu$ is the location parameter;
B. $\beta$ is the scale parameter;
C. $\xi$ is the shape parameter controlling tail heaviness;
D. $\xi > 0$ (Fréchet distribution): Heavy-tailed (e.g., t-distribution, financial crashes).
E. $\xi = 0$ (Gumbel distribution): Exponential decay (e.g., normal-like behavior).
F. $\xi < 0$ (Weibull distribution): Short-tailed (e.g., bounded losses, insurance claims).

2. **Non-parametric Methods**

(a) Historical Simulation (HS) method
The historical simulation (HS) method estimates VaR and CVaR by directly analyzing the empirical distribution of past returns without imposing parametric assumptions. This method assumes that historical return distributions reflect future risks.

Let $\left(\widetilde{d}_1, \widetilde{d}_2, \ldots, \widetilde{d}_T\right)^{\mathrm{T}}$ be the empirical time series data of all risk factors' actual/proportional changes $\widetilde{d}_t := (\widetilde{r}_{1,t}, \widetilde{r}_{2,t}, \ldots, \widetilde{r}_{n,t})^{\mathrm{T}}$over a sample period $t = 1, 2, \ldots, T$. Here, $\widetilde{r}_{i,t}$ is the empirical actual/proportional changes of $i$-th risk factor at time $t$. The sample frequency is identical to the time horizon of VaR and CVaR, e.g, one samples daily returns if the time horizon of VaR and CVaR is 1-Day. Then, for ***ex ante*** P&Ls/returns of the portfolio for time horizon $[t, t+\Delta t]$, one applies $\widetilde{d}_t$, $t = 1, 2, \ldots, T$ to the currnt risk factors one by one to the current portfolio value according to the time order of every single sampling time interval. In this way, there are $T$ scenarios of ***ex ante*** portfolio P&Ls/returns $(R_1, R_2, \ldots, R_T)^{\mathrm{T}}$ for the next time horizon interval. It is the loss distribution $L := -(R_1, R_2, \ldots, R_T)^{\mathrm{T}}$that is used to compute VaR and CVaR. For computing ***ex ante*** portfolio P&Ls/returns, there is a need for modeling the relationship between risk factors and portfolio values/returns. Normally, a multivariable linear regression technique is used while other models are still alternatives for this purpose.

Afterwards, one finds the loss data from loss distribution $L$ as the $\lfloor \alpha \rfloor T$-indexed loss as the $\mathrm{VaR}_{t,\tau,\alpha}$ with minor transformation mentione in REM.2.4 if it is needed. For CVaR, it is computed as the mean (or weighted mean for some particular purposes) of loss data expcess the loss therehold $\alpha$.

$$\mathrm{VaR}_{t,\tau,\alpha} = -\sqrt{\frac{\tau}{\Delta t}} R_{\lfloor \alpha \rfloor T}$$

$$\mathrm{CVaR}_{t,\tau,\alpha} = -\frac{1}{T - \lfloor \alpha \rfloor T} \sqrt{\frac{\tau}{\Delta t}} \sum_{i=\lfloor \alpha \rfloor T}^{T} R_i$$

(b) Bootstrapping method

Unlike the Historical Simulation (HS) method, which directly uses past observations, the bootstrapping method estimates risk by randomly resampling historical returns to generate synthetic data for computing ex ante portfolio Profit & Loss (P&L) distributions. This approach helps mitigate sampling bias and enhance statistical robustness in risk estimation.

In financial risk management, bootstrapping is applied to empirical marginal distributions of risk factor absolute or proportional changes. There are four main types of bootstrapping techniques:1) Naïve bootstrap or i.i.d. bootstrap; 2) Simple block bootstrap; 3) Moving block bootstrap; 4) Stationary bootstrap [10]. The first one reconstructs time series of risk factor actual/proportional changes without considering their series auto-correlation and cross-correlation. It is done by simply resampling the marginal distribution w/o replacement. The second and the third one reconstructs time series of risk factor actual/proportional changes by resampling the marginal distribution with blocks of time series data. The difference between 2) and 3) is that blocks in 2) do not overlapp but blocks in 3) overlapp. For method 4), resampling principle is similar to 2) and 3). However, blocks length in 4), they are determined randomly via designing a certain probability distribution and an average length for resampling blocks.

Generally speaking, the four major bootstrapping methods have their own featurs:

i. Naïve bootstrap: P&L marginal distribution shows thinner tail and smaller kurtosis comparing to the original price paths.

Simotaneously, its terminal price distribution is similar to the original one. As a result, VaR and CVaR are under-estimated. It may cause issues in risk management policies executions, due to compainsation oriented trading desk employees' behaviors.

ii. Simple block bootstrap, Moving block bootstrap and Stationary block bootstrap: P&L marginal distributions are similar to the original one. However, their terimal price distirbution shows heavier tail and larger kurtosis comparing to the original price paths. Consequentially, VaR and CVaR are relatively well-estimated, while max downside and upside risks are over-estimated and making them better for conservative risk management policies.

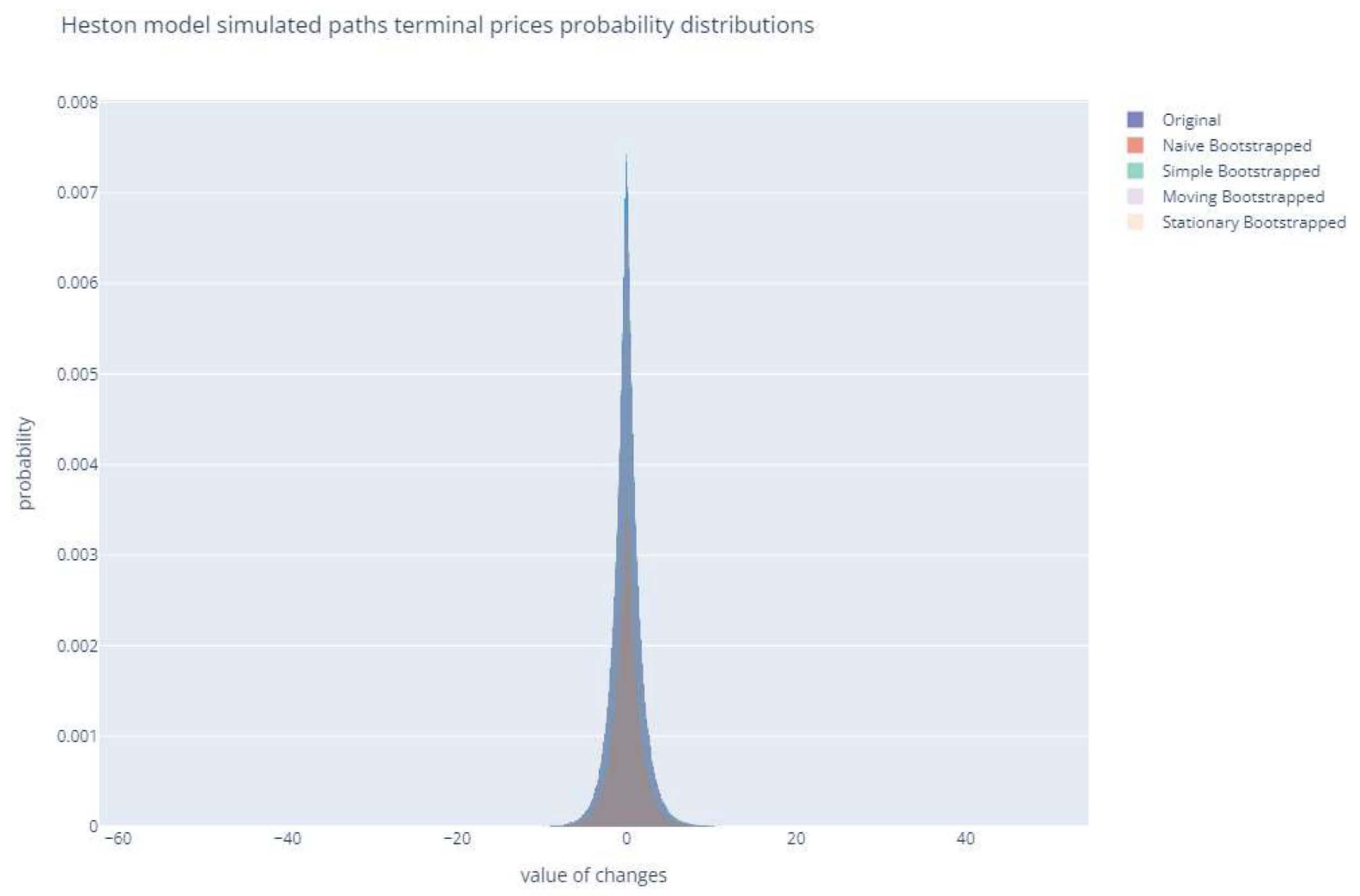

Figure 2.1: Performance of different bootstrap methods that resamples value of changes from Heston model simulated paths
Note: The smallest distribution is generated by Naïve bootstrap.

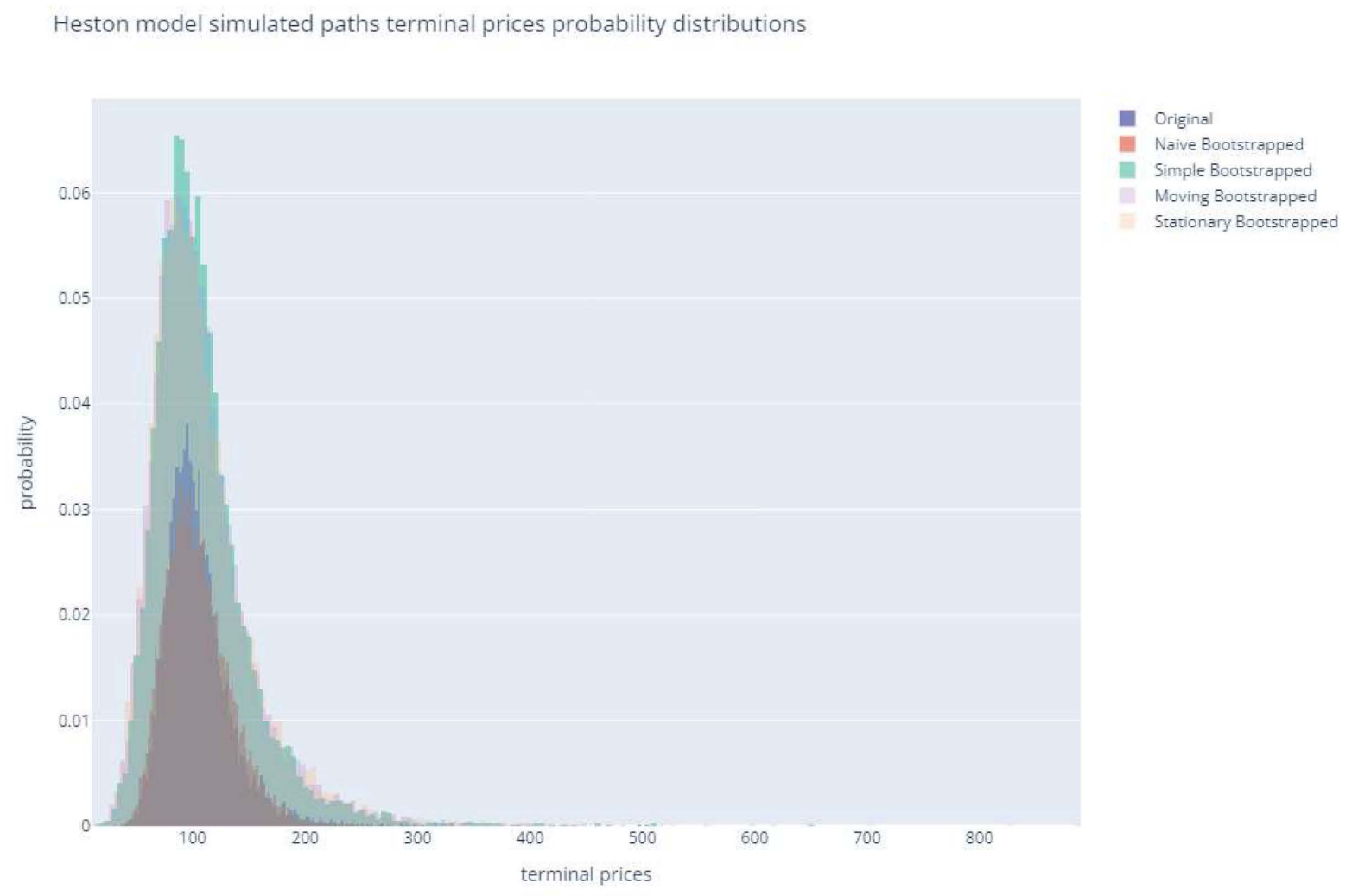

Figure 2.2: Performance of different bootstrap methods that resamples terminal prices from Heston model simulated paths
Note: The smallest distribution is generated by Naïve bootstrap.

The rest procedure computing VaR and CVaR is identical to the procedure in HS method.

3. **Simulation-based Methods**

   (a) *Monte-Carlo Method*
   Monte-Carlo method generates multiple random scenarios of asset returns or portfolio losses based on a specified stochastic process model. These random scenarios can follow various stochastic processes, including: 1) Geometric Brownian Motion (GBM); 2) Heston Model (stochastic volatility process); 3) Stochastic Alpha-Beta-Rho (SABR) model; 4) Cox–Ingersoll–Ross (CIR) process (interest rate modeling); 5) Generalized Autoregressive Conditional Heteroskedasticity (GARCH) process; 6) Jump-Diffusion models, among others.

   Unlike historical simulation method, Monte-Carlo method does not rely directly on past market data but instead uses historical or even real-time data to calibrate model parameters. This approach is particularly effective when returns deviate from normality or when risk factors exhibit nonlinear dependencies, making it a useful technique for complex financial instruments, options pricing, scenario simulation and tail risk estimation.

Mone-Carlo method follows steps below:
STEP1: Let $R_t^P$ be the ***ex ante*** portfolio return at time $t$, which

$$R_t^P = w_t^{\mathrm{T}} R_t^C$$

where $w_t$ is weights of component assets and $R_t^C$ is ***ex ante*** returns of component assets at time $t$. Then, Monte-Carlo method simulates $R_t^C$ via setting particular stochastic processes. Parameters in the stochastic processes are normally estimated from historical data, whereas there are still usages via setting parameters by specific requirements.

STEP2: Monte-Carlo method constructs loss distribution, according to the distribution of portfolio returns over $M$ asset paths simulated.

$$L_{t,i}^P = -R_{t,i}^P \quad i = 1, 2, \ldots, M$$

STEP3: After a loss distribution is constructed, then VaR and CVaR of a certain time horizon $[t, t+\tau]$ and confidence level $\alpha$ will be computed.

$$\mathrm{VaR}_{t,\tau,\alpha} = \mathrm{Quantile}_{(1-\alpha)}(L_t^P)$$

$$\begin{aligned} \mathrm{CVaR}_{t,\tau,\alpha} &= \mathbb{E}[L_t^P | L_t^P \geq \mathrm{VaR}_{t,\tau,\alpha}] \\ &= \frac{1}{M(1-\alpha)} \sum_{j:L_{t,j}^P \geq \mathrm{VaR}_{t,\tau,\alpha}} L_{t,j}^P \end{aligned}$$

i. *Copula Method*

The Copula method provides a flexible way to model dependencies between multiple financial assets or risk factors, allowing for better tail risk estimation than traditional correlation-based methods. One can regard copula method as an enhanced Monte-Carlo method on the asepect of modeling more complex correlation and interactions across risk factors. This approach is widely used in portfolio risk management, systemic risk modeling, and stress testing in both bank and regulator models [11].

By Sklar's Theorem, any multivariate distribution $F(X_1, X_2, \ldots, X_d)$, here the joint distribution of losses,can be separated into a link of individual marginal distribution $(F_1(X_1), F_2(X_2), \ldots, F_d(X_d))^{\mathrm{T}}$via a unique copula function $C : [0,1]^d \to [0,1]$ iif random variables $(X_1, X_2, \ldots, X_d)^{\mathrm{T}}$are continuous [9]:

$$F(X_1, X_2, ..., X_d) = C\left(F_1(X_1), F_2(X_2), \ldots, F_d(X_d)\right)$$

whre $F_i(X_i)$ are the marginal cumulative distribution functions (CDFs) of the individual assets.

A copula function $C$ is a transformation that models the dependency structure among random variables while preserving their individual marginal distributions. In real-world applications, empirical data often exhibit pathologies such as multimodal distributions, skewness, and heavy tails, making direct analytical expressions difficult to obtain. This issue extends to Monte Carlo-generated data, where complex stochastic models further complicate dependency construction.

To address this, the first step in copula modeling is to transform each marginal distribution into a well-behaved uniform space over $[0, 1]$, using their CDFs and percentiles respectively:

$$U_i = F_i(X_i) \sim \mathcal{U}(0, 1)$$

where $U_i$ is the mapped uniform random variables via percentile injective mapping. The joint dependency among these random variables of transformed marginal distribution is then modeled via the copula function, which can assume various forms such as the Gaussian copula, Student-t copula, Clayton copula, or Gumbel copula etc., depending on the nature of dependency (e.g., tail dependence, asymmetric correlations). Mathmatically, for any certain probability density function (PDF) $\mu$ and its CDF $\Phi$ corresponding to a specific copula function $C$, we have

$$Z_i = \Psi^{-1}(U_i), \quad Z_i \sim \psi(U_i)$$

Afterwards, one is able to model dependency by parameterizing joint probability distribution $J \in [0, 1]$ of $(Z_1, Z_2, \ldots, Z_d)^{\mathrm{T}}$ or via non-parametrical techniques. Both way, however and especially when the situation comes to high-dimensional modeling, it requires Monte-Carlo techniques to sample joint probability $P$ from the joint distribution $J$ in order to attain its statistical features.

$$P \sim J(Z_1, Z_2, \ldots, Z_d)$$

The final step involves inverting this joint distribution back to recover the original dependent variables. Similarly, one applies percentile injective mapping over the new individual marginal distributions $(J_i)$ of $J$.

$$X_i^{'} = J_i^{-1}(Z_i)$$

For the rest of utilizing copula method to compute VaR and CVaR, the procedure follows the aforementioned Monte-Carlo method.

Tail risk refers to the probability and magnitude of extreme losses that occur in the higher tail of a portfolio's loss distribution. Mathematically, it can be characterized using probability measures and risk functionals.

*Remark* 2.5. Let $X$ be a random variable representing portfolio losses, defined on a probability space $(\Omega, \mathcal{F}, \mathbb{P}; \{\mathcal{F}_t\}_{t\in[0,T]})$. The tail probability at level $\alpha$ over a time horizon $\tau$ form time $t$ is given by:

$$\mathbb{P}_{t+\tau}[X \geq \mathrm{VaR}_{t,\tau,\alpha}(X)\,|\mathcal{F}_t] = 1 - \alpha$$

This measures the likelihood of a loss exceeding the VaR threshold.

*Remark* 2.6. The expected portfolio loss occurs when a tail risk market event realizes over a time horizon $\tau$ form time $t$ at a confidence level $\alpha$ is measured by CVaR:

$$\mathrm{CVaR}_{t,\tau,\alpha}(X) = \mathbb{E}^{\mathbb{P}}_{t+\tau}\left[X|X_\lambda \geqq \mathrm{VaR}_{t,\tau,\alpha}(X)\right]$$

This measures the average loss in the worst $1 - \alpha$ proportion of cases, capturing the severity of tail risk events. Usually for measuring tail risks, $\alpha$ is 99% or 99.9%, and $\tau$ is 1 day.

**Definition 2.5.** A general tail risk measure $\rho(X)$ is defined as:

$$\rho(X) = \inf_{\mathbb{Q}\in\mathcal{M}_\alpha(\mathbb{P})} \mathbb{E}^{\mathbb{Q}}[X]$$

where $\mathcal{M}_\alpha(\mathbb{P})$ is the set of probability measure absolutely continuous w.r.t. $\mathbb{P}$ and constrained by a tail event condition.

# 3 Method

In this secsion, we exhibit our method that converts tail risk hedging problem into an optimization problem. We show the way of using our NN optimizer to address tail risk hedging problem. Firstly, we formulate our approch to conduct tail risk hedging problem.

**Definition 3.1.** Tail risk hedging problem in our work is formulated as a stochastic optimization problem by finding the optimal hedging strategy $n_t^{\mathrm{H}} := \left(n_{1,t}^{\mathrm{H}}, \ldots, n_{d,t+\tau}^{\mathrm{H}}\right)^{\mathrm{T}}$. A strategy here stands for the dynamics of number of component assets in a certain portfolio according to the filtration $\mathcal{F}_t$ for a particular purpose. Here, we assume that the primal portfolio contains $m$ component assets and the hedging portfolio contains $d$ component assets. The optimial tail risk hegding strategy in this paper are such one that minimizes tail risk measure $\rho(\cdot)$ while we are obmiting trading constrains, e.g., long-only, rebalancing order size limitation and etc., for simplicity:

$$\min_{n_t^{\mathrm{H}}} \mathbb{E}^{\mathbb{Q}}\left[\rho(X^\pi)|\mathcal{F}_t\right], \quad X_t^\pi := -V_{t+\tau}^\pi + V_t^\pi + C(\Delta n_t^{\mathrm{C}}, \Delta n_t^{\mathrm{H}}) \tag{3.1}$$

$$
\begin{aligned}
V_{t+\tau}^{\pi} &= V_t^{\pi} + \Delta V_t^{\mathrm{P}} + \Delta V_t^{\mathrm{H}} \\
&= V_t^{\pi} + \Delta n_t^{\mathrm{C},\mathrm{T}} \Delta S_t^{\mathrm{C}} + \Delta n_t^{\mathrm{H},\mathrm{T}} \Delta S_t^{\mathrm{H}}
\end{aligned} \tag{3.2}
$$

where $V_t^{\pi}$ is the market value of the hedged portfolio at time $t$, and $X^{\pi}$ is the ***ex ante*** L&P (minus value of P&L) of the hedge portfolio happening during time period $[t, t+\tau]$. $\Delta n_t$ is the change of number of component assets in the primal portfolio during time period $[t, t+\tau]$. Here $-V_{t+\tau}^{\pi} + V_t^{\pi}$ has incorporated implicit trading costs in Cor2.1 and Prop2.3. To introduce explict trading costs for customizability, e.g., trasaction fees, short selling interests, margin interests, dealers' compensation, management fees and stamp duties etc. Note that explicit trading costs are accounted in cash account, then we additionally put the explicit trading costs function $C(\cdot)$ into consideration along portfolio mark value changes.

In the other hand, for the manner of protfolio weights, we have

$$
\begin{aligned}
V_{t+\tau}^{\pi} &= (1 + R_t^{\pi})\, V_t^{\pi} = \left(1 + \omega_t^{\mathrm{T}} R_t^{\mathrm{C}} + \pi_t^{\mathrm{T}} R_t^{\mathrm{H}}\right) V_t^{\pi} \\
&= \frac{n_{t+\tau}^{\mathrm{C},\mathrm{T}} S_{t+\tau}^{\mathrm{C}} + n_{t+\tau}^{\mathrm{H},\mathrm{T}} S_{t+\tau}^{\mathrm{H}}}{n_t^{\mathrm{C},\mathrm{T}} S_t^{\mathrm{C}} + n_t^{\mathrm{H},\mathrm{T}} S_t^{\mathrm{H}}} V_t^{\pi} \\
&= \left[1 + \omega_t^{\mathrm{T}} \left(\frac{S_{t+\tau}^{\mathrm{C}}}{S_t^{\mathrm{C}}} - \mathbf{1}\right) + \pi_t^{\mathrm{T}} \left(\frac{S_{t+\tau}^{\mathrm{H}}}{S_t^{\mathrm{H}}} - \mathbf{1}\right)\right] V_t^{\pi}
\end{aligned} \tag{3.3}
$$

where $\omega_t$ is weights of component assets in the primal portfolio, and $\pi_t$ is weights of component assets in the hedging portfolio. $\frac{S_{t+\tau}^{\mathrm{C}}}{S_t^{\mathrm{C}}}$ and $\frac{S_{t+\tau}^{\mathrm{H}}}{S_t^{\mathrm{H}}}$ are element-wise division, and $\mathbf{1}$ is the 1 vector.

*Remark* 3.1. For measuring tail risks, in our approach, we take 1-Day 99%CVaR(or 1-Day 99%ES) of $X_t^{\pi}$ as the specific measure.

The next step in our apporach to conduct the optimization problem of finding the optimal tail risk hedging strategy is to construct marginal loss disstribution. For this purpose, we block-bootstrap (see 2b in section 2 or Def.3.2) from empirical marginal changes/returns distribution. In section 4, one will find that the optimal strategy works better in controling drawback risks by intentionally choosing market data during the period of realized tail risk events.

**Definition 3.2.** Block-bootstrap is defined as below. Assume we have a observed data $S \coloneqq (S_1, \ldots, S_t, \ldots, S_T)^{\mathrm{T}}$ where $X$ is a statistically stational time series data. Block-bootstrap is emploied by firstly dividing $X$ into $n$ data blocks $B \coloneqq (B_1, \ldots, B_i, \ldots, B_n)^{\mathrm{T}}$ of length $l$, either non-overlapping (Simple Block Bootstrap) or overlapping (Moving Block Bootstrap).

$$
B_i = (S_i, S_{i+1}, \ldots, S_{i+l-1})^{\mathrm{T}}, \quad i = 1, \ldots, n+l-1
$$

Then, using divided data blocks $B$ block bootstrapping resamples blocks with replacement and forms a new random data series $X^*$.

$$S^* \coloneqq B_{i_1} \cup B_{i_2} \cup \ldots \cup B_{i_k}, \quad i_k \sim \text{Uniform}\left(\{1, 2, , \ldots, n+l-1\}\right)$$

where $i_k$ are uniformly randomly sampled from indices set $\{1, 2, , \ldots, n+l-1\}$ of $B$.

Under stationary assumption, block-bootstrap is asymptotic unbiased (consistent)[7]. There are multiple ways to test stational condition, including but not limited to, augmented Dickey-Fuller test (ADF) tests, Kwiatkowski-Phillips-Schmidt-Shin (KPSS) tests and Phillips-Perron (PP) tests. Although unbiased estimation is an ideal goal to achieve, block-bootstrap on nonstational time series data is still a powerful tool to create synthetic data. On the other hand, for length $l$, there is a heuristically reasonable choice as $l = \left\lfloor n^{\frac{1}{3}} \right\rfloor$[5].

The third step is to form ***ex ante*** loss marginal distribution via block-bootstrapping the empirical market marginal changes/returns distribution. In this way, the block-bootstrapped data $S^*$ anticipates a real market filtration $\mathcal{F}_t$. This is done by a manner of monte-carlo method. To be more specific, we draw $m$ counterfactual scenarios of market dynamics at each time point on empirical market time series data using block-bootstrapped synthetic data $S^*$. In this way, it is possible for us to keep heavy tail features in the real market data and generate ***ex ante*** tail risk measure, i.e., ***ex ante*** 1-Day 99%CVaR of the hedged portfolio. Since our method does not imposed an adoption of a specific market dynamics model, it is a model-free method to conduct tail risk hedging problem.

*Remark* 3.2. One can also construct the counterfactural market dynamics by using a particular market model, e.g, BSM, Heston, SABR and etc. Our method has such customizability, but a callibration procedure is required in this use case.

**Definition 3.3.** ***Ex ante*** tail risk is measured by the expectation value of 1-Day 99%CVaR in the $m$ counterfactual scenarios, i.e. simulating $m$ paths at every time point $t$ of empirical market data. The filtration $\mathcal{F}_t^{\text{empirical}}$ is constructed from empirical marginal changes/returns distribution.

$$\begin{aligned}\rho(X^\pi) &= \mathbb{E}\left[\text{CVaR}_{t,\tau,99\%}\left(S^*\right) | \mathcal{F}_t^{\text{empirical}}\right] \\ &= \frac{1}{m-1}\sum_{j=1}^{m} \text{CVaR}_{t,\tau,99\%}\left(S^{*,(j)}\right)\end{aligned}$$

where $S^{*,(j)}$ is the synthetic market dynamics data in $j$-th scenario. The ***ex ante*** tail risk measure here is computed by using loss marginal distribution obtained in the way that is given in Def.3.1 and Eq.3.2 or Eq.3.3. We underline that the computation method as Eq.3.2 and Eq.3.3 introduces implicit and explicit trading costs into consideration, therefore, our method has stronger customizability.

The forth step is to determine the optimal tail hedging strategy via minimizing our tail risk measure. For the sake of simplicity, we adopt Multilayer Perceptron (MLP) as our NN optimizer. It is the simplest form of deep neural networks, and it can avoid those details about the structure of neural networks that do not help the discussion about tail risk hedging problem.

**Definition 3.4.** MLP is defined as below. Mathmatically, it is a sequential series of linear algebra calculations. For a $n$ layer MLP, it is constructed as 1 input layer, $n-2$ input layer and 1 output layer. Every layer has its own activation function $\phi$ and its own bias $b$. Assume that there are input data $X$ of $d$-dimensions, then they are inputed into a MLP to obtain output $Y$. Dimensions refers to the number of features of $X$.

$$Y^{(1)} = \phi^{(1)} \left.\left[ \begin{pmatrix} \omega_{1,1}^{(1)} & \cdots & \omega_{1,j}^{(1)} & \cdots & \omega_{1,d}^{(1)} \\ \vdots & \ddots & \vdots & \ddots & \vdots \\ \omega_{i,1}^{(1)} & \cdots & \omega_{i,j}^{(1)} & \cdots & \omega_{i,d}^{(1)} \\ v & \ddots & \vdots & \ddots & \vdots \\ \omega_{I_1,1}^{(1)} & \cdots & \omega_{I_1,j}^{(1)} & \cdots & \omega_{I_1,d}^{(1)} \end{pmatrix} \begin{pmatrix} X_1 \\ \vdots \\ X_j \\ \vdots \\ X_d \end{pmatrix} + b^{(1)} \right]\right\} \text{Input}$$

$$Y^{(2)} = \phi^{(2)} \left.\left[ \begin{pmatrix} \omega_{1,1}^{(2)} & \cdots & \omega_{1,j}^{(2)} & \cdots & \omega_{1,d}^{(2)} \\ \vdots & \ddots & \vdots & \ddots & \vdots \\ \omega_{i,1}^{(2)} & \cdots & \omega_{i,j}^{(2)} & \cdots & \omega_{i,d}^{(2)} \\ v & \ddots & \vdots & \ddots & \vdots \\ \omega_{I_2,1}^{(2)} & \cdots & \omega_{I_2,j}^{(2)} & \cdots & \omega_{I_2,d}^{(2)} \end{pmatrix} Y^{(1)} + b^{(2)} \right]\right\} \text{Hidden 1}$$

$$\vdots$$

$$Y^{(n-1)} = \phi^{(n-1)} \left.\left[ \begin{pmatrix} \omega_{1,1}^{(n-1)} & \cdots & \omega_{1,j}^{(n-1)} & \cdots & \omega_{1,d}^{(n-1)} \\ \vdots & \ddots & \vdots & \ddots & \vdots \\ \omega_{i,1}^{(n-1)} & \cdots & \omega_{i,j}^{(n-1)} & \cdots & \omega_{i,d}^{(n-1)} \\ v & \ddots & \vdots & \ddots & \vdots \\ \omega_{I_{n-1},1}^{(n-1)} & \cdots & \omega_{I_{n-1},j}^{(n-1)} & \cdots & \omega_{I_{n-1},d}^{(n-1)} \end{pmatrix} Y^{(n-2)} + b^{(n-1)} \right]\right\} \text{Hidden n}$$

$$Y = \phi^{(n)} \left.\left[ \begin{pmatrix} \omega_{1,1}^{(n)} & \cdots & \omega_{1,j}^{(n)} & \cdots & \omega_{1,d}^{(n)} \\ \vdots & \ddots & \vdots & \ddots & \vdots \\ \omega_{i,1}^{(n)} & \cdots & \omega_{i,j}^{(n)} & \cdots & \omega_{i,d}^{(n)} \\ v & \ddots & \vdots & \ddots & \vdots \\ \omega_{I_n,1}^{(n)} & \cdots & \omega_{I_n,j}^{(n)} & \cdots & \omega_{I_n,d}^{(n)} \end{pmatrix} Y^{(n-1)} + b^{(n)} \right]\right\} \text{Output}$$

In computational practice, there are usually $m$ batches of input data $X$. Batches refers to independent input data points that are imputed into the same MLP to output multiple independent outputs $Y$.

The MLP optimizer's parameters are iterative approximated by the Adam (Adaptive Moment Estimation) optimizer, which is commonly used and performance stable in NN practices [6].

We apply MLP optimizer to minimize Eq.3.1, the loss function, by inputting $X = S^*$, and outputting $Y := \text{MLP}(\Delta S_t^{\text{C}}, \Delta S_t^{\text{H}}) = \left(n_t^{\text{C}}, n_t^{\text{H}}\right)^{\text{T}}$.

$$\text{LOSS} = \mathbb{E}^{\text{corss section}}\left[\mathbb{E}^{\text{pathwise}}\left[\text{CVaR}_{t,\tau,99\%}\left(\begin{array}{c} -\Delta n_t^{*,\text{C},\text{T}}\Delta S_t^{*,\text{C}} \\ +\text{MLP}(\Delta S_t^{*,\text{C}}, \Delta S_t^{*,\text{H}}) \circ \Delta S_t^{*,\text{H}} \end{array}\right)\Big|\mathcal{F}_t^{\text{empirical}}\right]\right] \tag{3.4}$$

where $\circ$ is Hadamard product. In this way, we compute the LOSS via Eq.3.2 or Eq.3.3. Then, by appling automatic differentiation, we iteratively find the optimal hedging stategy. The optimal hedging stategy is theoretically unique, because CVaR is a convex risk measure (Prop.2.4)[2]. Numerically, for stochastic optimization, one can merely have the optimal expected solution.

# 4 Numerical Experiment

In this section, we demonstrate the performance of our approach to minimize tail risks. For simplisity, we choose SPX as our primal portfolio instead of constructing portfolio from individual assets. SPX is the ticker name of S&P500 index, which is basing on a market-value-weighted portfolio of largest 500 listed stocks in the United States. Tail risk events have happened multiple time in SPX, e.g., 2001 dot-com recession, 2007-2008 Global Financial Crisis (GFC), 2020 Russia–Saudi Arabia Oil Price War and 2020 Covid-19 Stock Market Crash and etc.

We held 1 unit of SPX by its quoted price as the initial market value of our primal portfolio. Our hedging portfolio here was the short-selling SPX portfolio. We did not change the unit over our numerical backtesting, and we did not use SPX futures (CME ticker: ES) as our hedging instrument. Note that futures price is impacted by not only risk-free rate curves but also other market factors, e.g., contango and backwardation term structures, the ease of financing transactions and etc. Therefore, for the purpose of controling variables to test our approach, we set our numerical experiment in this way to avoid introducing extraneous interference. We conducted tail risk hedging problem by outputting hedge ratio of SPX via NN optimizer and backtest the performance of the hedged portfolio. Finally, we exhabite the graphs of backtesting to show the effectiveness of our approach on tail risk hedging problem in the end of this scetion.

For NN optimizer, we choose ReLU function as the activation functions of all layers excepting the output layer. Activation function in output layer is the identical mapping function. We tested NN optimizer with 32 nodes in input layer and 32 nodes in output layer. In hidden layers we set four types of NNs. They are

1. **None_Hidden_Layer_MLP**: no hidden layer;
2. **32_MLP**: one hidden layer with 32 nodes;
3. **32x32_MLP**: two hidden layers with 32 and 32 nodes;
4. **32x32x32_MLP**: three hidden layers with 32 and 32 and 32 nodes.

On the other hand, we tested the impact of choosing training data. We selected periods w/o tail risk events for our block-bootstrapping simulations. Those simulated data were used to train the NN optimizers. NN optimizers were trained to minimize 1-Day 99%CVaR generated from block-bootstrapping simulations at every time point of empirical market data.

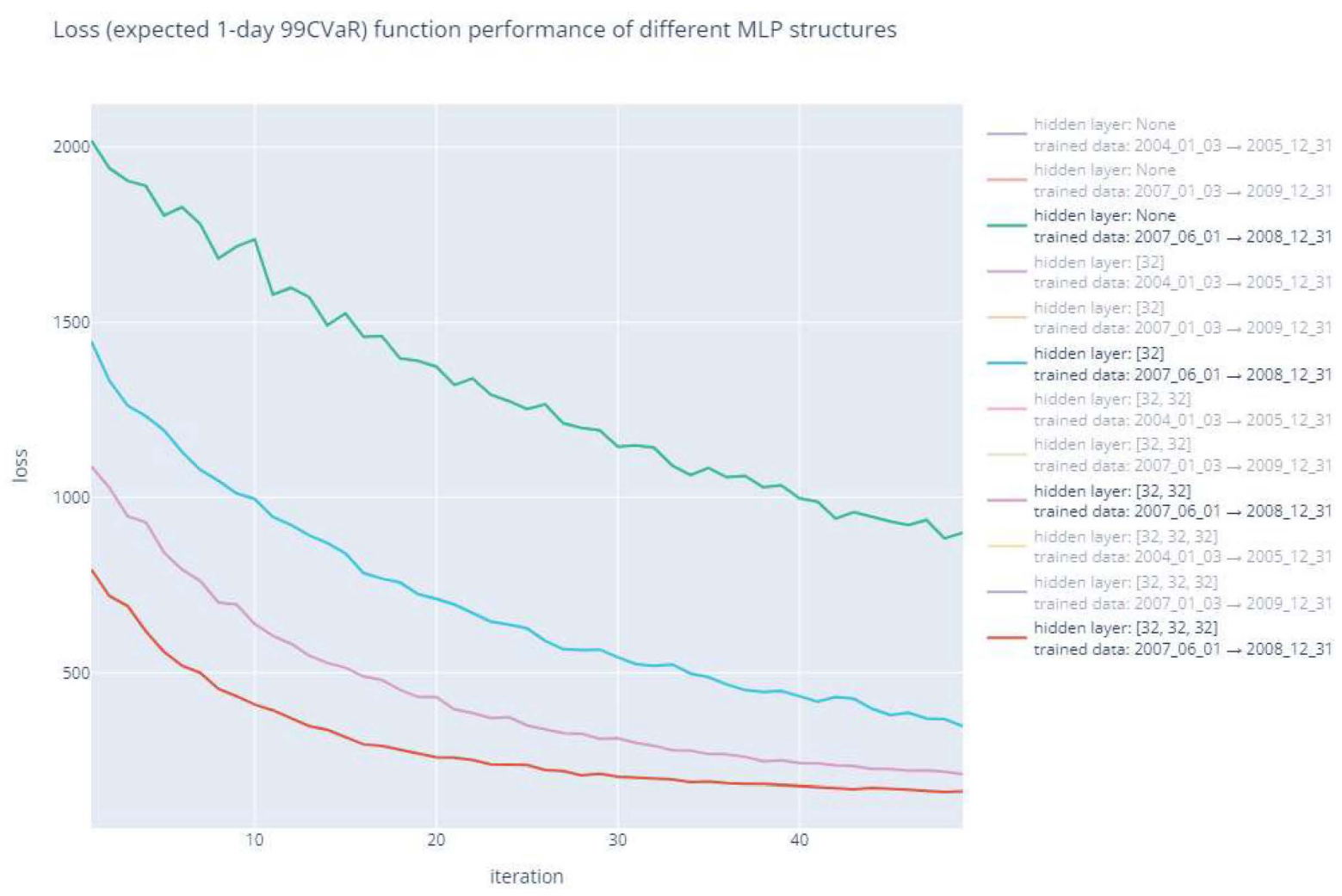

Figure 4.1: One example of loss function decreasing performance of MLP with different hidden layer structures. Learning iteration number is 50.

Fig.4.1 shows that MLP with more hidden layers generated a more repaid decrease in loss function. Loss function here is the tail risk measure, 1Day 99CVaR. For loss function convergence, MLP optimizer in our approach reaches its stable loss function state after a short learning process. Merely 50 iterations produce a useful MLP optimizer. One can check it with numerical experiment results below. It can be inferred that our approach is a computation resource saving method.

## 4.1 Back-test Performance via Training MLPs with Empirical Data from 2004/01/03 to 2005/12/31

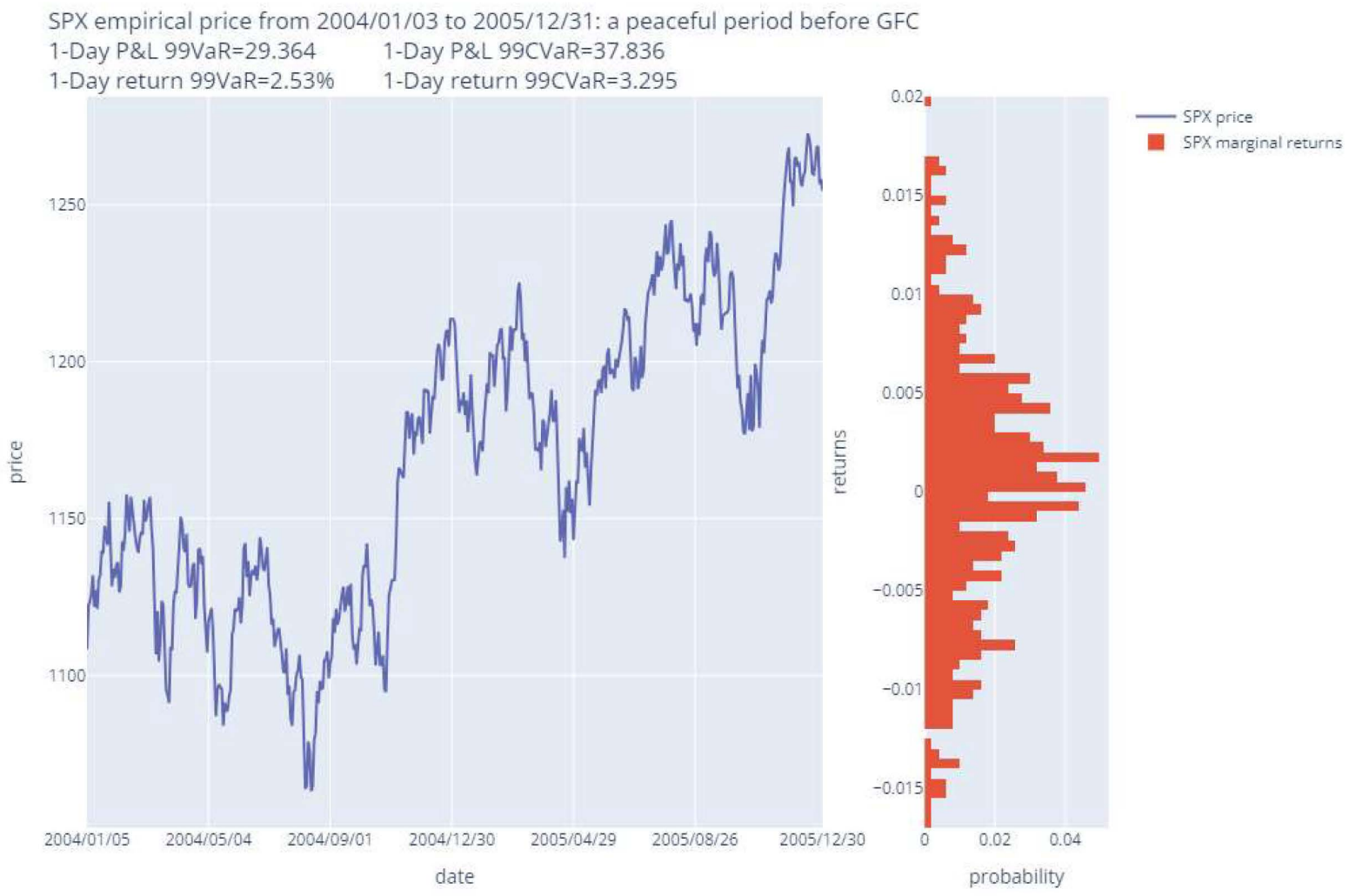

Figure 4.2: SPX close price and its marginal return distribution from empirical data of time period 2004/01/03 to 2005/12/31

Here we used SPX close price and its marginal return distribution from 2004/01/03 to 2005/12/31 to train those four MLPs. Empirical data shows a narrow and multi-mode feature. There are three modes centered at 0.5%, 0% and -0.8%. One can find this period is relative peaceful by undermarking the price volatile range without many tail risk events.

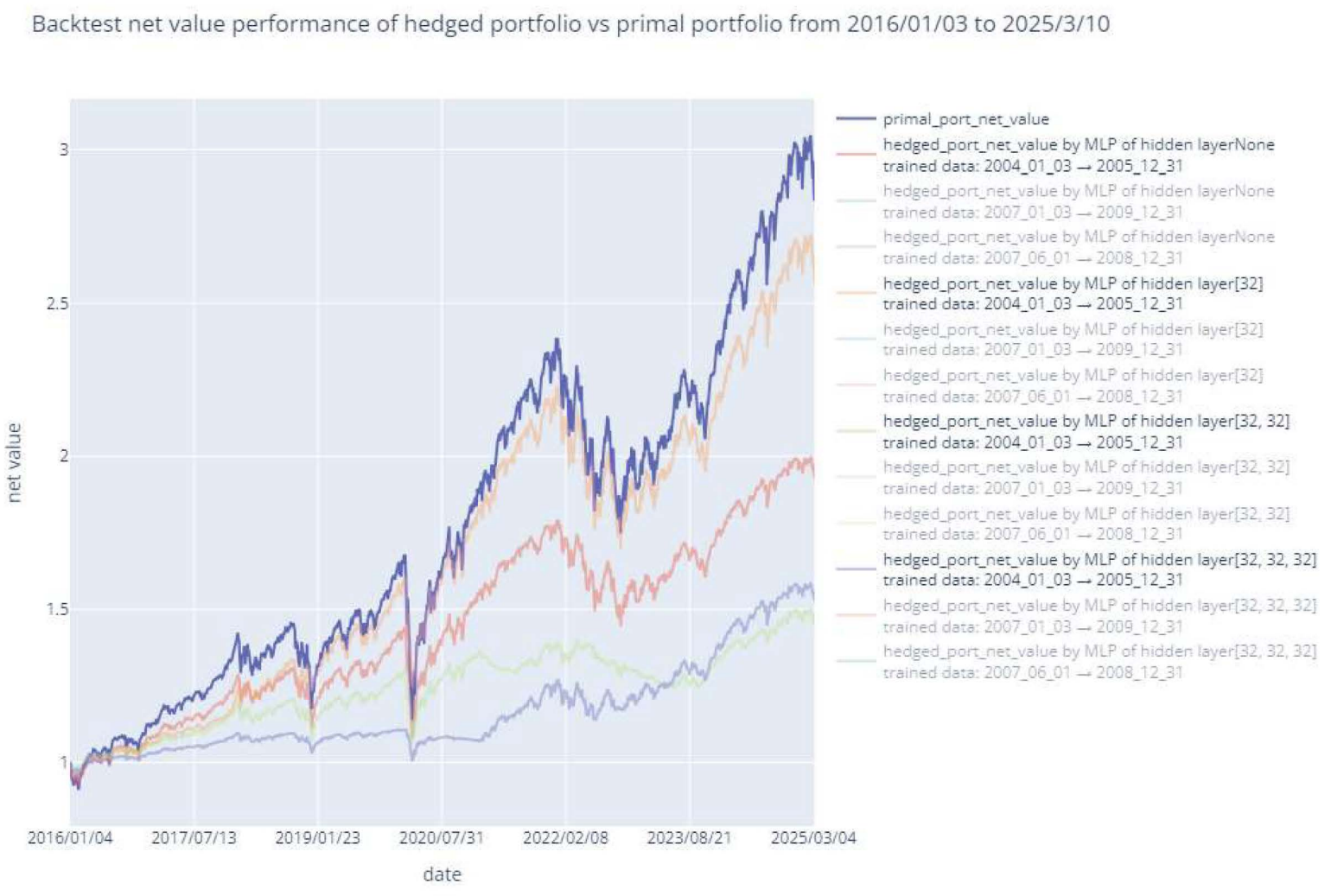

Figure 4.3: Back-test result of different MLP layer structures. Hedged portfolio net value benchmars at 2016/01/03. The backtest time horizon is from 2016/01/03 to 2025/03/10. Traning data is time horizon is from 2004/01/03 to 2005/12/31.

We trained MLP optimizers with empirical data from 2004/01/03 to 2005/12/31. Therefore, we completely avoid introducing future data issues. It is shown on Fig.4.3 that the more hidden layers, the relatively better tail risk hedge performance.

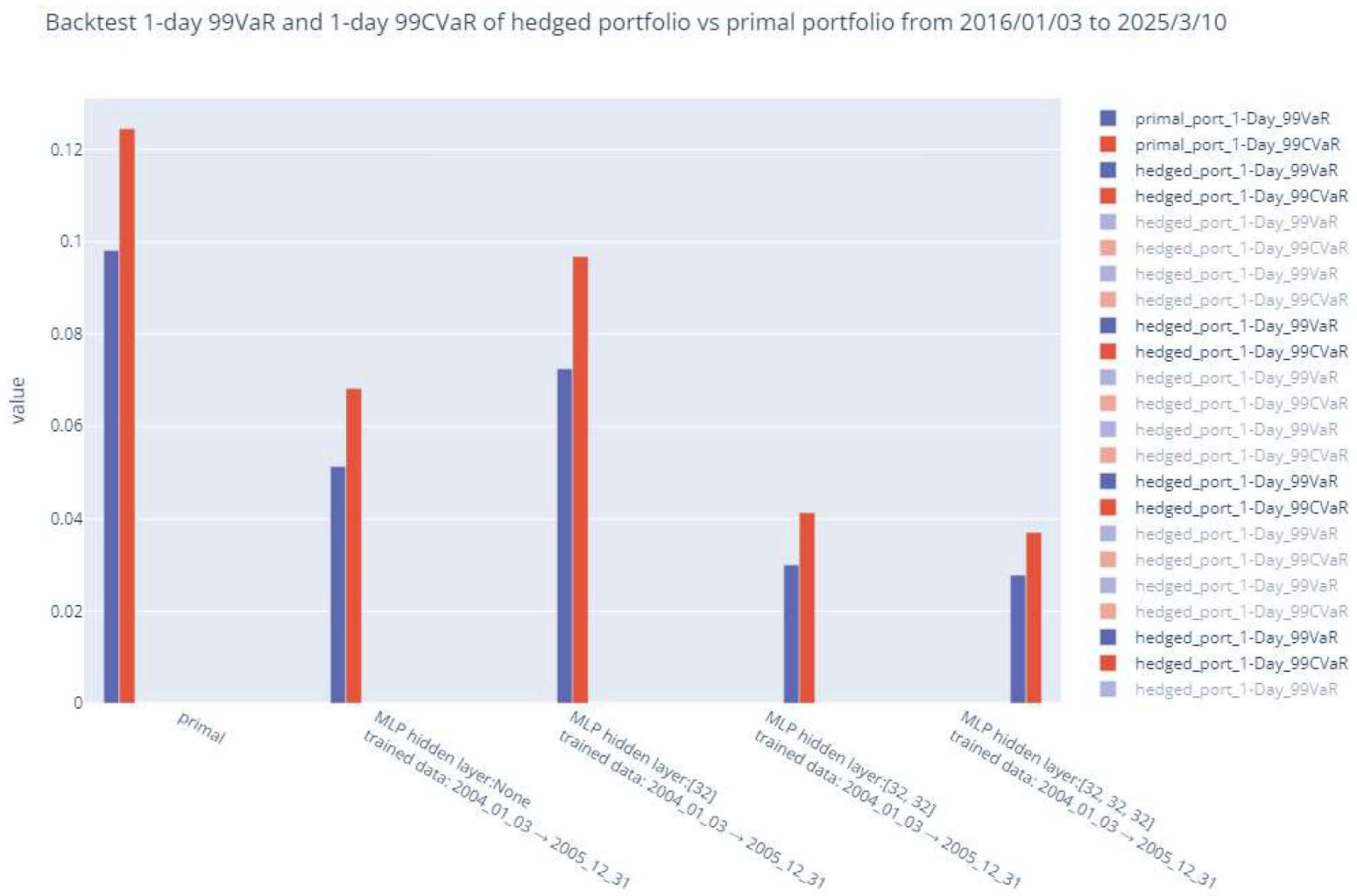

Figure 4.4: 1-Day 99VaR and 1-Day 99CVaR of primal portfolio and hedged portfolios net value changes. Training data set is from 2004/01/01 to 2005/12/31. Tail risk hedge results by the four MLP optimizers are shown on the for bar groups on the left.

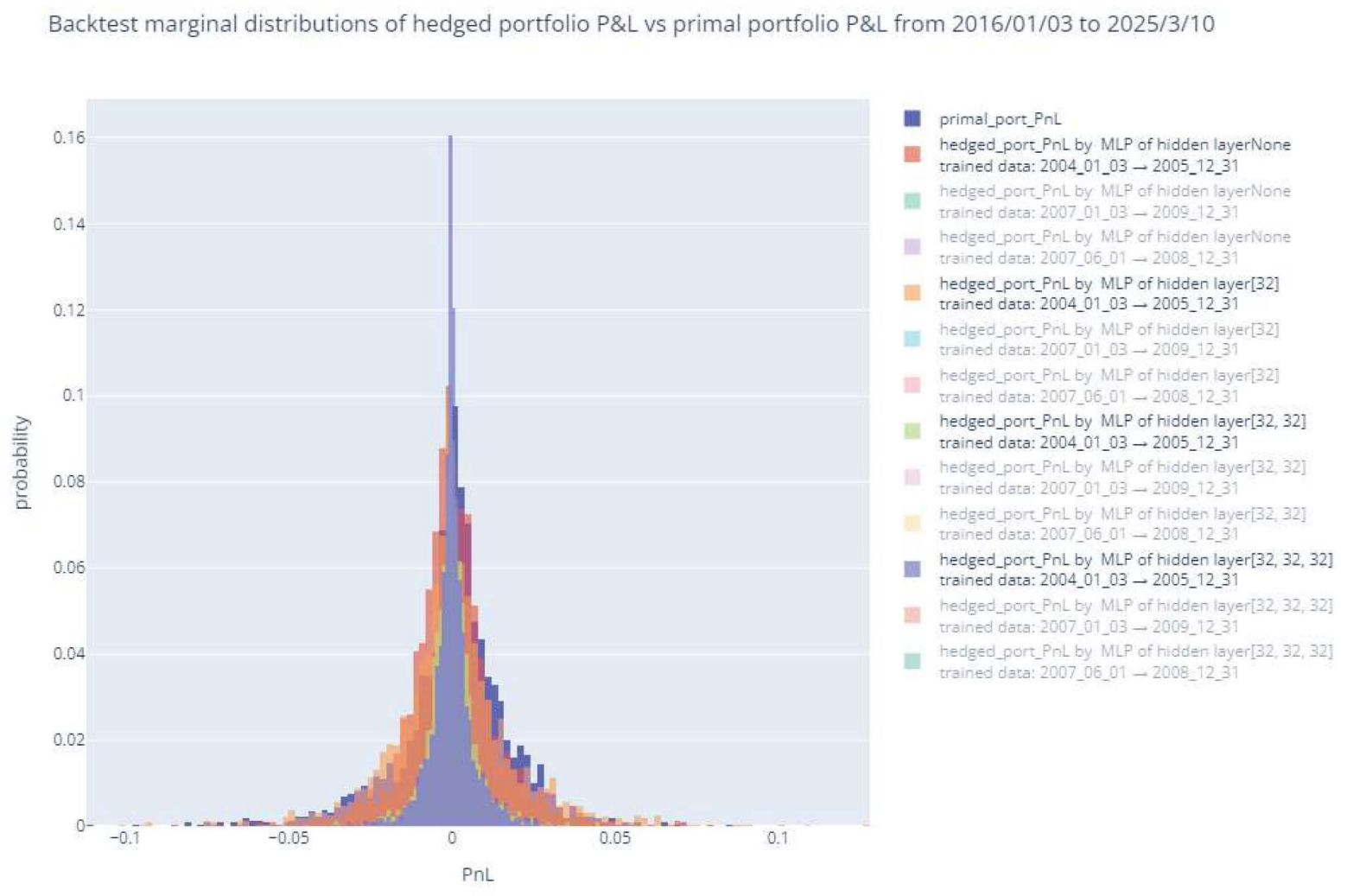

Figure 4.5: Histogram of primal and hedged portfolio P&Ls. The performance of MLP optimizers with different layer structures but learning from data of 2004/01/03 to 2005/12/31 is shown.

Fig.4.5 shows that with the number of hidden layers increasing the hedged portfolio net value P&L distribution narrows. MLP optimizer with more hidden layers results in a thinner tail and a lower kurtosis. In other words, it conducts a better tail risk hedging performance.

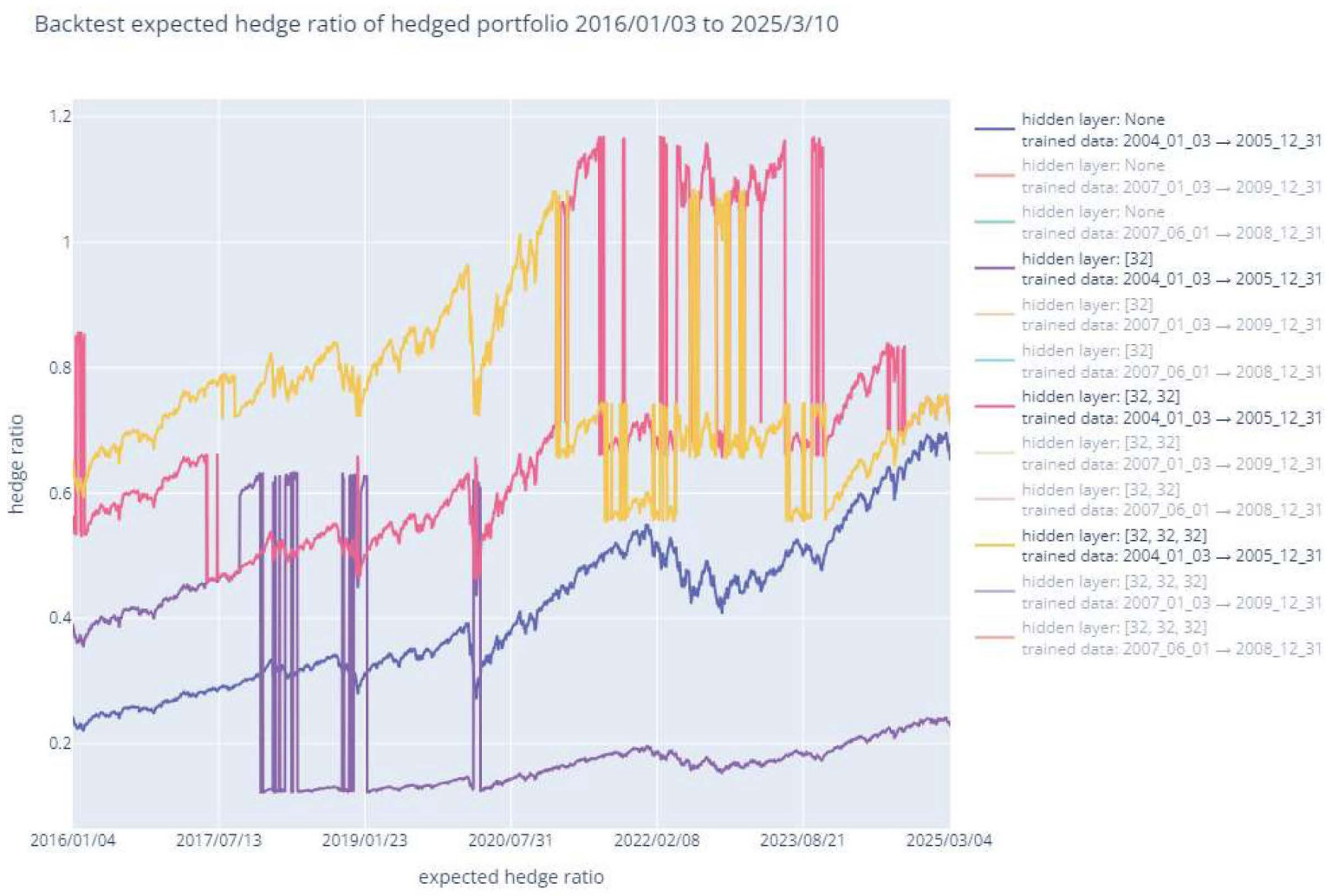

Figure 4.6: Expected hedge ratios are shown on this figure. The value on this figure means the amount of hedging portfolio, i.e., the position of shorting SPX index in this paper.

Fig.4.6 shows MLP optimizer with more hidden layers results in a higher hedge ratio during tail risk events period, i.e., hedged portfolio becomes more stable during tail risk events.

## 4.2 Back-test Performance via Training MLPs with Empirical Data from 2007/01/03 to 2009/12/31

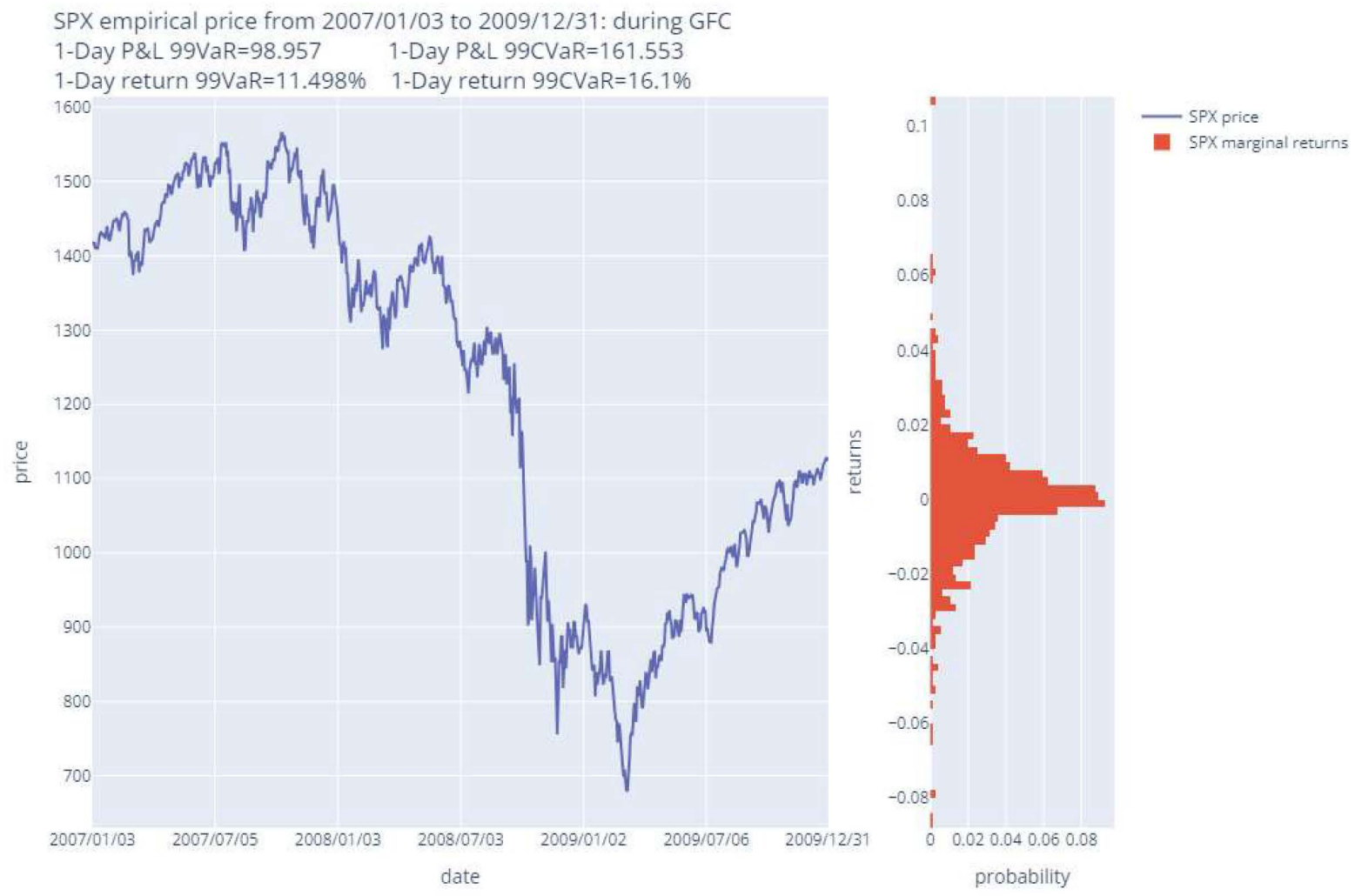

Figure 4.7: SPX close price and its marginal return distribution from empirical data of time period 2007/01/03 to 2009/12/31

Here we used SPX close price and its marginal return distribution from 2007/01/03 to 2009/12/31 to train those four MLPs. Empirical data shows a wide and multi-mode feature. There are two modes centered at 0% and -1%. One can find this period is violent due to the GFC, one can check this tail risk event around SPX price dynamics from 2008/07/03 to 2009/01/02.

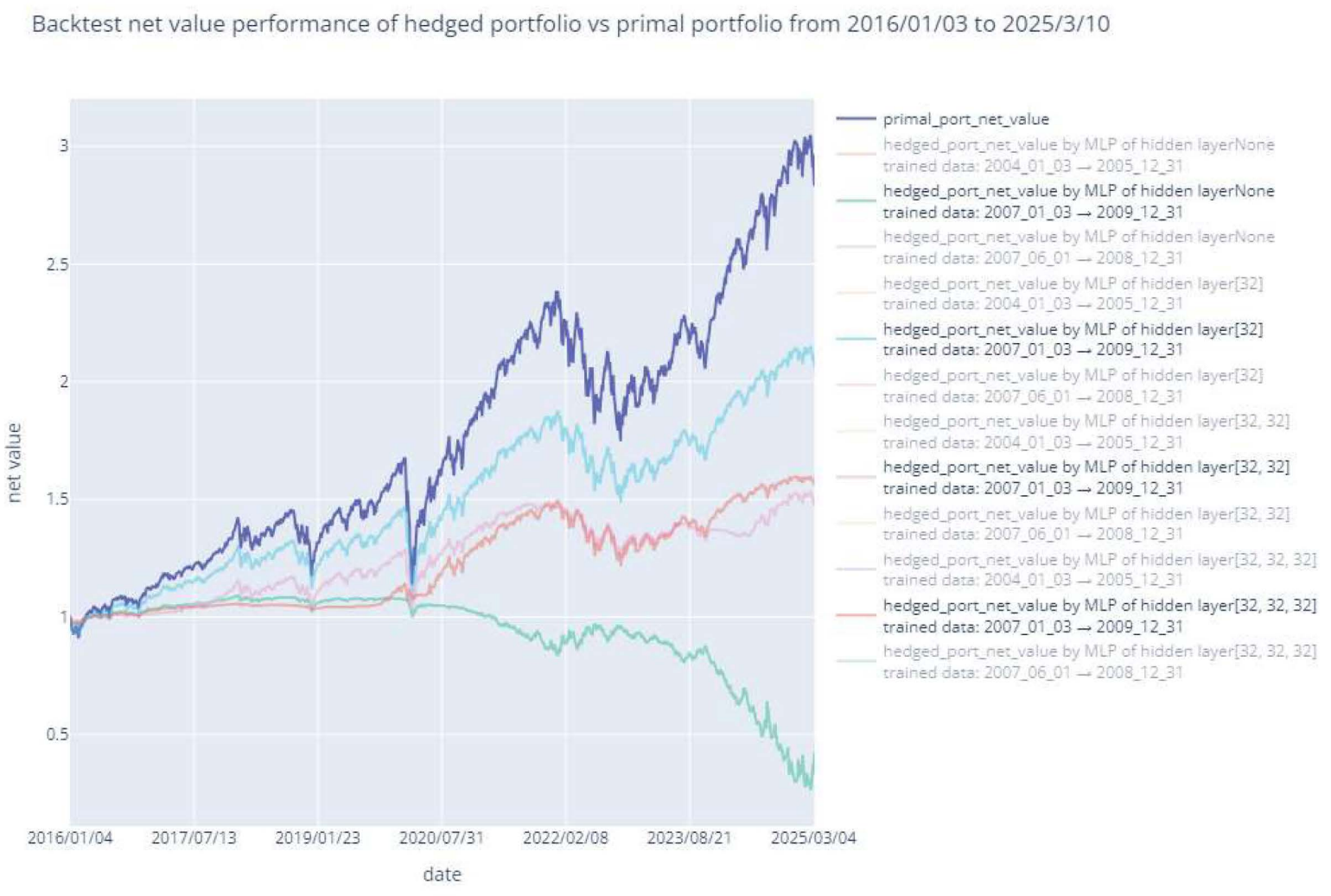

Figure 4.8: Back-test result of different MLP layer structures. Hedged portfolio net value benchmars at 2016/01/03. The backtest time horizon is from 2016/01/03 to 2025/03/10. Traning data is time horizon is from 2007/01/03 to 2009/12/31.

By learning from a relatively volatile market empirical data, MLP optimizers better while number of hidden layers is increasing. One can check it from the hedged portfolio net value dynamics. 32x32_MLP and 32x32x32_MLP with pink and orange color shows the most stable performance during 2020 market crashes. Nevertheless, None_Hidden_Layer_MLP performance the troublesome due to its continuous net value decrease.

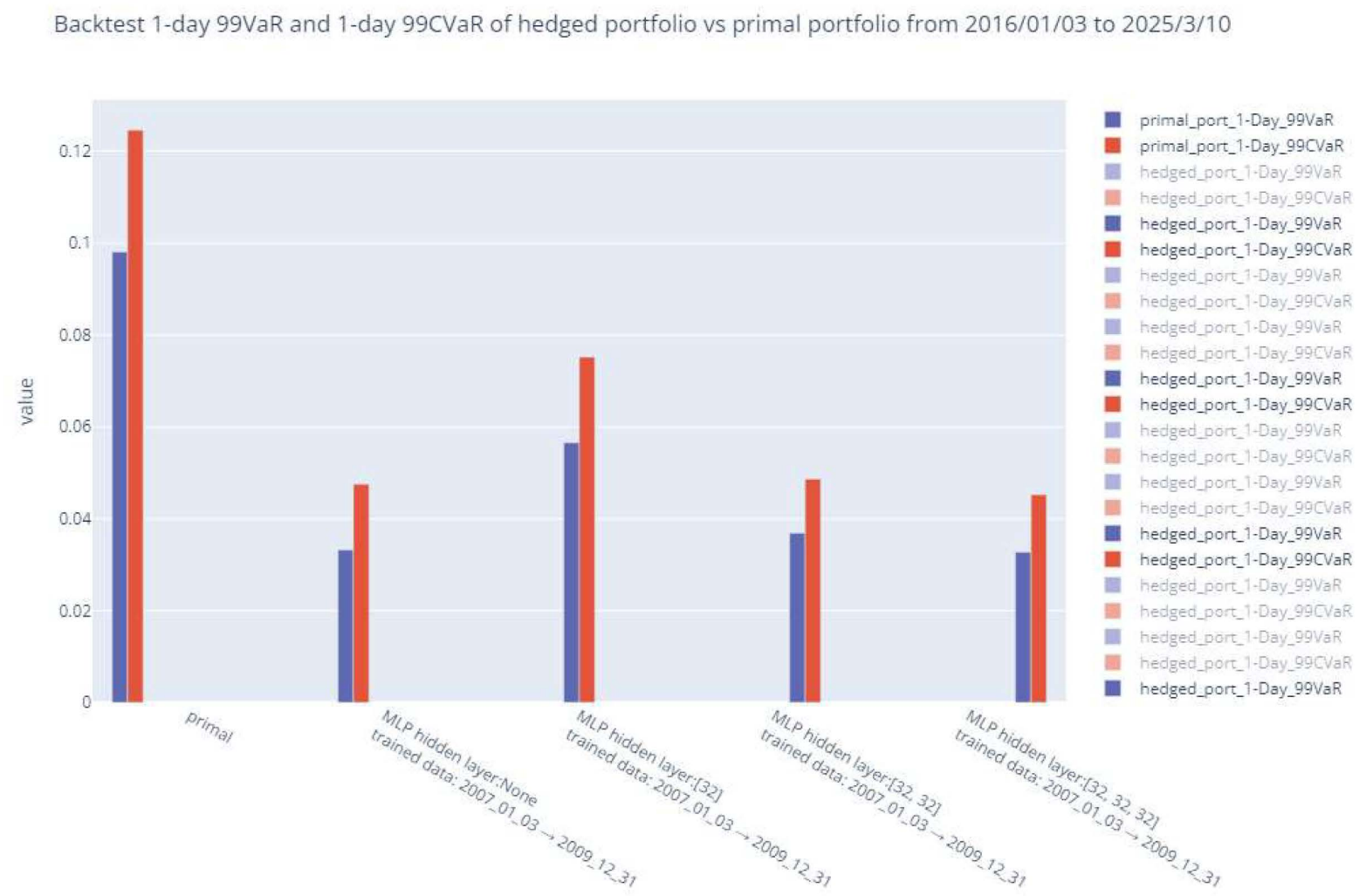

Figure 4.9: 1-Day 99VaR and 1-Day 99CVaR of primal portfolio and hedged portfolios net value changes. Training data set is from 2007/01/03 to 2009/12/31. Tail risk hedge results by the four MLP optimizers are shown on the for bar groups on the left.

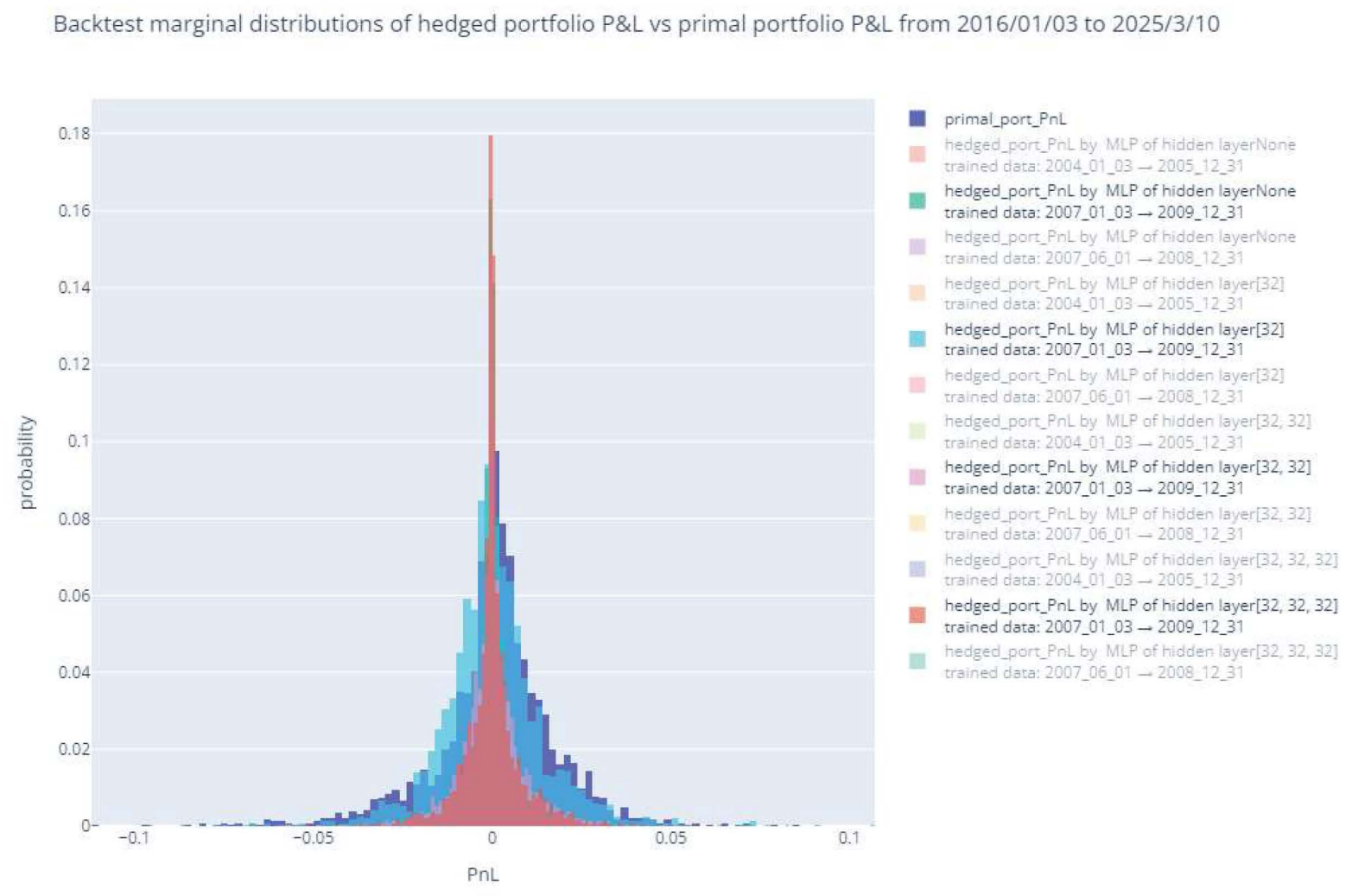

Figure 4.10: Histogram of primal and hedged portfolio P&Ls. The performance of MLP optimizers with different layer structures but learning from data of 2007/01/03 to 2009/12/31 is shown.

Fig.4.10 rechecks the result from Fig.4.5.

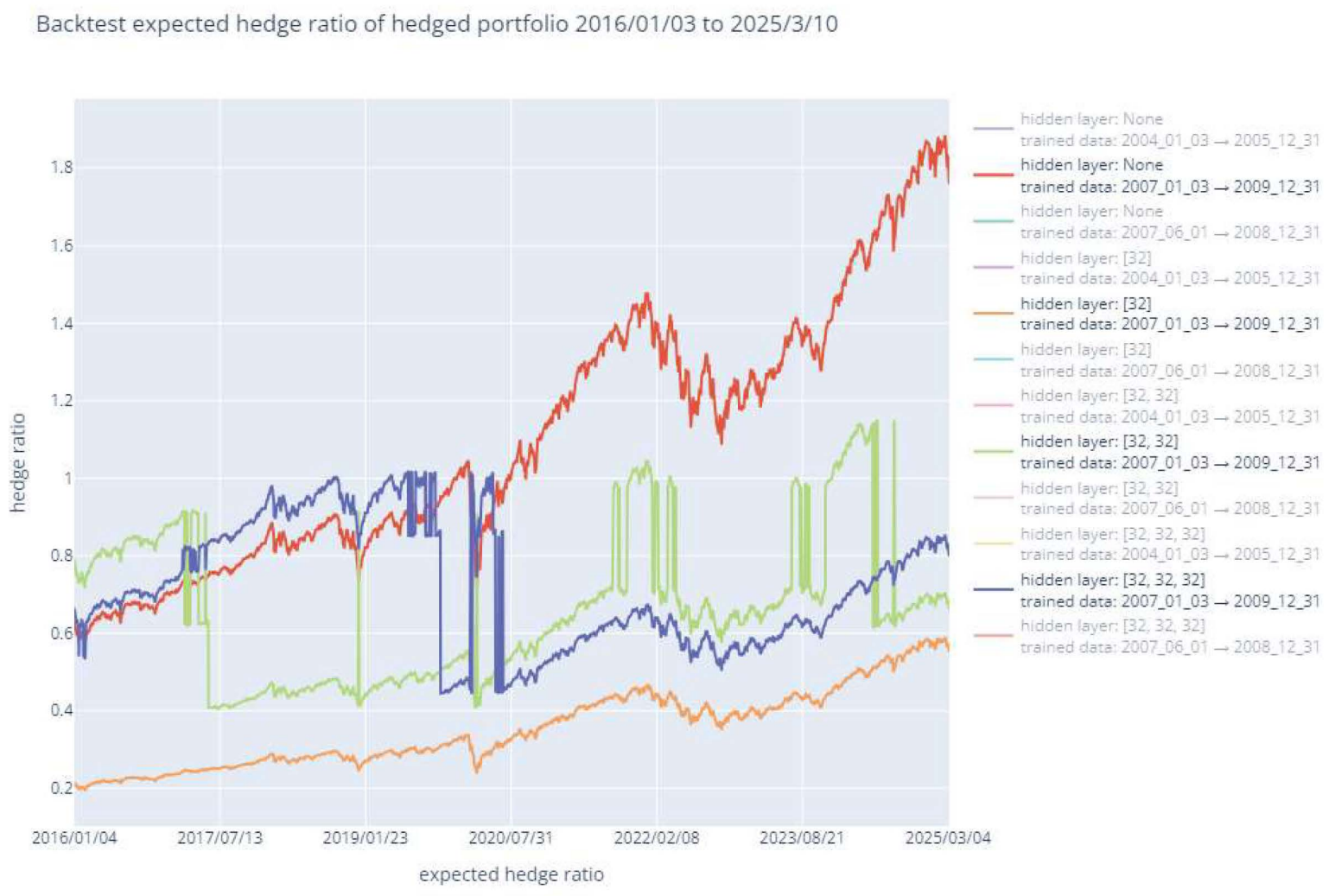

Figure 4.11: Expected hedge ratios are shown on this figure. The value on this figure means the amount of hedging portfolio, i.e., the position of shorting SPX index in this paper.

## 4.3 Back-test Performance via Training MLPs with Empirical Data from 2007/06/01 to 2008/12/31

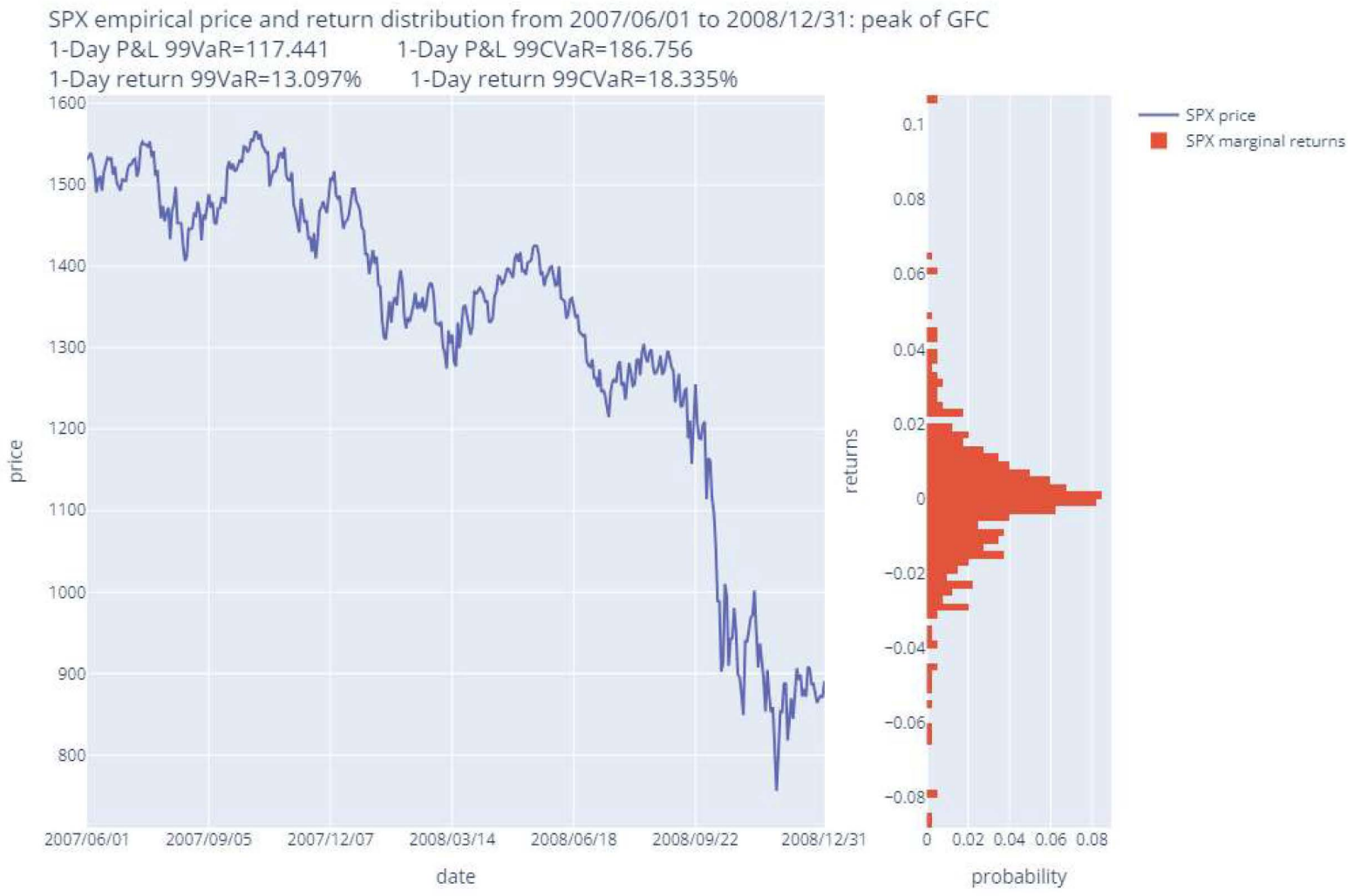

Figure 4.12: SPX close price and its marginal return distribution from empirical data of time period 2007/06/01 to 2008/12/31

Here we used SPX close price and its marginal return distribution from 2007/06/01 to 2008/12/31 to train those four MLPs. Empirical data shows a narrow and multi-mode feature. There are two modes centered at 0% and -1.5%. This figure shows the peak of GFC that an intensive tail risk event was undergoing.

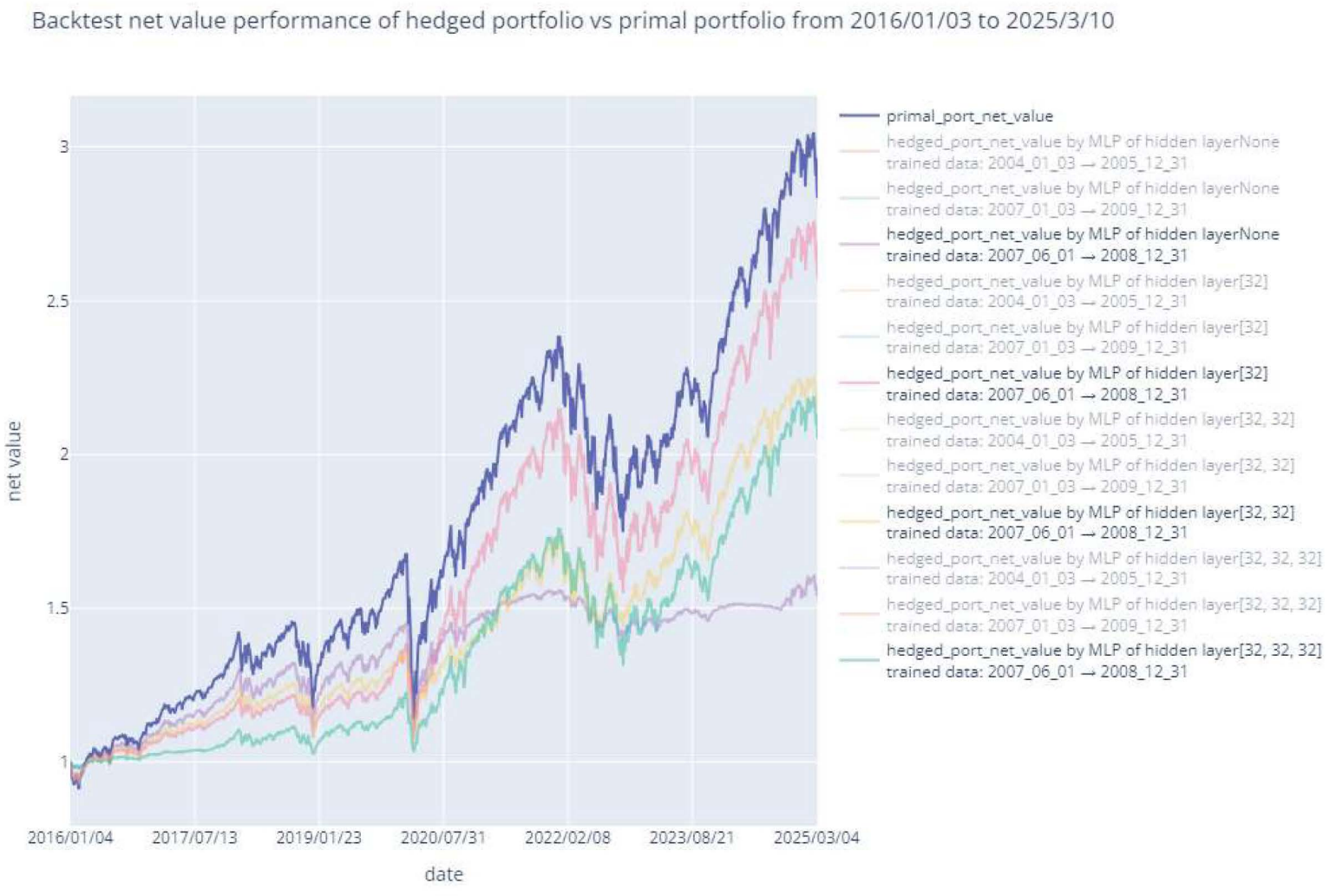

Figure 4.13: Back-test result of different MLP layer structures. Hedged portfolio net value benchmars at 2016/01/03. The backtest time horizon is from 2016/01/03 to 2025/03/10. Traning data is time horizon is from 2007/06/03 to 2008/12/31.

By learning from a the most volatile market empirical data during the peak of GFC, MLP optimizers better while number of hidden layers is still generally increasing. One outlier is the None_Hidden_Layer_MLP optimizer. Its hedged portfolio net value evolved the most stable with 1-Day 99VaR of 0.04 and 1-Day 99CVaR of 0.06 in net value changes, Fig.4.14. Although None_Hidden_Layer_MLP worked the best in this comparison, it shows still the more layers, the better MLP optimizer performance in tail risk hedging, Fig.4.14.

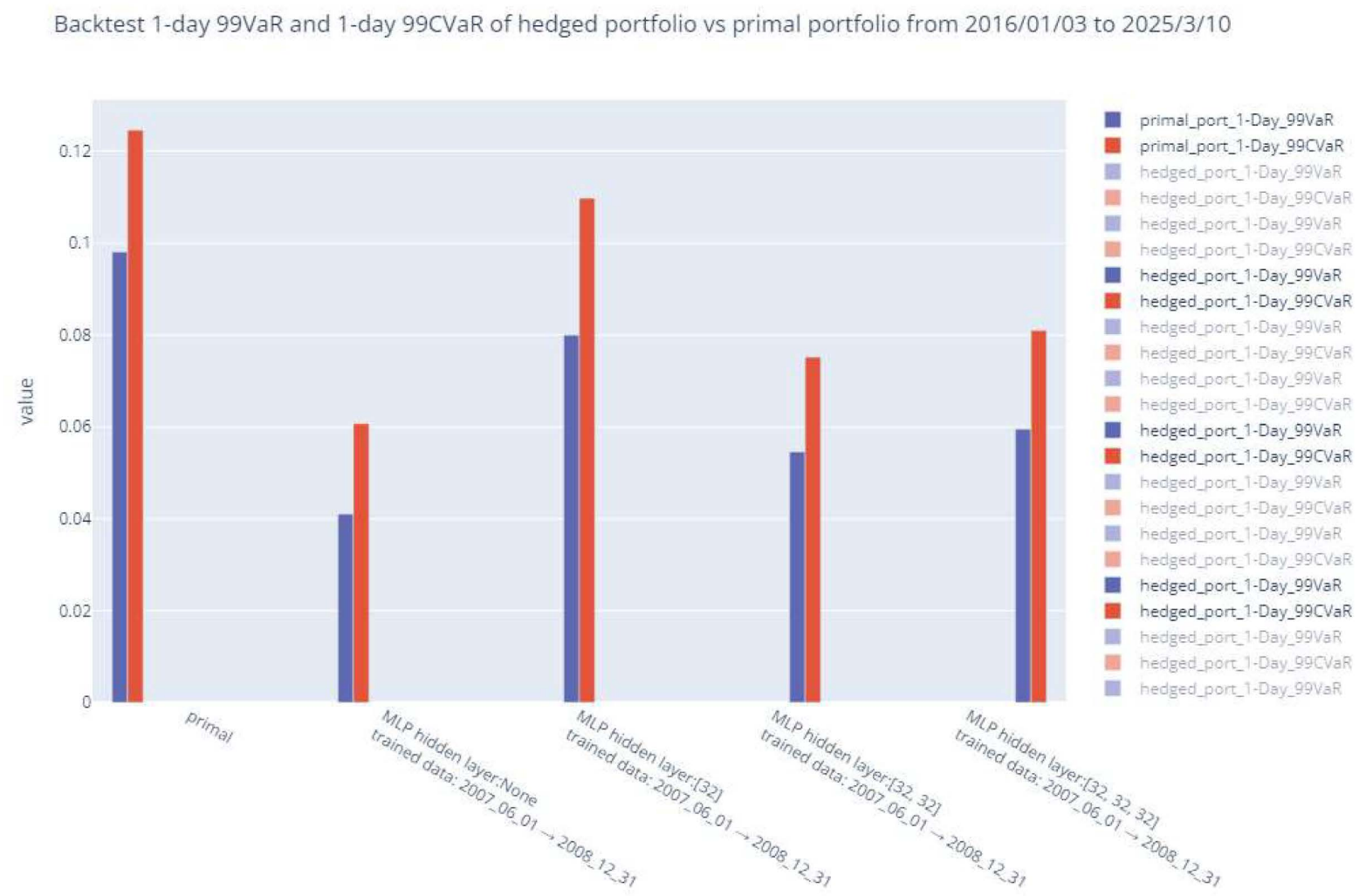

Figure 4.14: 1-Day 99VaR and 1-Day 99CVaR of primal portfolio and hedged portfolios net value changes. Training data set is from 2007/06/01 to 2008/12/31. Tail risk hedge results by the four MLP optimizers are shown on the for bar groups on the left.

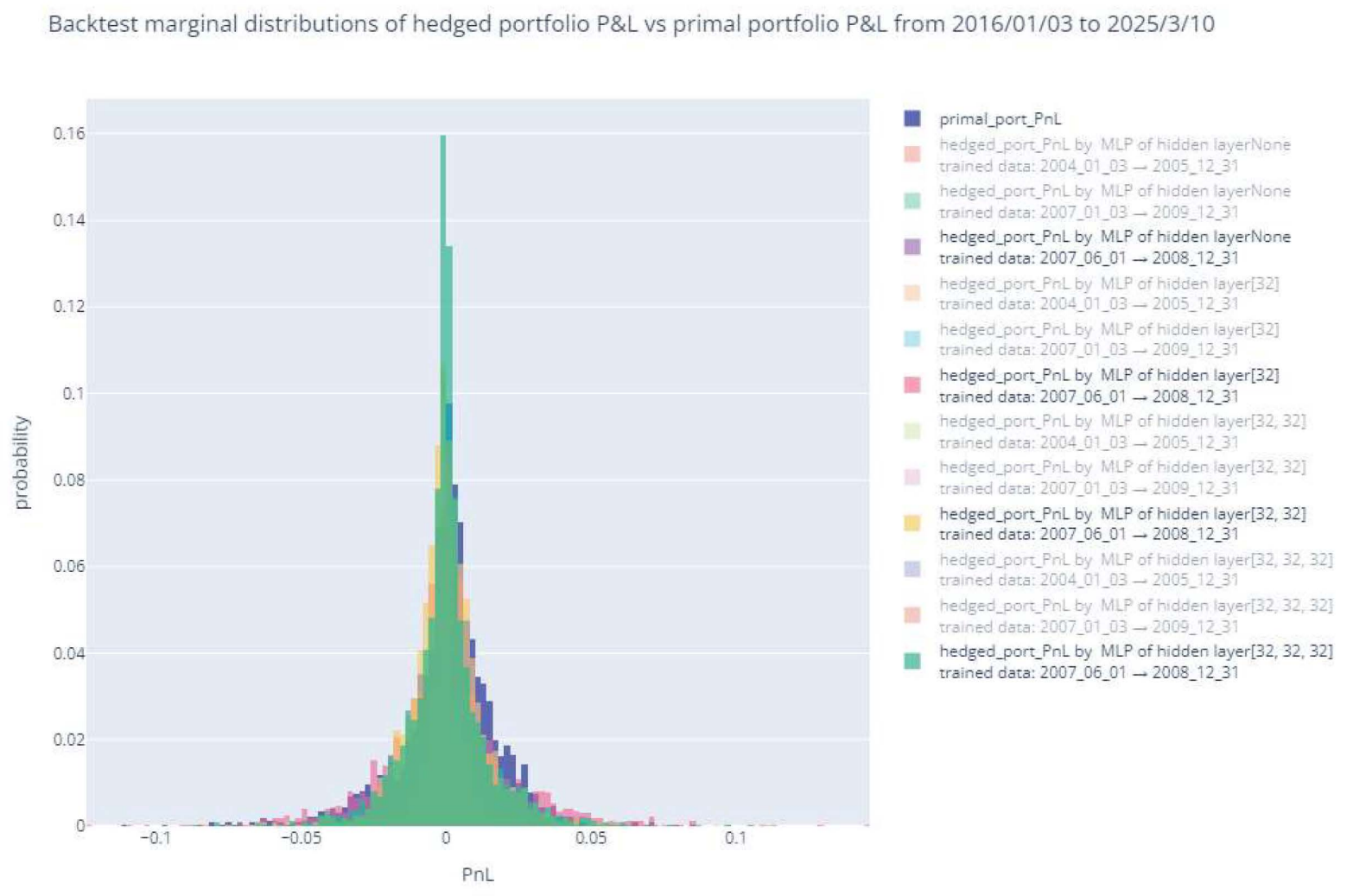

Figure 4.15: Histogram of primal and hedged portfolio P&Ls. The performance of MLP optimizers with different layer structures but learning from data of 2007/06/01 to 2008/12/31 is shown.

Fig.4.15 shows a similar result as the result from Fig.4.5 and Fig.4.10. However, remarking the distribution from the hedging performance of 32x32x32_MLP optimizer in green color, we found that learning from a severely volatile market regime worsen the performance of MLP optimizers on tail risk hedging problem.

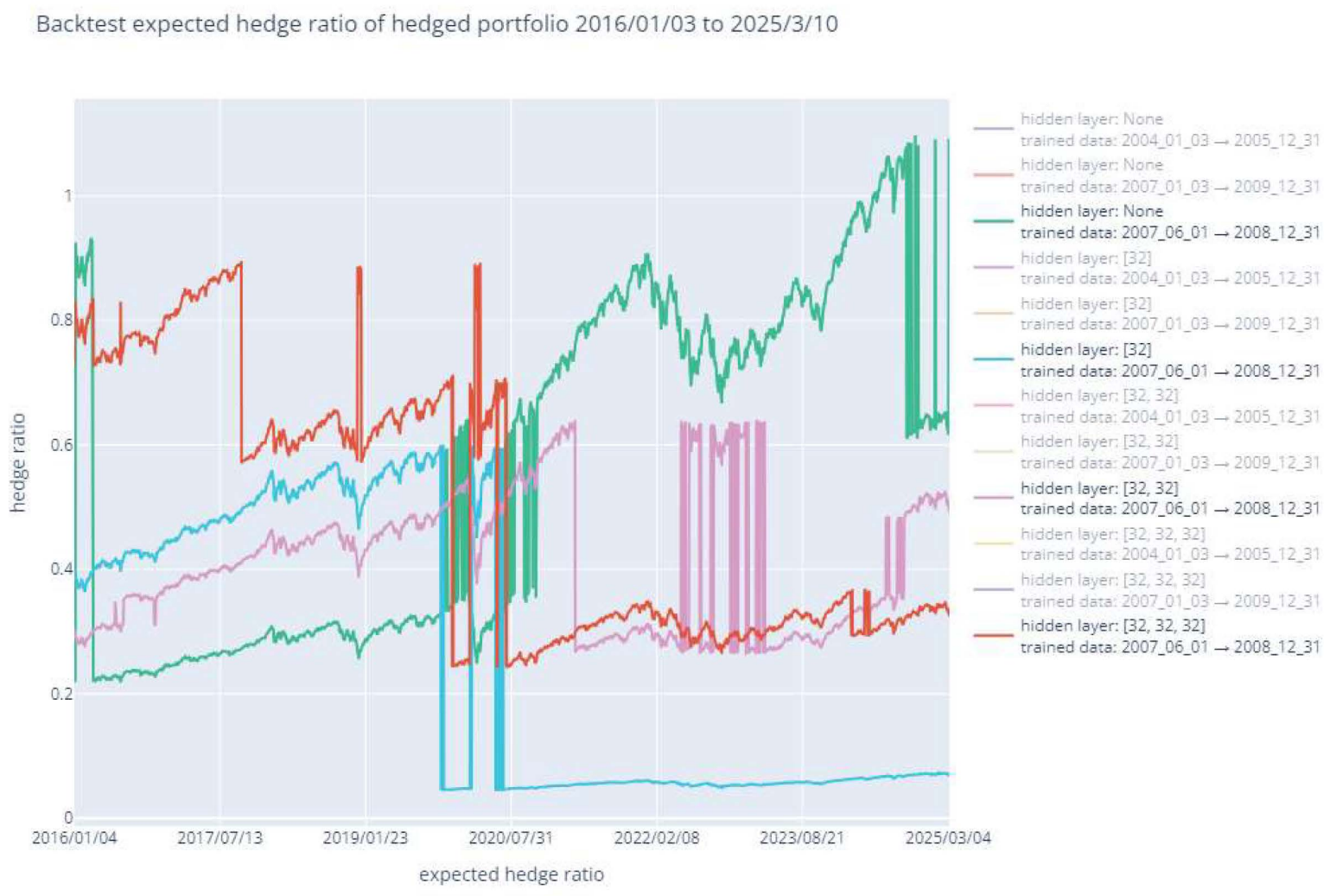

Figure 4.16: Expected hedge ratios are shown on this figure. The value on this figure means the amount of hedging portfolio, i.e., the position of shorting SPX index in this paper.

Fig.4.16 shows a similar result as it is given by Fig.4.6. Further, we found 32x32x32_MLP changed its hedge ratio to a low level when the tail risk event had passed. It is a desired feature for quantitative portfolio risk management.

## 4.4 Back-test Performance via None_Hidden_Layer_MLP w.r.t. Different Training Data Periods

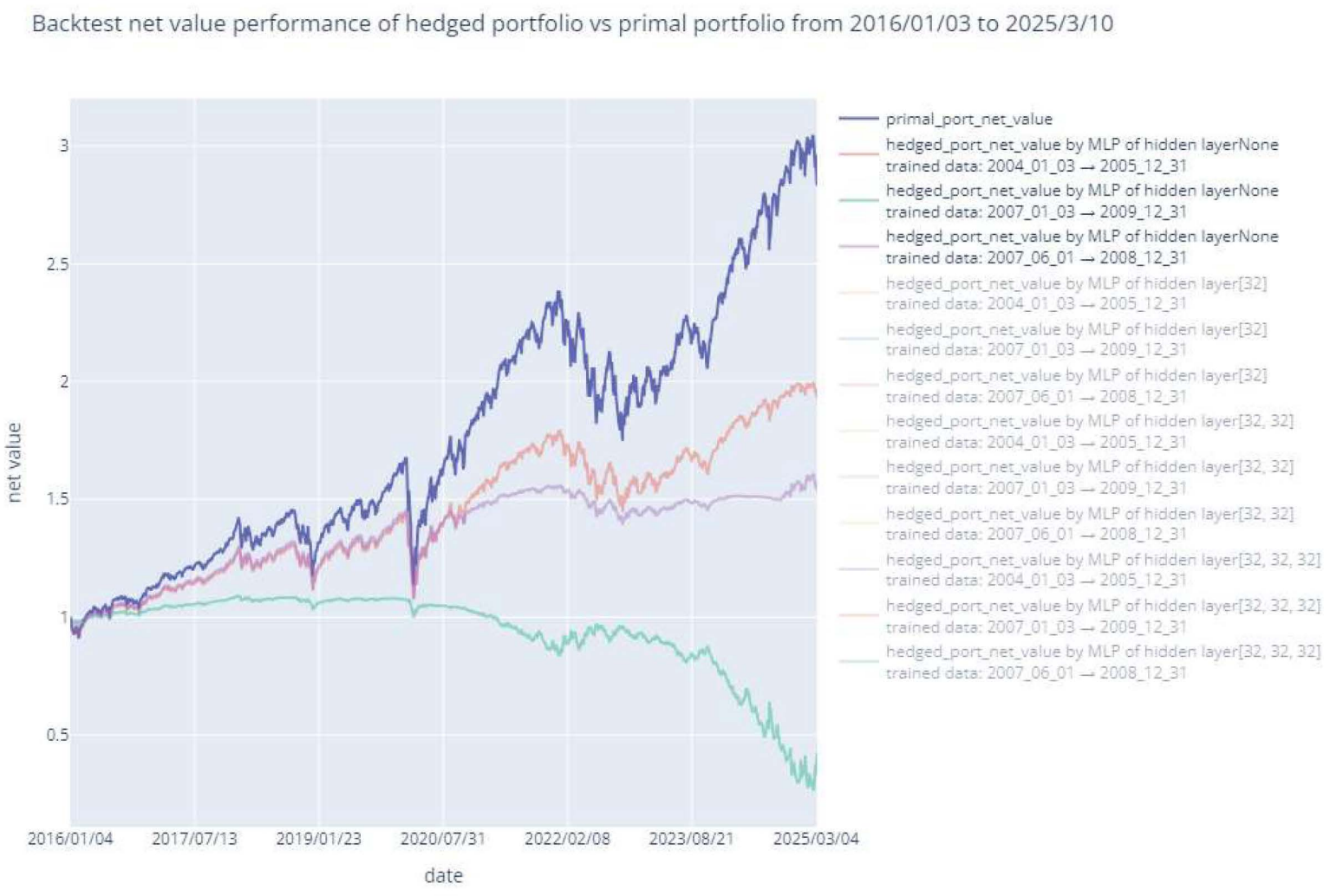

Figure 4.17: Back-test result of different traing data from different time period. Hedged portfolio net value benchmars at 2016/01/03. The back-test time horizon is from 2016/01/03 to 2025/03/10. It shows performance of None_Hidden_MLP by learning from different empirical data of periods from 2004/01/03 to 2005/12/31, from 2007/01/03 to 2009/12/31 and from 2007/06/01 to 2008/12/31.

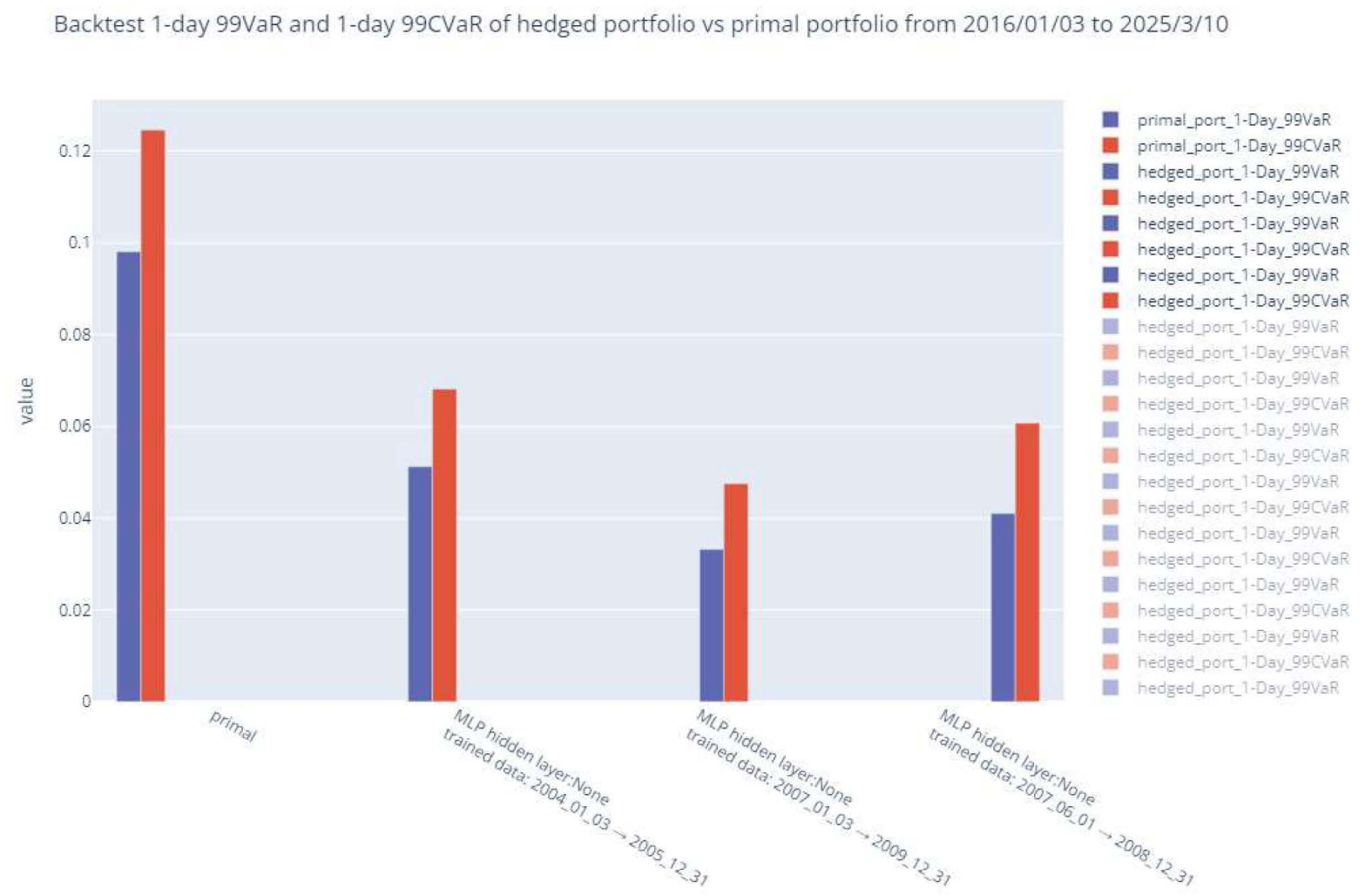

Figure 4.18: 1-Day 99VaR and 1-Day 99CVaR of primal portfolio and hedged portfolios net value changes. Training data sets are 2004/01/03 to 2005/12/31, 2007/01/03 to 2009/12/31 and 2007/06/01 to 2008/12/31. Tail risk hedge results by None_Hidden_Layer_MLP optimizers are shown.

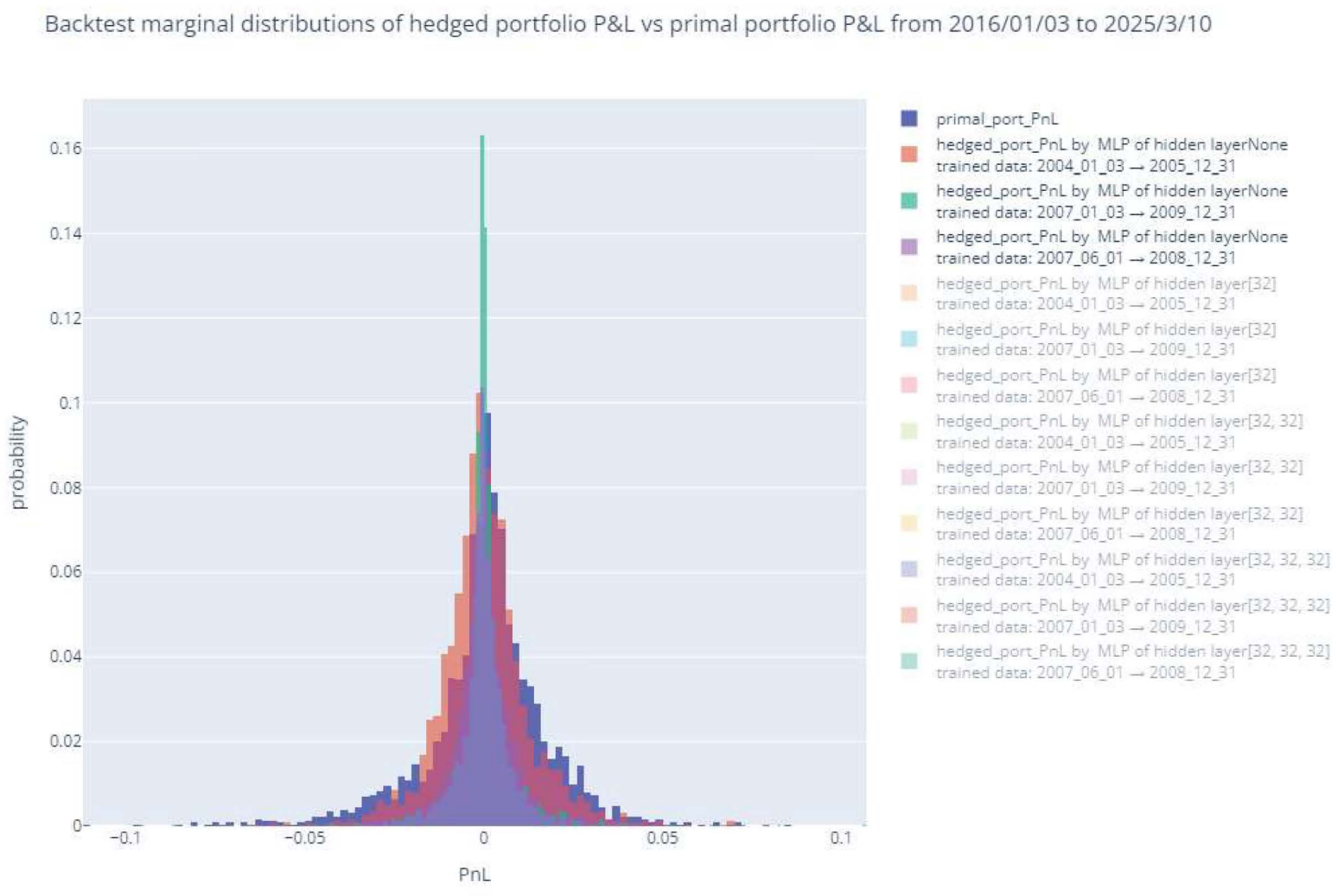

Figure 4.19: Histogram of primal and hedged portfolio P&Ls. The performance of MLP optimizers with the none hidden layer structure but learning from differnt empirical data of is shown.

Fig.4.19 shows that None_Hidden_Layer_MLP produces the thinnest tail on marginal distribution of hedged portfolio P&L. However, we also found that, comparing Fig.4.17, Fig.4.18, Fig.4.19 and Fig.4.20, a thinner-tailed marginal distribution may not be a sufficient condition to a useful MLP optimizer on tail risk hedging problem.

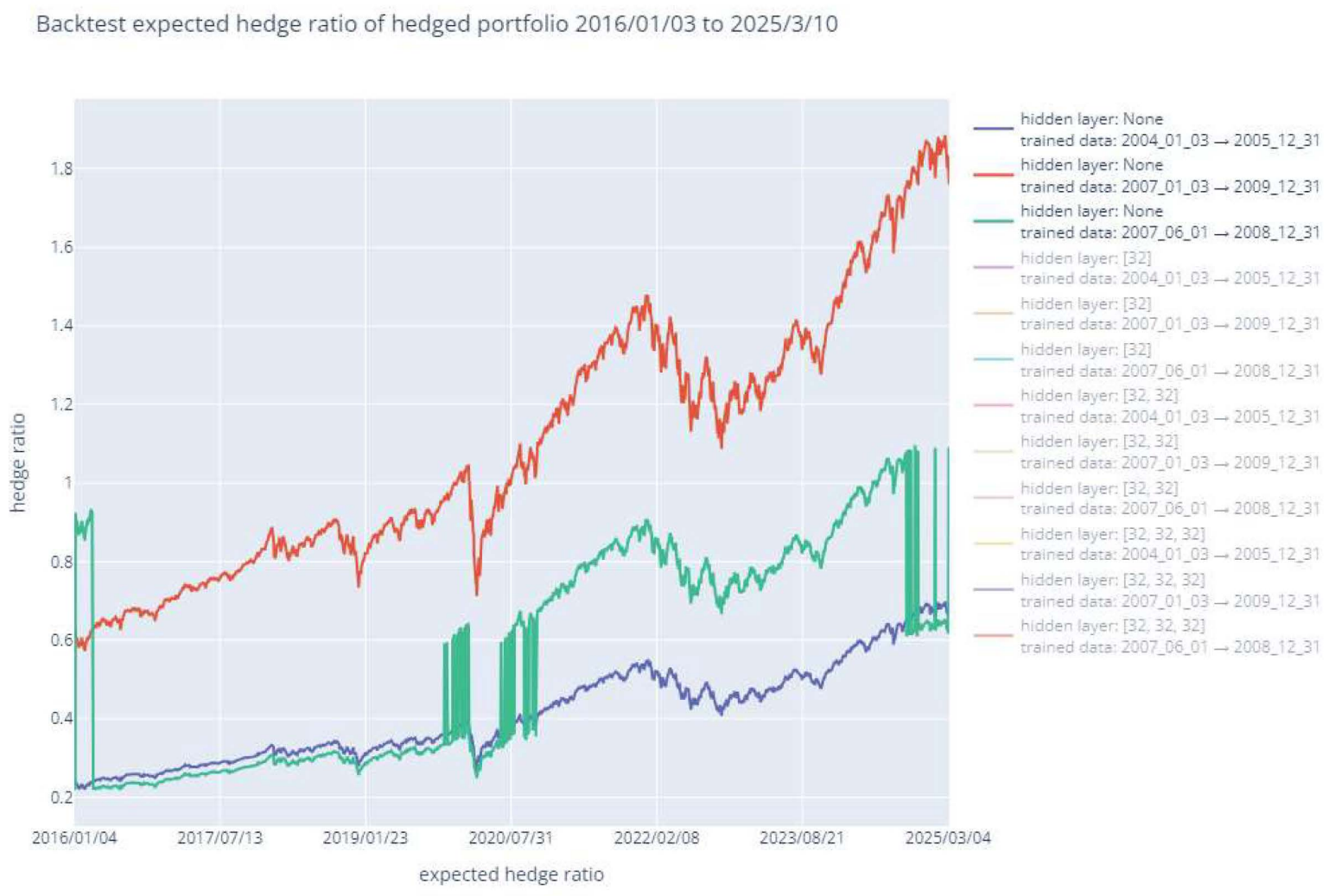

Figure 4.20: Expected hedge ratios are shown on this figure for None_Hidden_Layer_MLPs of learning different empirical data.

We found that None_Hidden_Layer_MLPs works the best by learning a relatively high volatile market empirical data, because it hedged the most when tail risk event realizes. Data from a peaceful period results in a under hedge result.

## 4.5 Back-test Performance via 32_MLP w.r.t. Different Training Data Periods

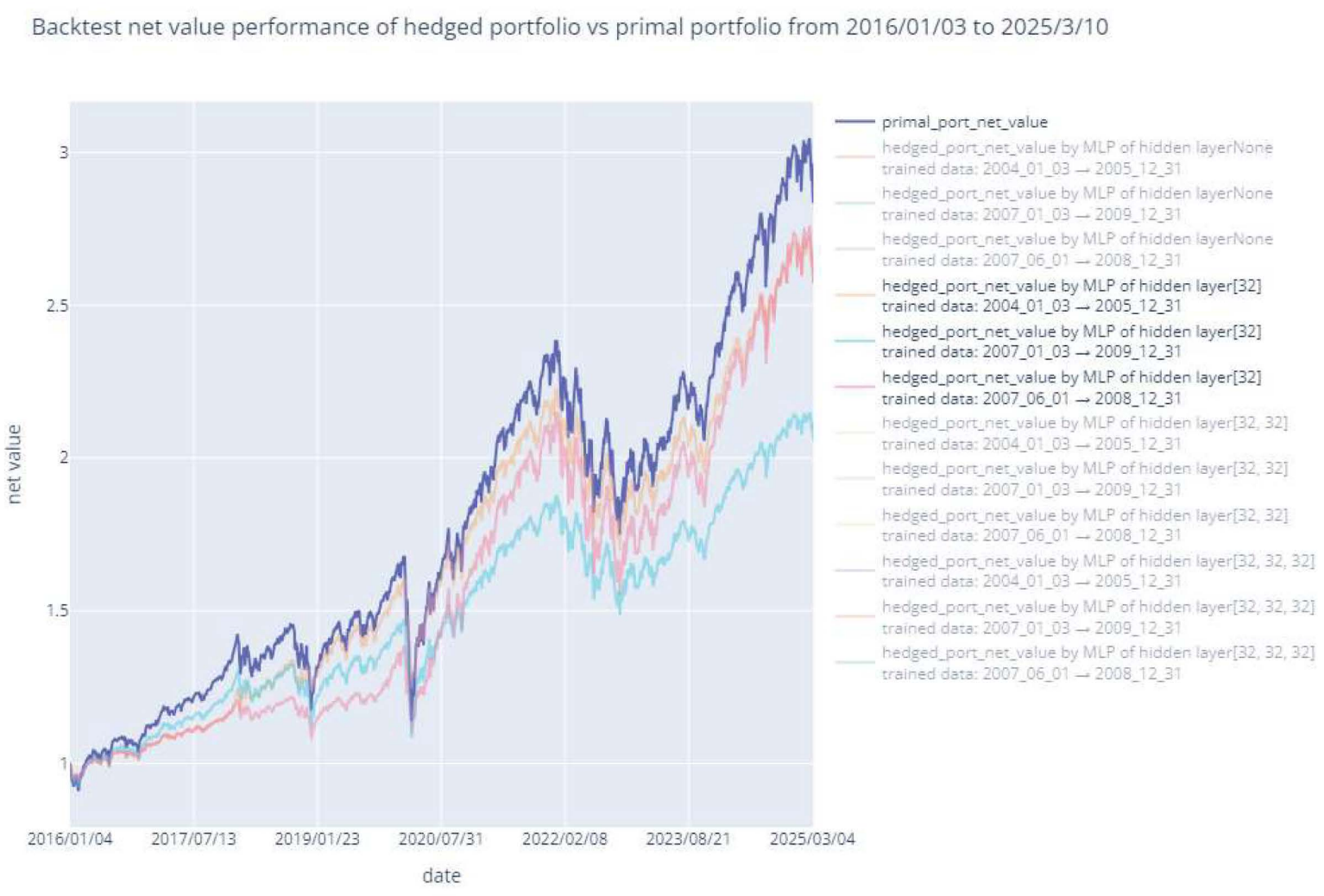

Figure 4.21: Back-test result of different traing data from different time period. Hedged portfolio net value benchmars at 2016/01/03. The backtest time horizon is from 2016/01/03 to 2025/03/10. It shows performance of 32_MLP by learning from different empirical data of periods from 2004/01/03 to 2005/12/31, from 2007/01/03 to 2009/12/31 and from 2007/06/01 to 2008/12/31.

The performance of applying 32_MLP is not efficient to reduce drawbacks during tail risk events. Fig.4.22 shows that 1-Day 99VaR and 1-Day 99CVaR are only reduced a relatively small amount.

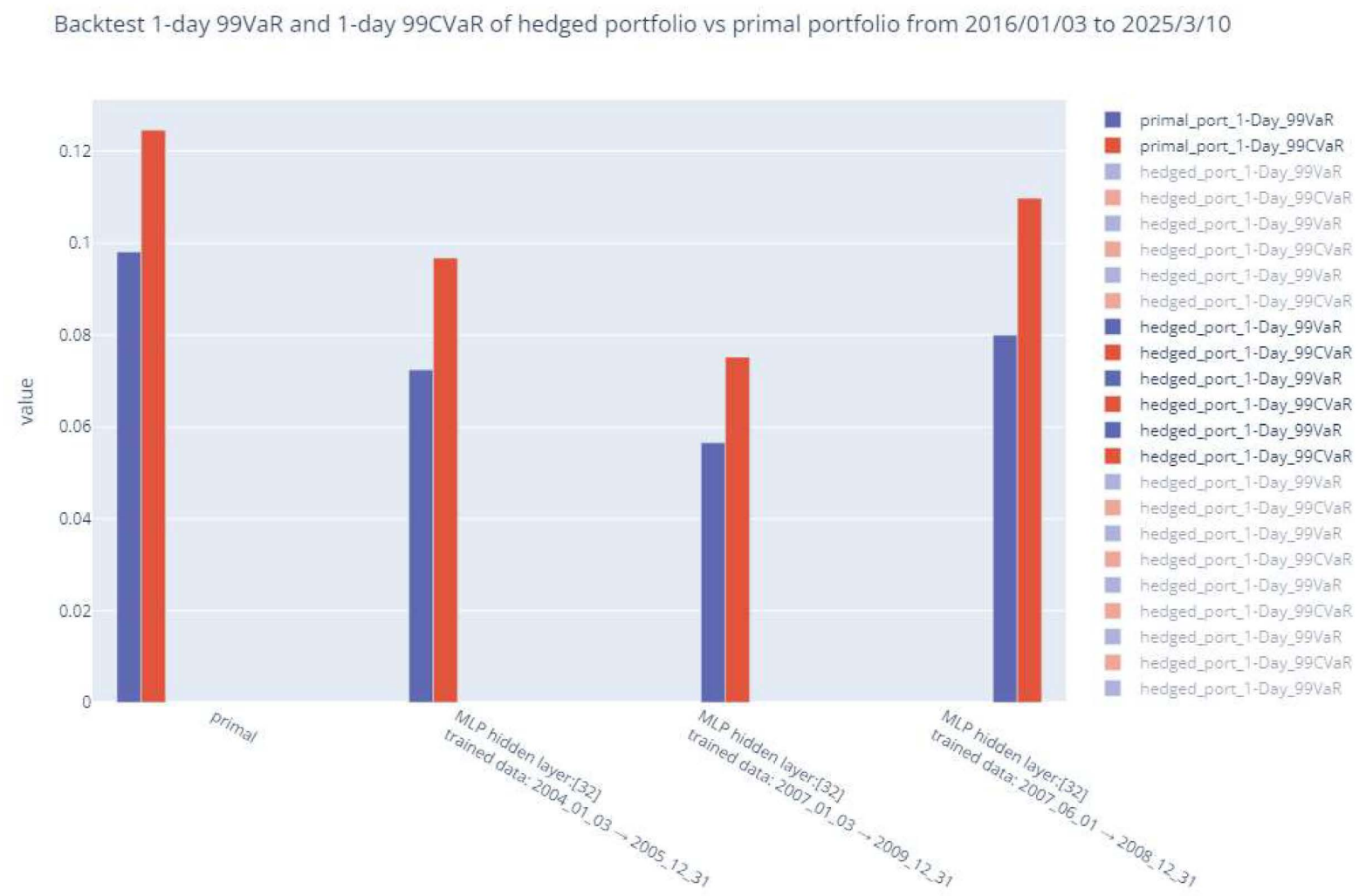

Figure 4.22: 1-Day 99VaR and 1-Day 99CVaR of primal portfolio and hedged portfolios net value changes. Training data sets are 2004/01/03 to 2005/12/31, 2007/01/03 to 2009/12/31 and 2007/06/01 to 2008/12/31. Tail risk hedge results by 32_MLP optimizers are shown.

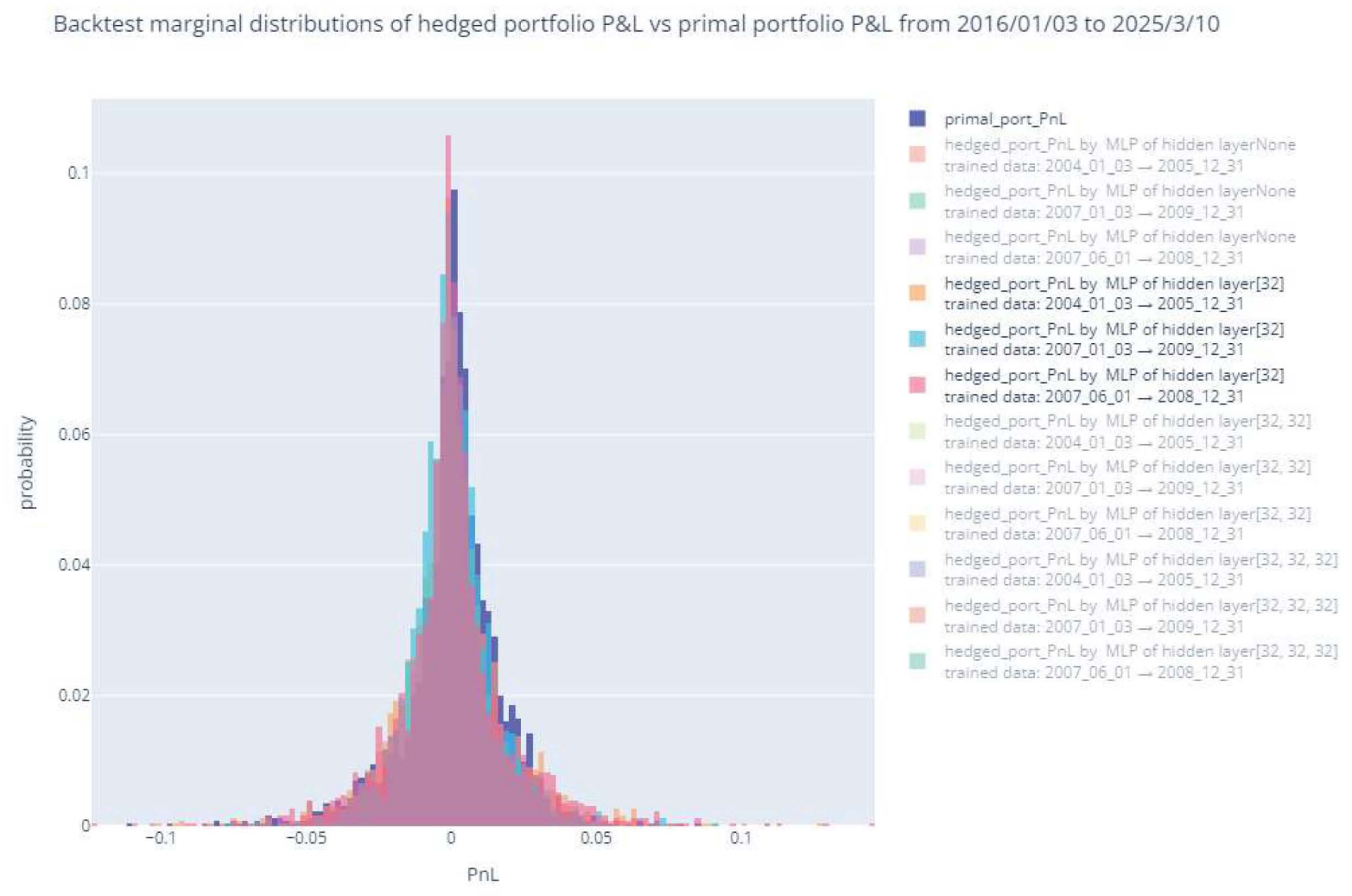

Figure 4.23: Histogram of primal and hedged portfolio P&Ls. The performance of MLP optimizers with the one 32-node hidden layer structure but learning from differnt empirical data of is shown.

We rechecked the result from Fig.4.22 by Fig.4.23. Marginal distributions of hedged portfolio P&L are not changed largely from the one of primal portfolio net value P&L.

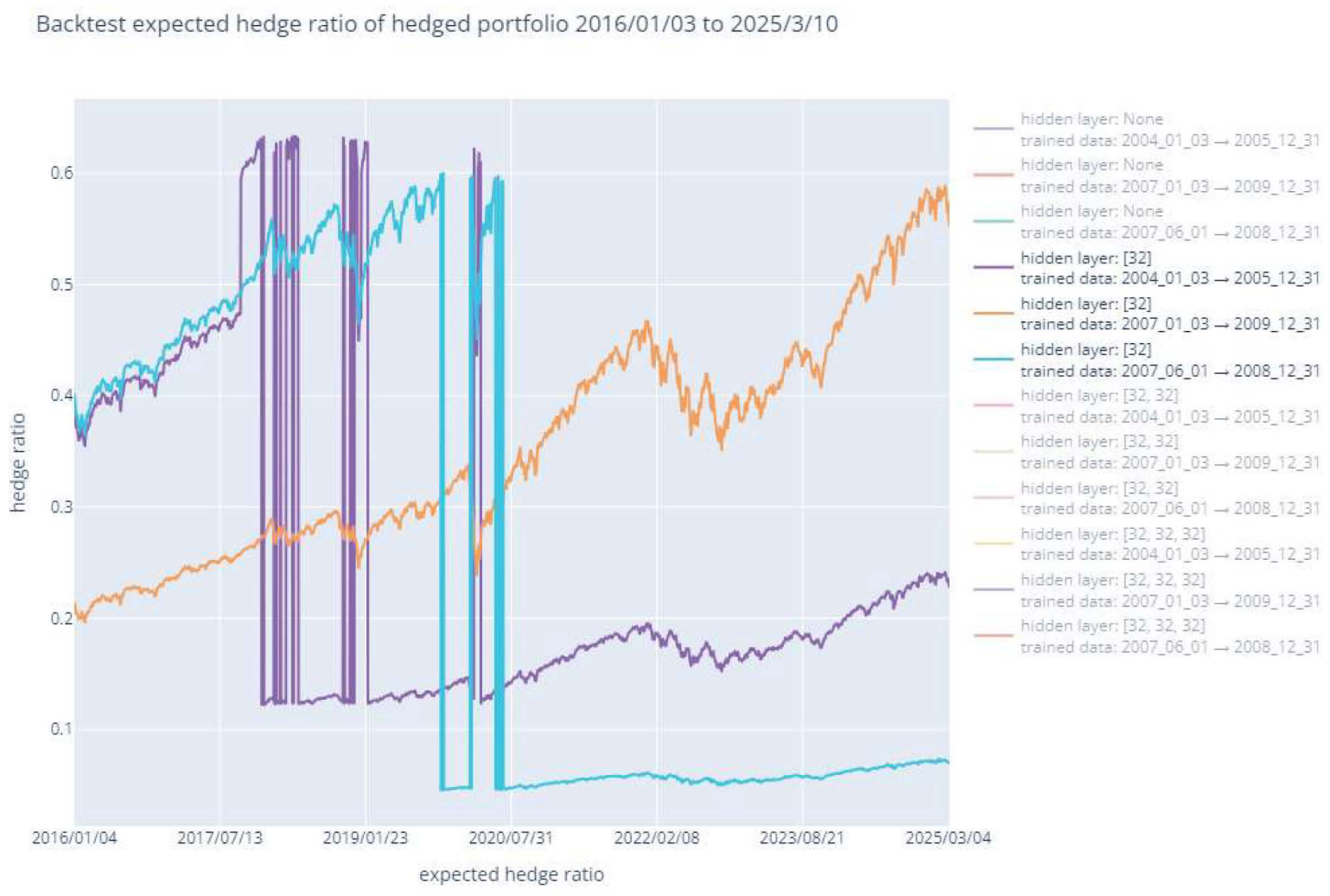

Figure 4.24: Expected hedge ratios are shown on this figure. The value on this figure means the amount of hedging portfolio, i.e., the position of shorting SPX index in this paper.

Although 32_MLP did not work efficiently, Fig.4.24 gives the same experimental result that the more volatile empirical market data is learnt, the more hedge ratio during tail risk events and the more swift change on hedge ratio according to market regime changes.

## 4.6 Back-test Performance via 32x32_MLP w.r.t. Different Training Data Periods

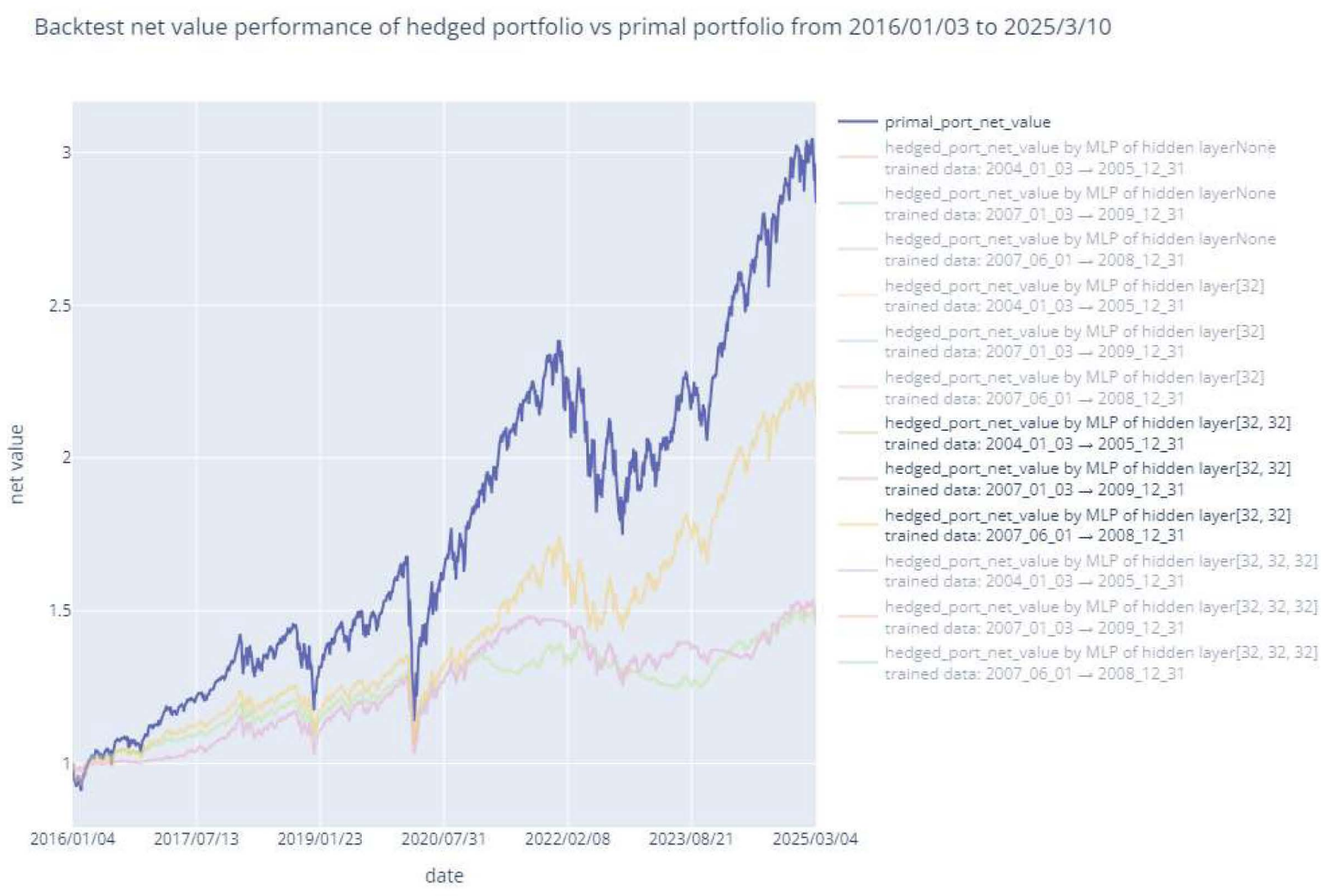

Figure 4.25: Back-test result of different traing data from different time period. Hedged portfolio net value benchmars at 2016/01/03. The backtest time horizon is from 2016/01/03 to 2025/03/10. It shows performance of 32x32_MLP by learning from different empirical data of periods from 2004/01/03 to 2005/12/31, from 2007/01/03 to 2009/12/31 and from 2007/06/01 to 2008/12/31.

Fig.4.25 reconfirms the result that, by learning from a more volatile market data, hedged portfolio becomes more stable.

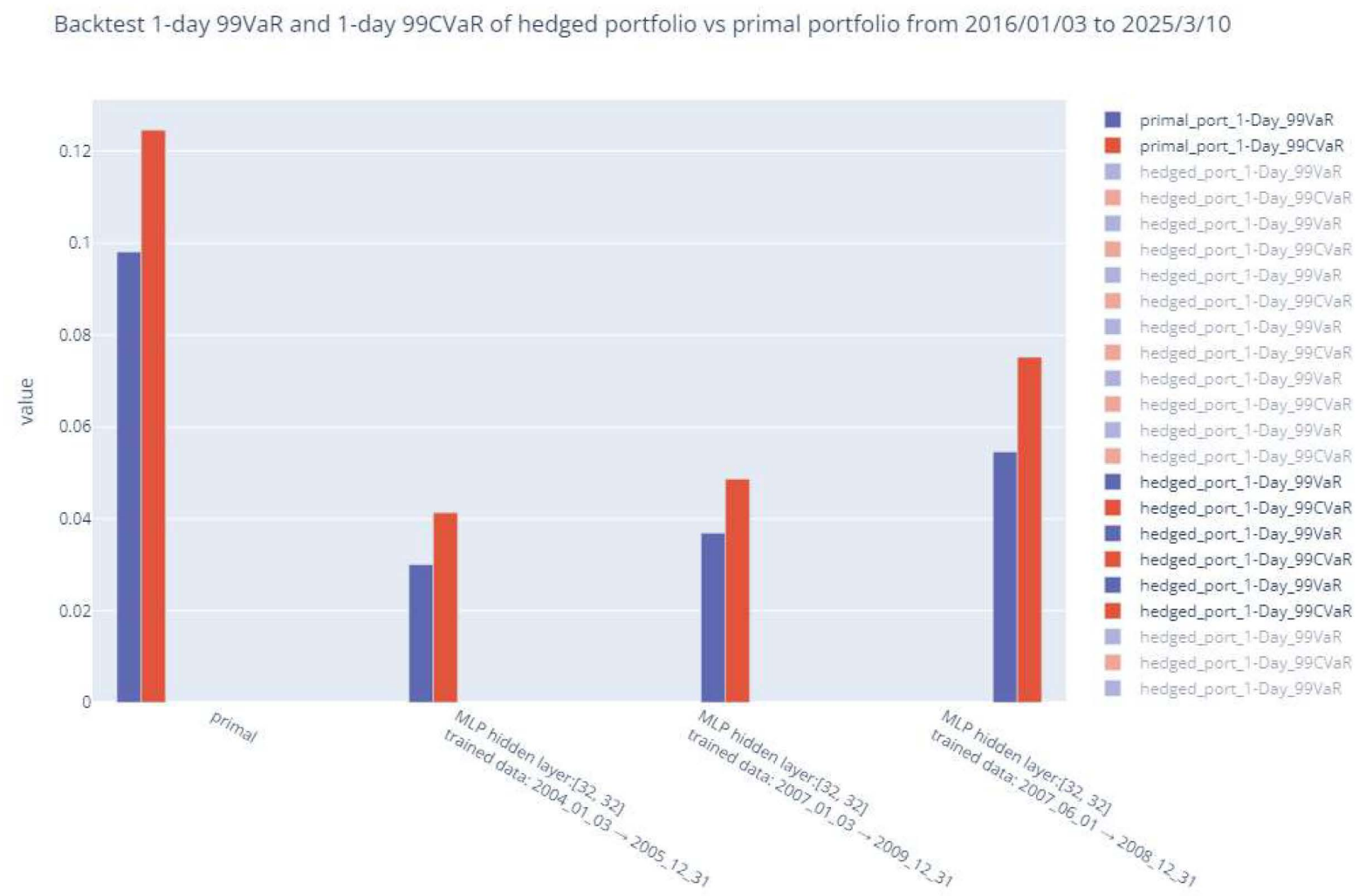

Figure 4.26: 1-Day 99VaR and 1-Day 99CVaR of primal portfolio and hedged portfolios net value changes. Training data sets are 2004/01/03 to 2005/12/31, 2007/01/03 to 2009/12/31 and 2007/06/01 to 2008/12/31. Tail risk hedge results by 32x32_MLP optimizers are shown.

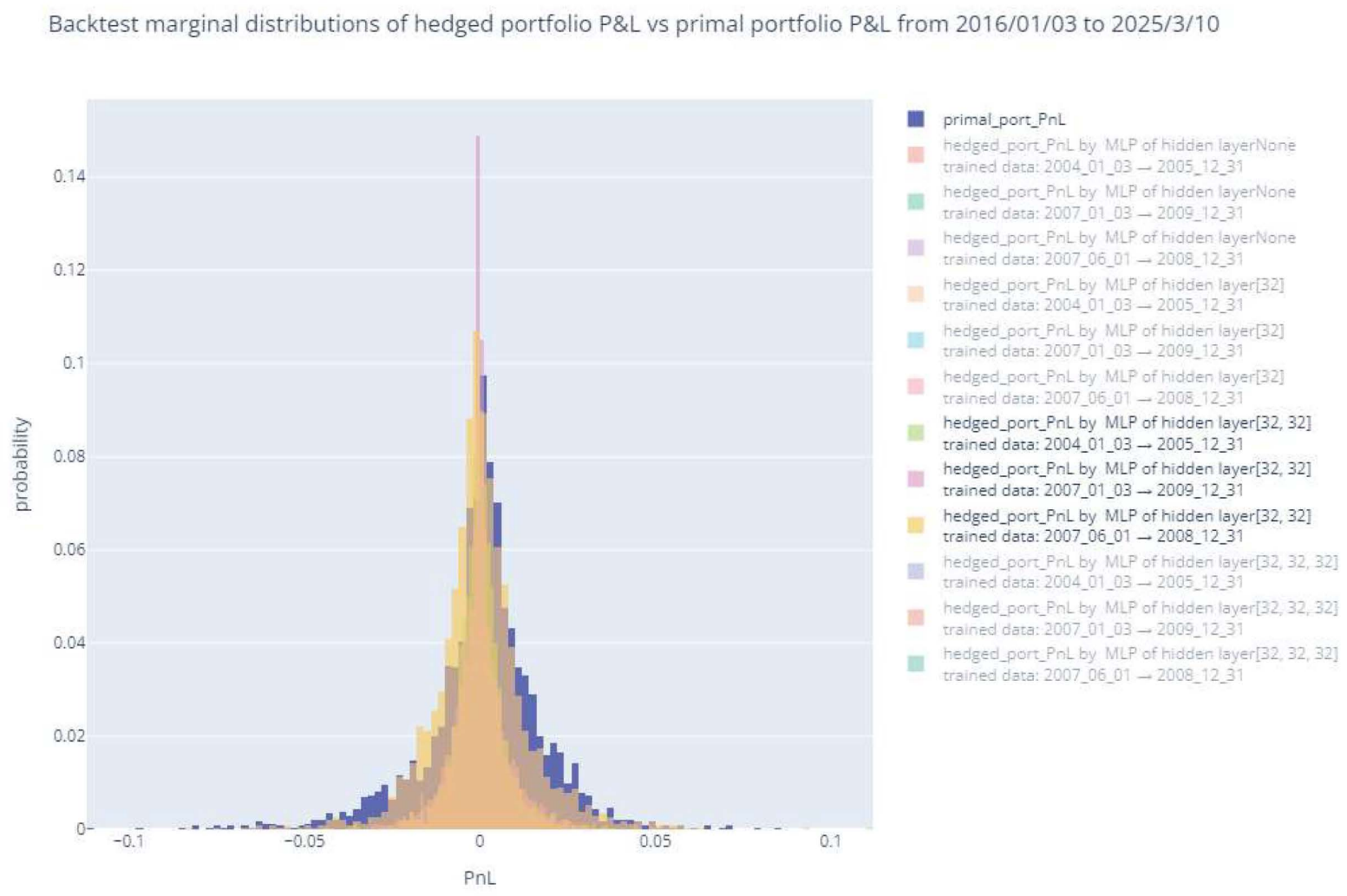

Figure 4.27: Histogram of primal and hedged portfolio P&Ls. The performance of MLP optimizers with the two 32-node hidden layer structure but learning from differnt empirical data of is shown.

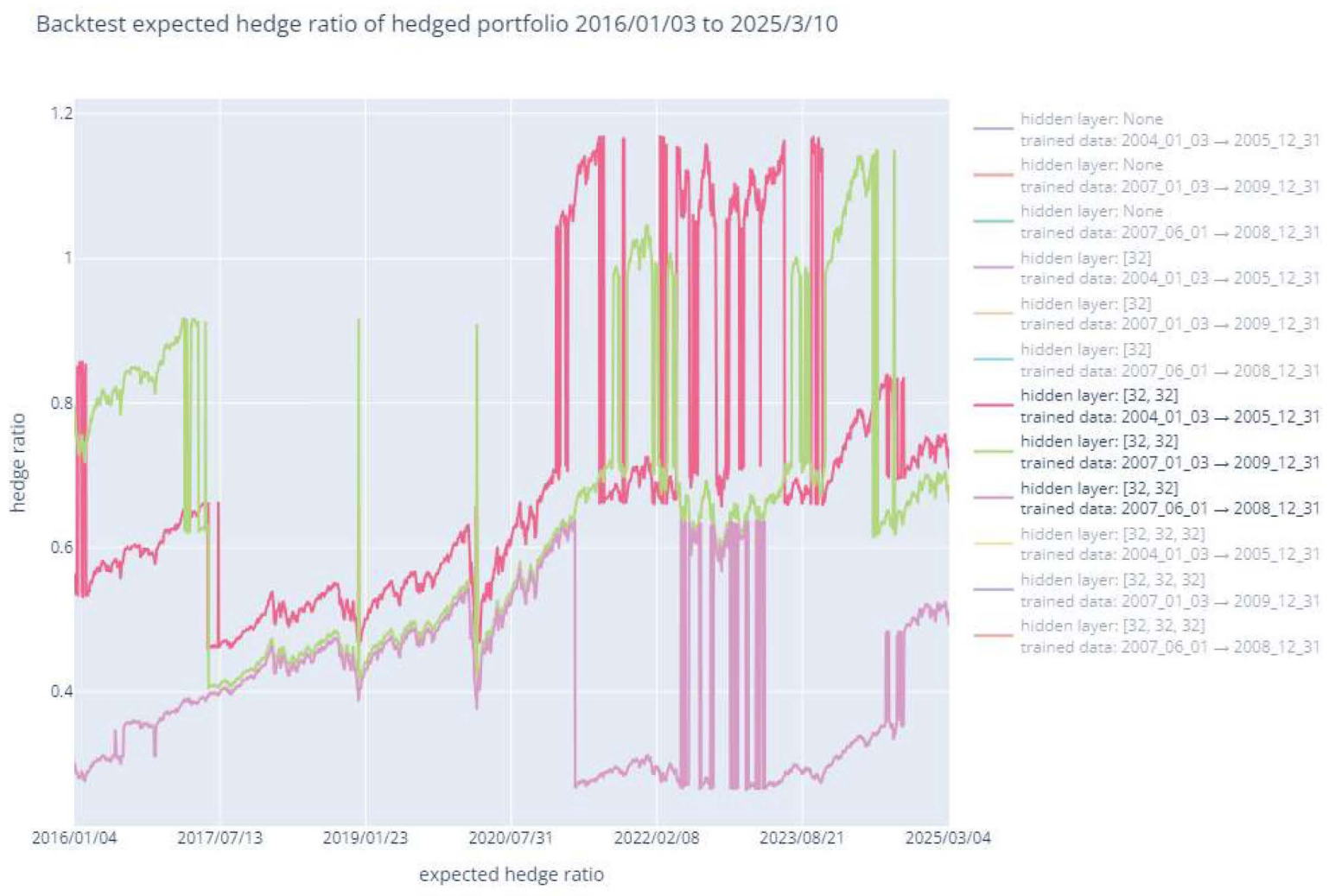

Figure 4.28: Expected hedge ratios are shown on this figure. The value on this figure means the amount of hedging portfolio, i.e., the position of shorting SPX index in this paper.

## 4.7 Back-test Performance via 32x32x32_MLP w.r.t. Different Training Data Periods

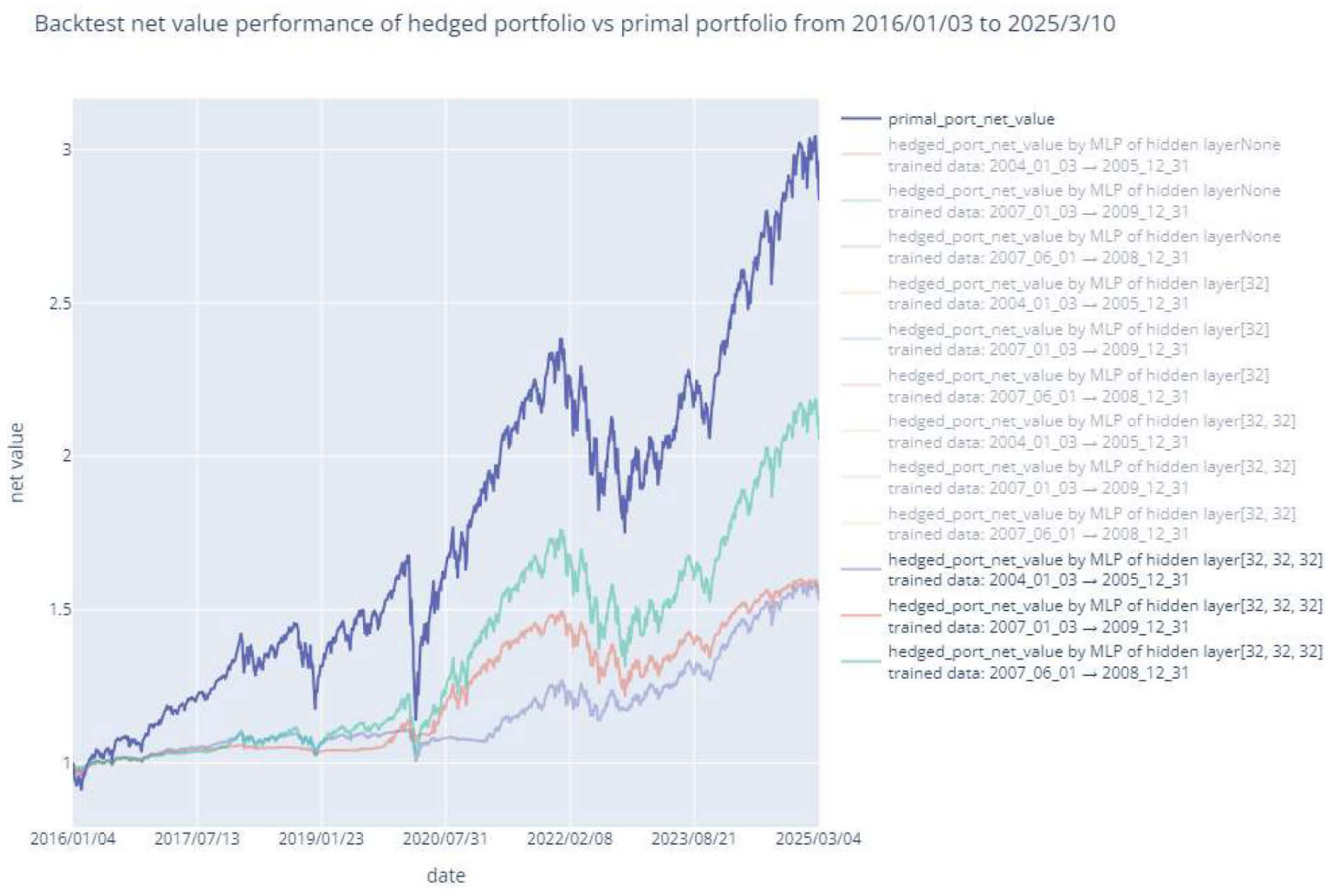

Figure 4.29: Back-test result of different traing data from different time period. Hedged portfolio net value benchmars at 2016/01/03. The backtest time horizon is from 2016/01/03 to 2025/03/10. It shows performance of 32x32x32_MLP by learning from different empirical data of periods from 2004/01/03 to 2005/12/31, from 2007/01/03 to 2009/12/31 and from 2007/06/01 to 2008/12/31.

Fig.4.29 shows that with learning from more volatile market data, 32x32x32_MLP works more profit generative while keeping tail risk hedge performance.

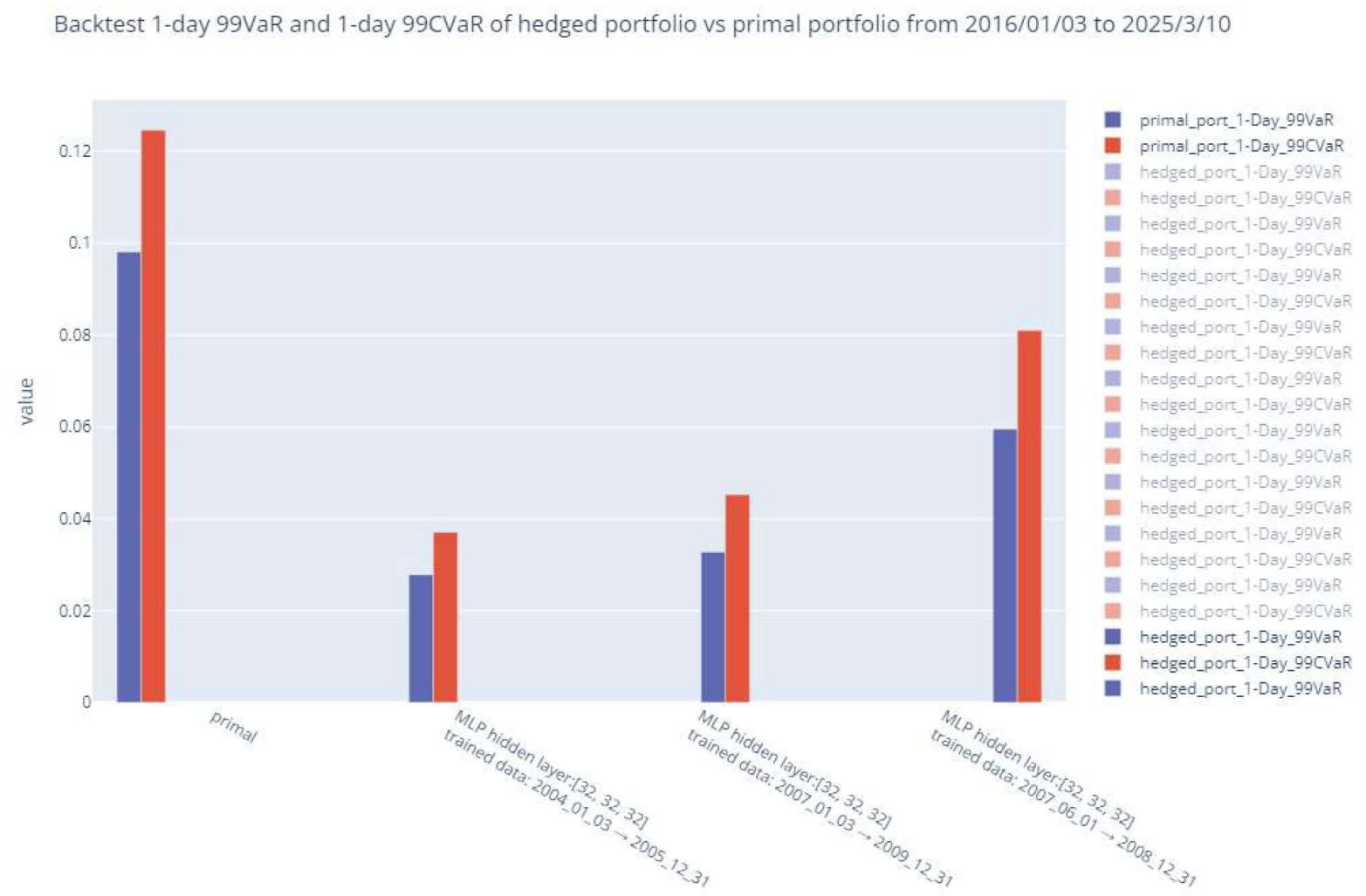

Figure 4.30: 1-Day 99VaR and 1-Day 99CVaR of primal portfolio and hedged portfolios net value changes. Training data sets are 2004/01/03 to 2005/12/31, 2007/01/03 to 2009/12/31 and 2007/06/01 to 2008/12/31. Tail risk hedge results by 32x32x32_MLP optimizers are shown.

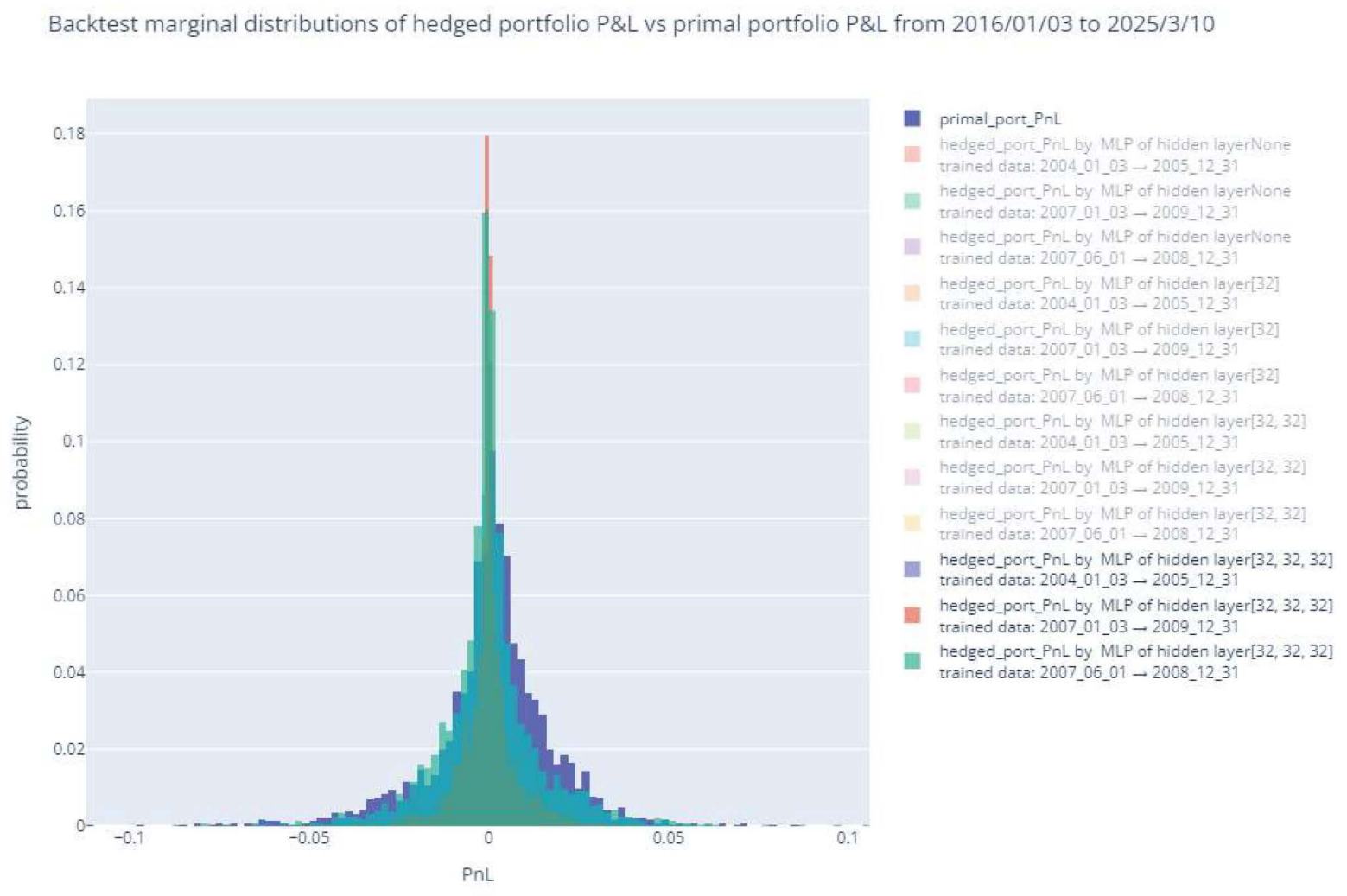

Figure 4.31: Histogram of primal and hedged portfolio P&Ls. The performance of MLP optimizers with the three 32-node hidden layer structure but learning from differnt empirical data of is shown.

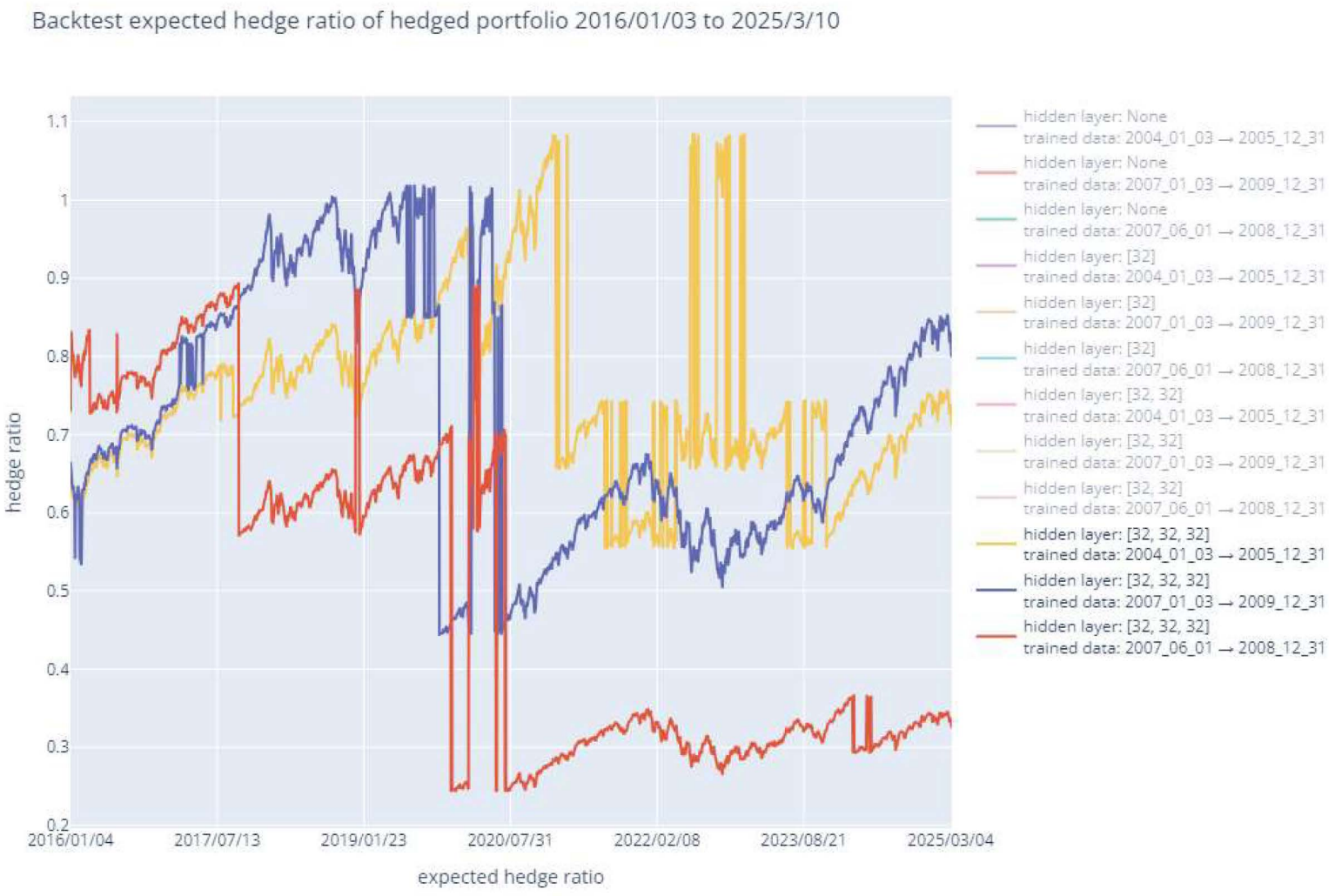

Figure 4.32: Expected hedge ratios are shown on this figure. The value on this figure means the amount of hedging portfolio, i.e., the position of shorting SPX index in this paper.

# 5 Conclusion and Discussion

We concluded from numerical experiments that our approach reduces tail risk measure, 1-Day 99CVaR, efficiently. Generally, performance of MLP optimizer increases on tail risk hedging problem with more hidden layers. On the other hand, MLP optimizer learning from a relatively more volatile market empirical data leads to a stronger profit generative performance while remaining tail risk hedging performance without deterioration.

For financial practitioners, it is guided by numerical experiments that choosing empirical data with proper volatility is essential to enhance tail risk hedging performance of our approach. Adding more hidden layers in MLP optimizer may output a better hedged portfolio performance, however, it is not a sufficient condition to achieve a goal. Therefore, it is needed for practitioners to numerically test MLP optimizers comprehensively before employing our approach into real trading and portfolio management procedure. Model risk is still a vital concern, but traditional quantitative financial methods face the same risk in training desk workflows.

There is a disadvantage of our NN approach to tail risk hedging problem. If one chooses two highly co-integrated portfolios as its primal portfolio and hedging portfolio, it is obvious that the minimum convex tail risk measure will be achieved by adjusting hedge ratio to offset the primal portfolio market value changes and hedging portfolio changes. Intentionally, a full hedge strategy would be the best tail risk hedging policy. However, it is not what we are expecting. It is ideal to us, financial market practitioners, that the method outputs a solution balancing tail risk hedging and profit generating.

The solution to this disadvantage is trivial. Choose a hedging portfolio from lower-co-integrated portfolios to the primal portfolio. For example, hedge SPX with NASDAQ Index (Tiker: IXIC) or Dow Jones Index (Ticker: DJI) instead of SPX itself, as well as put options. For using options to hedge tail risks, in our approach, one needs to calculate dummy put options on bootstrapped path space and set their strike price as, e.g., 95% of the initial price. Alternatively, cut training iteration rounds before the NN optimizer reaches the pitfall mentioned above. One can choose both or either of them.

To be summarized, our approach is efficient on addressing tail risk hedging problem. It is model-free via bootstrapping method to create real-world-featured stochastic paths. It is salable due to low computation resource requirement. We did our numerical experiment with only one RTX4090 GPU. Our approach is highly customization, one can introduce enhancing features, e.g., trading costs, tax policies, internal risk appetite policies, market regulator's guidelines and etc.

Future works are extending our approach on portfolios with multiple assets and hedged by lower-co-integrated portfolios, i.e., proxy hedging. Additionally, correlation features into bootstrapping workflow can be refined when it comes to multiple asset portfolio cases. Tail risk hedging performance of our approach on strategy adopted with non-linear products, e.g., put options, basket options, look-back options or other exotic derivatives, is remained to be numerically investigated.